%% file: paper.tex
\documentclass[letterpaper,twocolumn,10pt]{article}
\usepackage{usenix}

\usepackage{hyperref}
\usepackage{graphicx}
\usepackage{amsmath}
\usepackage[font=bf]{caption}
\usepackage{subcaption} 
\usepackage{enumitem}
\usepackage{xurl}
\usepackage{xspace} 
\newcommand{\sysname}{ReVo}

\newcommand{\netlayer}{L3}
\newcommand{\applayer}{L7}

\newcommand{\grace}{\emph{GRACE}}
\newcommand{\tambur}{\emph{Tambur}}

\begin{document}

\date{}

\title{\Large \bf ReVo: A Cross-Layer Reliable Volumetric Videoconferencing System}

\author{
{\rm Ankur Aditya$^{1*}$ \quad Diptyaroop Maji$^{1*}$ \quad Lingdong Wang$^{1}$} \\
{\rm Bhavya Ramakrishna$^{2}$ \quad Ramesh Sitaraman$^{1,3}$ \quad Prashant Shenoy$^{1}$} \\[1ex]
$^{1}$University of Massachusetts Amherst \quad $^{2}$Dolby Labs \quad $^{3}$Akamai Tech \\[1ex]
{\normalsize \{aaditya, dmaji, lingdongwang, ramesh, shenoy\}@cs.umass.edu} \\
{\normalsize Bhavya.Ramakrishna@dolby.com}
}

\maketitle
\begingroup
\renewcommand\thefootnote{*}
\footnotetext{Student authors with equal contribution.}
\endgroup
\input{0-abstract}

\input{1-introduction-new}

\input{2-background}

\input{3-system-design-new}

\input{4-implementation}
\input{5-evaluation}
\input{6-related-work}

\input{8-conclusions}

\bibliographystyle{plain}
\bibliography{paper}
\newpage
\appendix

\input{A2-appendix}

\input{A3-appendix}
\input{A4-appendix}

\input{A5-appendix}

\end{document}

%% file: 0-abstract.tex
\begin{abstract}
Volumetric videoconferencing enables immersive six Degrees of Freedom interactions by jointly transmitting visual appearance and 3D geometry. However, delivering volumetric video over today’s networks remains challenging due to high bandwidth demands, strict real-time latency constraints, and frequent packet loss. Packet loss not only degrades visual quality but also corrupts geometric structure, leading to severe artifacts and video freezes that significantly degrade Quality of Experience. Existing solutions either optimize volumetric videos assuming reliable networks or focus on loss recovery for 2D video, and are insufficient for volumetric videoconferencing.
In this paper, we present \sysname{}, a loss-resilient volumetric videoconferencing system that jointly recovers RGB and depth content under packet loss while meeting real-time constraints on desktop-grade hardware. \sysname{} leverages the insight that effective recovery requires a cross-layer, modality-aware design. It decouples volumetric video into RGB and depth streams, selectively protects critical content using network-layer FEC, and reconstructs corrupted non-critical frames using a post-decode neural recovery module. \sysname{} is implemented end-to-end over WebRTC and supports both traditional and neural video codecs. Our evaluations using real-world loss traces show that \sysname{} improves median SSIM by up to $32\%$ (resp. 13\%) for RGB (resp. depth) content and reduces video freezes by up to $95.7\%$ compared to existing techniques.

\end{abstract}

%% file: 1-introduction-new.tex
\section{Introduction}
\label{sec:introduction}
Volumetric videos represent scenes in three dimensions, enabling viewers to freely change their viewpoint with six Degrees of Freedom (6-DoF)~\cite{volumetric-video, volumetric-video2, volumetric-video3}. Unlike traditional 2D videos, which render a fixed viewpoint, volumetric videos allow viewers to observe scenes from arbitrary angles and directions. For example, participants in a volumetric videoconferencing session can feel as if they are seated in an actual physical space, facing or sitting next to each other. They can also move around during a session, experiencing natural viewpoint changes that resemble reality. Together, these allow volumetric videos to enable more immersive and natural interactions than 2D videos. As a result, applications such as volumetric videoconferencing have begun to emerge rapidly in recent years~\cite{lawrence2024project, google-beam, hp-google-beam, msr-volume}.

Despite this promise, volumetric videoconferencing over today’s networks remains challenging. Volumetric videos can be represented in several forms~\cite{point-cloud, nerf, 3dvoxel, 3dmesh, rgbd}. One widely used representation is point clouds~\cite{point-cloud, point-cloud2}, which consist of collections of 3D points describing the surface of an object in space, with each point also encoding color (RGB) information. These representations typically contain substantially more data than 2D video and remain bandwidth-intensive even after compression~\cite{ghosh2025livo, lee2020groot}. Consequently, transmitting volumetric videos is often hindered by insufficient and unstable bandwidth, leading to packet loss or delayed delivery. Because volumetric frames are substantially larger and jointly encode color and geometry, packet loss is more likely to corrupt frames or critical regions than in 2D video. Real-time interactive applications further impose strict latency constraints (typical end-to-end latency requirements being 150--200 ms~\cite{tambur, ghosh2025livo, fouladi2018salsify, lee2022r, lee2021demystifying, dhawaskar2023converge}), under which delayed packets are effectively treated as lost~\cite{grace, meng2024hairpin}. As a result, packet loss is common in practice and can lead to frame corruption or video freezes. Moreover, in volumetric video, packet loss affects not only visual appearance but also geometric structure; distortions in depth can produce spatial artifacts that are especially noticeable under 6-DoF viewing. Consequently, even low packet loss rates can significantly degrade the Quality of Experience (QoE) of volumetric videoconferencing.

\textbf{Limitations of Prior Works.} Prior works have explored related aspects of this problem, but important gaps remain. Existing works on volumetric videoconferencing primarily focus on reducing bandwidth and latency requirements~\cite{guan2023metastream, ghosh2025livo, lawrence2024project, tu2024tele, cheng2024magicstream} (for example, representing volumetric videos as RGB and depth streams instead of point clouds), but typically assume reliable networks without packet loss. Conversely, prior works on loss-resilient videoconferencing have developed techniques to mitigate packet loss using network-layer mechanisms (e.g., redundancy or retransmissions) or application-layer recovery (e.g., extrapolation or reconstruction of corrupted frames), but these approaches primarily target 2D video~\cite{tambur, grace, reparo, meng2024hairpin, hu2023dynamic, Gemino}. 

Extending such techniques to volumetric videoconferencing is non-trivial. While network-layer mechanisms can effectively recover both RGB and depth information, retransmissions are not feasible under real-time latency requirements~\cite{tambur}, and Forward Error Correction (FEC)-based redundancy incurs significant bandwidth overhead~\cite{tambur, wu2025nevo}. Network-layer techniques also struggle to accurately predict the loss rate to add sufficient redundancy~\cite{grace}, resulting in video freezes when frames cannot be recovered. In contrast, application-layer techniques are bandwidth-efficient, but frame reconstruction is typically ineffective if entire frames or critical video content are lost~\cite{grace, kang2022error}, and extrapolation performance degrades when loss propagates across multiple frames. Moreover, techniques designed to correct color artifacts cannot be directly applied to depth artifacts, which have different error characteristics and perceptual impact~\cite{ghosh2025livo}.

\textbf{Our Goal.} Motivated by these challenges, our goal is to design and implement a volumetric videoconferencing system that maintains high QoE even under packet loss while meeting the strict latency requirements of real-time interaction. Specifically, we aim to simultaneously (i) tolerate bursty packet loss commonly observed in networks, (ii) meet real-time deadlines, and (iii) jointly recover RGB and depth content to preserve both visual appearance and geometric consistency. Because videoconferencing is a widely-used application, we further target a practical system that operates at interactive frame rates on desktop-grade hardware.

\textbf{Research Contributions.} To achieve this goal, we present \sysname{}, a \emph{loss-resilient, one-to-one volumetric videoconferencing system} designed to operate with a wide range of video codecs under real-time constraints and packet loss. Our work is driven by a key observation: neither network-layer nor application-layer recovery alone is sufficient for volumetric videoconferencing; effective recovery requires a \emph{cross-layer, modality-aware approach} that combines the complementary strengths of both. First, \sysname{} decouples volumetric video into RGB and depth streams, enabling modality-aware processing and recovery while also lowering bandwidth requirements. Then, it applies selective network-layer FEC protection to critical content, while using application-layer \emph{neural} frame reconstruction to correct non-critical RGB and depth content at the receiver. It is implemented end-to-end over WebRTC~\cite{webrtc}, supports multiple codecs, and is evaluated using real-world packet loss traces. Our results demonstrate that a cross-layer, modality-aware recovery can significantly improve QoE for volumetric videoconferencing under lossy network conditions. To the best of our knowledge, \emph{\sysname{} is the first work to handle packet loss in volumetric videoconferencing}. Our specific contributions follow.

\noindent\textbf{(1) Cross-Layer, Modality-Aware System Design.} We present \sysname{}, a volumetric videoconferencing system that employs a cross-layer loss-recovery design to maintain high QoE even under packet loss. \sysname{} selectively applies FEC to protect critical video content at the network layer, while using a \emph{post-decode neural loss recovery module} to reconstruct corrupted non-critical frames in the application layer. Our recovery module is modality-aware and uses separate ML models based on Video Vision Transformers (ViViT)~\cite{videomae} to recover both RGB and depth content effectively at the receiver while meeting latency constraints.

\noindent\textbf{(2) End-to-End System Implementation.} We implement an end-to-end system prototype for one-on-one volumetric conferencing to address practical challenges associated with our design. Our implementation runs over WebRTC and supports multiple codecs (H.264~\cite{h264}, H.265~\cite{h265}, DCVCRT~\cite{dcvcrt}), demonstrating its generalizability.

\noindent\textbf{(3) Evaluation Under Real-World Conditions.} We extensively evaluate \sysname{} using real-world network loss traces spanning Ethernet, WiFi, and cellular networks. Our experiments show that \sysname{} improves RGB (resp. depth) SSIM by up to 32\% (resp. 13\%) over baselines designed for 2D videos or that rely solely on application-layer recovery, and reduces video freezes by 95.7\% compared to network-layer recovery approaches. In addition, an AWS deployment achieves a median RGB (resp. depth) SSIM of up to $0.92$, and a user study shows up to a $1.3\times$ higher Mean Opinion Score than other baselines, demonstrating practical feasibility.

\noindent\textbf{(4) Open Source.} We release our entire codebase, including the ViViT-based loss recovery models, as open source to foster further research in this area. \sysname{} can be accessed at \textbf{\url{https://umassos.github.io/revo-website/}}.

%% file: 2-background.tex
\section{Background}
\label{sec:background}
\noindent\textbf{Volumetric Videos.} 
Volumetric videos enable immersive 6-DoF experiences by allowing users to freely change viewpoints within a 3D scene. Unlike 2D videos, volumetric content can be represented using point clouds~\cite{point-cloud, point-cloud2}, meshes~\cite{3dmesh}, voxels~\cite{3dvoxel}, RGB-D~\cite{rgbd}, Neural Radiance Fields (NeRFs)~\cite{nerf}, and Gaussian splatting~\cite{4dgs, li2025gifstream4dgaussianbasedimmersive}. Neural representations (e.g., NeRF, Gaussian splatting) offer high visual quality and photorealistic rendering, but incur substantial computation and generation overhead~\cite{wu2025nevo, li2025gifstream4dgaussianbasedimmersive, yuan20251000fps4dgaussian, 4dgs}, making them impractical for interactive videoconferencing.
In contrast, RGB-D offers a practical trade-off between visual quality and latency. Each frame pairs an RGB image with a depth map, enabling 3D reconstruction by projection into the Cartesian space. Crucially, RGB-D data can be compressed using standard 2D video codecs, resulting in significantly lower encoding and decoding latency. These properties make RGB-D well-suited for volumetric videoconferencing.

\noindent\textbf{Compression Codecs.} Video compression codecs span both traditional and more recent neural codecs. Traditional codecs (e.g., H.265~\cite{h265}) separate video frames into keyframes (I-frames), which exploit spatial redundancy within a frame, and predicted frames (P/B-frames), which exploit both spatial and temporal redundancy. In this paper, we consider only I- and P-frames. Under low-bandwidth or lossy network conditions, these codecs are susceptible to packet loss, frame corruption, and visual artifacts, particularly in real-time settings. Recently, neural codecs (e.g., DCVC-RT~\cite{dcvcrt}) have shown higher compression efficiency while remaining real-time on desktop-grade GPUs. While promising in low-bandwidth settings, these approaches still rely on correct packet delivery and remain vulnerable to packet loss in interactive applications.

\noindent\textbf{Interactive Video Streaming.} Interactive applications, such as cloud gaming~\cite{xbox-gaming, nvidia-gaming} and videoconferencing~\cite{zoom, google-meet, ms-teams}, are used at a global scale~\cite{zoom-vc} and impose significantly strict latency constraints. For example, videoconferencing requires end-to-end latency of $\sim$150--200 ms~\cite{tambur, fouladi2018salsify, ghosh2025livo} to maintain a high QoE. Unlike on-demand streaming, delayed packets in interactive settings are often treated as lost~\cite{grace}, making robustness to packet loss critical. These constraints limit the use of retransmissions and necessitate efficient, low-latency loss recovery mechanisms, particularly for volumetric video.


\noindent\textbf{Loss Recovery Techniques}. Loss recovery techniques broadly fall into network- and application-layer approaches. 

\noindent \emph{(1) Network-Layer (\netlayer) Loss Recovery.}
These include Forward Error Correction (FEC)~\cite{reed-solomon, tambur, webrtc-rfc} and retransmissions~\cite{retransmit}. FEC adds redundancy using parity packets, enabling recovery as long as a sufficient number of data or parity packets are delivered. However, selecting the right redundancy level requires an accurate estimation of the network loss rate. Underestimation leads to insufficient recovery, while overestimation increases bandwidth overhead. Retransmissions incur additional latency, making them unsuitable for interactive applications with tight delay constraints~\cite{meng2024hairpin}.

\noindent \emph{(2) Application-Layer (\applayer) Loss Recovery.}
These include error concealment~\cite{kang2022error} and frame extrapolation~\cite{hu2023dynamic}, and neural recovery approaches~\cite{reparo, grace, wu2025nevo}. Error concealment aims to compensate for lost data by exploiting spatial and temporal correlations in previously decoded frames (e.g., by estimating motion vectors to predict pixels), while extrapolation techniques use statistical methods to estimate missing frames. Neural approaches leverage deep learning models to reconstruct corrupted frames. While these methods can produce high-quality recovery, their effectiveness depends on codec behavior, frame dependencies, and availability of I-frames, making robust recovery under severe loss challenging.

\vspace{-1em}

%% file: 3-system-design-new.tex
\section{\sysname{} System Design}
\label{sec:system-design}

\begin{figure*}[ht]
    \centering
    \includegraphics[width=0.9\linewidth]{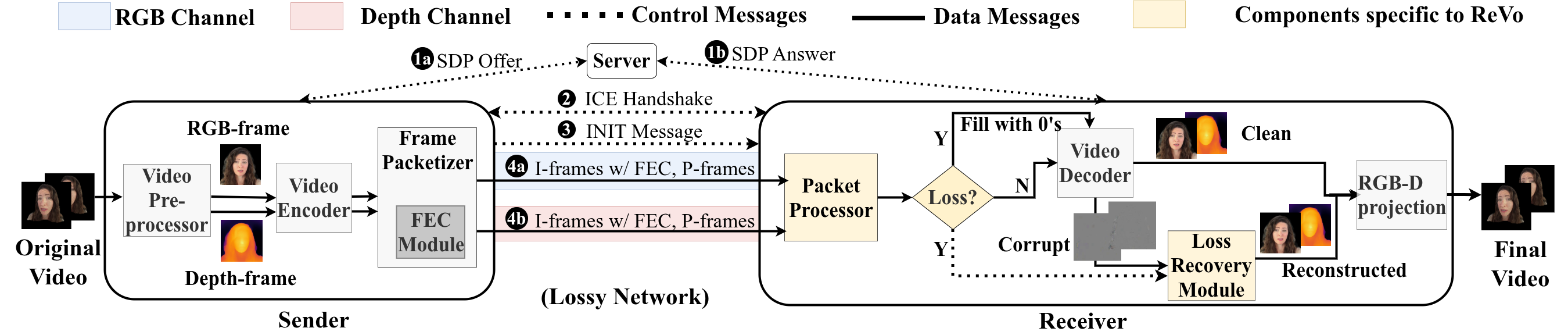}
    \caption{\sysname{} end-to-end design. During videoconferencing, the sender packetizes RGB and depth frames and sends them over the corresponding channels. The receiver processes the packets and decodes the video. If packet loss is detected, the respective decoded frame is recovered before display; otherwise, it is displayed directly.}
    \label{fig:volustream-system-design}
\vspace{-1em}
\end{figure*}

\vspace{-0.5em}
\subsection{Challenges and End-to-End Overview}
\label{sec:design-overview}

Packet loss during volumetric videoconferencing can severely degrade QoE by causing visual artifacts or video freezes. 
Recovering lost content before display can mitigate this degradation.
Loss recovery can be performed at the network (\netlayer{}) or application (\applayer{}) layer, but each has limitations when applied in isolation. \netlayer{} techniques are codec-agnostic but incur bandwidth overhead or excessive latency under real-time constraints~\cite{tambur, meng2024hairpin}. \applayer{} techniques are bandwidth-efficient, but are often codec-specific~\cite{grace, Gemino, reparo}. Codec-agnostic \applayer{} recovery typically operates post-decode and struggles to recover lost I-frames due to codec semantics, leading to severe QoE degradation~\cite{grace}. Moreover, recovering depth differs significantly from recovering RGB~\cite{ghosh2025livo}, requiring separate handling.

The goal of \sysname{} is to mitigate QoE degradation under packet loss regardless of the codec. \sysname{} is motivated by the observation that effective loss recovery for volumetric videoconferencing requires a \emph{cross-layer} design: codec-agnostic \applayer{} recovery is powerful but \emph{conditional} on the availability of critical information (e.g., I-frames). \sysname{} therefore combines both approaches by applying \netlayer{} protection to critical frame information to enable \applayer{} recovery, and then using \applayer{} neural reconstruction to efficiently recover corrupted RGB and depth content. We next outline the key challenges and how \sysname{} addresses them. 

\noindent\textbf{Challenge 1: Bandwidth-efficient and low-latency representation.}
Many volumetric representations (e.g., point clouds or NeRF) are unsuitable for real-time interactive videoconferencing due to high bandwidth demands and encoding-decoding latency~\cite{ghosh2025livo, draco, lee2020groot}. \sysname{} addresses this challenge by using RGB-D representation, which has relatively lower bandwidth and latency at the cost of slightly reduced visual quality. \sysname{} decouples video frames into RGB and depth components, and transmits them separately as 2D projections of the underlying 3D scene. This enables \sysname{} to leverage existing 2D video codecs that have high compression efficiency and lower encode-decode latency, allowing smooth playback under real-time constraints and high frame rates. Decoupling also allows interleaving RGB and depth packets, improving bandwidth balance and reducing the likelihood of losing an entire modality within a frame.

\noindent\textbf{Challenge 2: Enabling real-time L7 recovery.}
\applayer{} recovery is only feasible for interactive volumetric videoconferencing if it (i) meets strict real-time deadlines with desktop-grade hardware, and (ii) gets decoded frames from the codec even when frame contents are partially lost.

\noindent\textit{(i) Real-time feasibility.} Because recovery occurs sequentially after decoding, the combined latency of decoding, recovery, and rendering must meet real-time constraints to avoid buffering (e.g., 33 ms at 30 fps). While traditional codecs decode frames in 1--2 ms, neural codecs incur higher latency (e.g., $14.69$ ms for DCVC-RT on NVIDIA RTX 4070 GPU). \sysname{} ensures that model inference fits within the remaining budget by limiting the temporal context used for recovery. Based on empirical analysis, we select a history length that preserves reconstruction quality while keeping inference latency within budget for 30 fps playback. Decoupling RGB and depth further allows decoding and recovery to run in parallel using independent CUDA streams~\cite{cuda-parallel}, preventing depth processing from adding critical-path latency.

\noindent\textit{(ii) Decoded frame outputs.} Post-decode recovery requires decoded outputs even for partially lost frames. I-frame loss is especially damaging: missing I-frame packets cause decoders to discard the frame or produce artifacts that are challenging to recover, rendering all dependent P-frames within the same Group of Pictures (GoP) unusable. \sysname{} therefore protects I-frames using \netlayer{} FEC, ensuring their reliable delivery without the prohibitive overhead of protecting all frames (\S~\ref{sec:iframe-reliability}). P-frames, in contrast, are delivered best-effort and must remain decodable even under partial loss. Across all evaluated codecs, we observe that a P-frame remains decodable as long as its header and encoded frame length are available. Accordingly, \sysname{} attaches the encoded frame length metadata to each packet and applies FEC only to the header packet, so that partially received P-frames can still be decoded.
Simultaneously, a receiver-side module tracks missing portions of each frame in real time and sends partially available frames for decoding once deadlines expire (\S~\ref{sec:loss-detection}).

\noindent\textbf{Challenge 3: Robust \applayer{} loss recovery.}
Unlike prior codec-specific loss recovery techniques~\cite{grace, Gemino, reparo}, \sysname{} performs recovery after decoding to support multiple codecs. This increases generality but introduces two problems: (i) identical loss patterns can produce different artifacts across codecs, and (ii) RGB and depth frames exhibit distinct corruption characteristics even under the same loss. 

\sysname{} addresses these problems by using separate, modality-specific ML models for RGB- and depth recovery. The models are trained to minimize distinct loss functions tailored to visual quality for RGB and geometric consistency for depth, addressing their different error characteristics. Training also undergoes a two-stage process: (i) \emph{pre-training}, where the models learn general spatio-temporal correlations using synthetically generated loss patterns, improving robustness across codecs; and (ii) \emph{fine-tuning}, where each model specializes in recovering from loss patterns generated by a target codec, improving reconstruction quality for that codec (\S~\ref{sec:neural-loss-recovery}).

\vspace{-1em}
\subsubsection{\textbf{End-to-end Overview}} 
Figure~\ref{fig:volustream-system-design} illustrates the end-to-end workflow of \sysname{}. \sysname{} consists of a sender and a receiver that first use a common signaling server to establish a connection (1a, 1b). \textcolor{black}{The sender and receiver also negotiate information such as the bitrate, frame rate, and frame resolution at session startup (2)}. After a videoconferencing session begins (3), each volumetric video frame is decoupled into RGB- and D-frames, encoded using a pre-specified codec, packetized, and transmitted over separate data channels for RGB and depth over a lossy network (4a, 4b). The receiver reconstructs encoded frames using frame metadata while detecting packet loss in real time. Intact frames are decoded and forwarded directly for playback, whereas corrupted frames are routed to the loss-recovery module for reconstruction before playback.

We now discuss the core components of \sysname{}.
\begin{figure}
    \centering
    \includegraphics[width=0.85\linewidth]{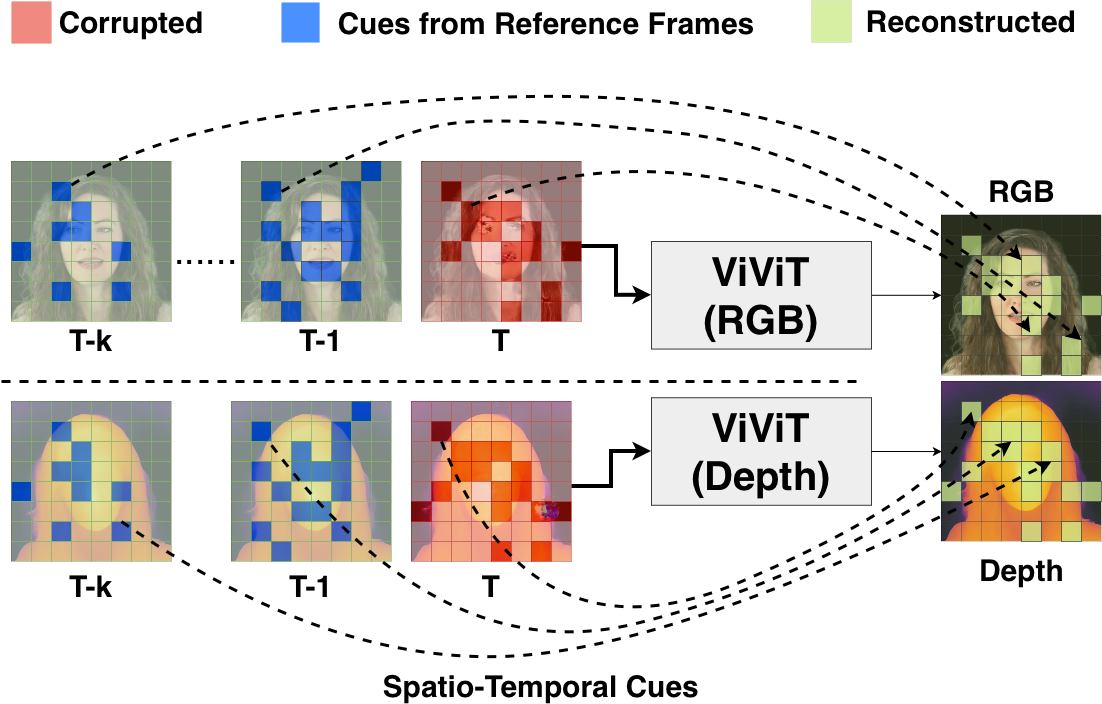}
    \caption{ViViT-based loss recovery models that exploit spatio-temporal cues to infer motion trajectories and reconstruct lost regions in both RGB and depth.}
    \label{fig:loss-recovery-module}
    \vspace{-2em}
\end{figure}
\vspace{-1em}

\subsection{FEC Reliability for Critical Frame Content}
\label{sec:iframe-reliability}

Our loss recovery module requires reliable delivery of I-frames and P-frame headers.
To meet this requirement under packet loss, \sysname{} applies \netlayer{} Forward Error Correction (FEC) using Reed–Solomon (RS) codes~\cite{reed-solomon}. 

Prior work~\cite{tambur} shows that although packet loss has long tails, the 95th-percentile loss rate remains below 30\% across networks.
We use this empirical observation to choose our redundancy rate. Specifically for I-frames, our goal is to ensure reliable delivery under typical loss conditions while keeping bandwidth overhead low enough to support real-time volumetric streaming. To apply FEC, an I-frame is divided into $n$ data packets, and \sysname{} generates $r$ parity packets using RS coding. At the receiver, the I-frame can be reconstructed as long as any 
$n$ packets out of the $n+r$ total packets are received, yielding a maximum tolerable loss rate of $loss_{max} = r/(n+r)$. We set $r = 0.5\ n$, corresponding to a 50\% parity overhead. This means that I-frames can be recovered as long as the packet loss rate is below 33\%, which covers more than 95\% of the scenarios observed in practice~\cite{tambur}. 

\sysname{} applies 50\% redundancy only to I-frames. For P-frame headers, we simply retransmit the header packet, which is equivalent to a 100\% redundancy rate. We protect only these critical components because the loss recovery module depends on them: I-frame loss causes GoP collapse, while loss of a P-frame header makes the frame undecodable.

Although the fixed parity rates may appear high, their overall bandwidth impact is modest because I-frames and P-frame headers constitute only a small fraction of the total transmitted bytes, with P-frame payload carrying the bulk of the data. Consequently, this results in an average bandwidth overhead of $8.3\%$ (\S \ref{sec:l3_comp}). This overhead can be further reduced by dynamically adapting the parity rate based on current or predicted loss rates. For example, Rudow et al.~\cite{tambur} predict loss rates using recent loss history; similar techniques can be incorporated into the \sysname{} with minimal changes.

\vspace{-1em}
\subsection{Packet Processing and Loss Detection}
\label{sec:loss-detection}
As \sysname{} operates over lossy networks, packets may be delayed, lost, or arrive out of order. At the receiver, a packet processor reassembles incoming packets into frames and tracks missing portions using per-frame \emph{DESC} metadata.

\sysname{} performs loss detection using \emph{deadline-based frame assembly}, driven by the playback schedule. We assume that the first frame ($f_0$, an I-frame) of a videoconferencing session is delivered without loss. Let $t_0$ denote the time at which all packets of $f_0$ have arrived. At $t_0$, the receiver forwards $f_0$ for decoding and starts a timer.
Since playback proceeds at a fixed frame rate ($fps$), $f_0$ must be decoded and ready for display at $t_0 + 1/fps$. As discussed in \S~\ref{sec:design-overview}, decode and loss recovery are designed to complete within $\frac{1}{fps}$ seconds.

Deadlines for subsequent frames follow the same schedule. Specifically, for the $i^{th}$ frame $f_i$, its decoding deadline is:
\setlength{\abovedisplayskip}{1pt}
\setlength{\abovedisplayshortskip}{1pt}
\setlength{\belowdisplayskip}{1pt}
\setlength{\belowdisplayshortskip}{1pt}
\begin{equation}
    \text{Deadline}_\text{decode} (f_i) = t_0 + (i/fps) 
\end{equation}
At this deadline, the receiver forwards $f_i$ for decoding regardless of whether all packets have arrived. If a frame is fully assembled before its deadline, \sysname{} forwards it immediately rather than waiting.

This schedule naturally pipelines assembly, decoding, recovery, and display. For example, at $fps = 30$, the receiver forwards $f_1$ for decoding by $t=33$ ms. Between $t=33$ ms and $t=66$ ms, it continues assembling frame $f_2$ while decoding and, if necessary, recovering $f_1$ in the background. At $t=66$ ms, it displays $f_1$ and forwards $f_2$ for decoding.

Frames are handled differently depending on what data is available at the deadline. If a frame is fully reassembled, the receiver decodes it and buffers it for display.
If a P-frame is partially available, we fill the missing parts with zeros to preserve the encoded frame length, forward it for decoding, and then perform loss recovery. If the full P-frame, or its header packet (even after FEC), is missing, the frame is dropped and declared lost, as it cannot be decoded. In contrast, I-frames are declared lost unless fully reassembled. 

If an I-frame is lost, the entire GoP is considered lost. The receiver temporarily freezes playback and waits for the next I-frame to resume. In contrast, when a P-frame is fully lost, the receiver freezes playback only for that frame and resumes once the next frame is available. This is because, although losing a P-frame corrupts subsequent decoder outputs in the GoP, the loss-recovery module can still reconstruct these frames. 

Thus, by coupling real-time decoding and loss recovery with deadline-driven loss detection, \sysname{} enables real-time playback of volumetric video despite packet loss.

\subsection{Neural Loss Recovery Module} 
\label{sec:neural-loss-recovery}
\input{./3b-neural-loss-recovery}


%% file: 3b-neural-loss-recovery.tex
The goal of our neural loss recovery module is to reconstruct partially lost RGB and depth P-frames at the receiver. To support multiple codecs, we place this module after decoding so that it operates independently of codec semantics. 

We make three key observations while designing the module. First, packet loss in the encoded frame bitstream can corrupt multiple pixels in the decoded frame, leading to structured artifacts. Second, RGB and depth exhibit different corruption characteristics under the same packet loss. Third, the mapping from bitstream loss to pixel-level artifacts is codec-dependent, leading to different distortions across codecs even under identical loss (see Figure~\ref{fig:loss_across_codecs} in Appendix \ref{app:bitstream-map}).

\sysname{} addresses these as follows. It leverages temporal dependencies across frames, using past clean frames to reconstruct the current corrupted frame in the pixel domain. It employs modality-specific neural models tailored for RGB and depth reconstruction. Finally, it adopts a two-stage training pipeline: pretraining on pixel-level loss patterns to learn the spatial-temporal context, followed by codec-specific fine-tuning to improve reconstruction quality.

\vspace{-1em}
\subsubsection{\textbf{Neural Architecture Design}}

Figure \ref{fig:loss-recovery-module} illustrates the architecture and workflow of the module. Our design is based on Video Vision Transformers (ViViT) \cite{videomae}, a state-of-the-art class of models for capturing long-range spatio-temporal dependencies in video. Our models use context from preceding clean frames to infer missing information in corrupted frames. Specifically, they take the closest $k$ clean frames (frames with blue patches in Figure~\ref{fig:loss-recovery-module}) and the current corrupted frame as input and output a reconstructed frame. The parameter $k$ is tunable and controls the trade-off between reconstruction quality and inference latency: a higher $k$ improves accuracy but increases latency.

As codecs use context from previous frames to decode the current P-frame, corruption in a P-frame can propagate across subsequent frames within a GoP~\cite{grace, wu2025nevo}. Thus, once a frame is corrupted, all following P-frames in the GoP require recovery. To handle this, we use positional embeddings in ViViT to encode the temporal distance between reference frames and corrupted targets. This allows our models to effectively reuse the same set of clean reference frames to reconstruct all affected frames in a GoP, mitigating error propagation.


\begin{figure}[t]
    \centering

    \begin{subfigure}[t]{0.4\columnwidth}
        \centering
        \includegraphics[width=\linewidth]{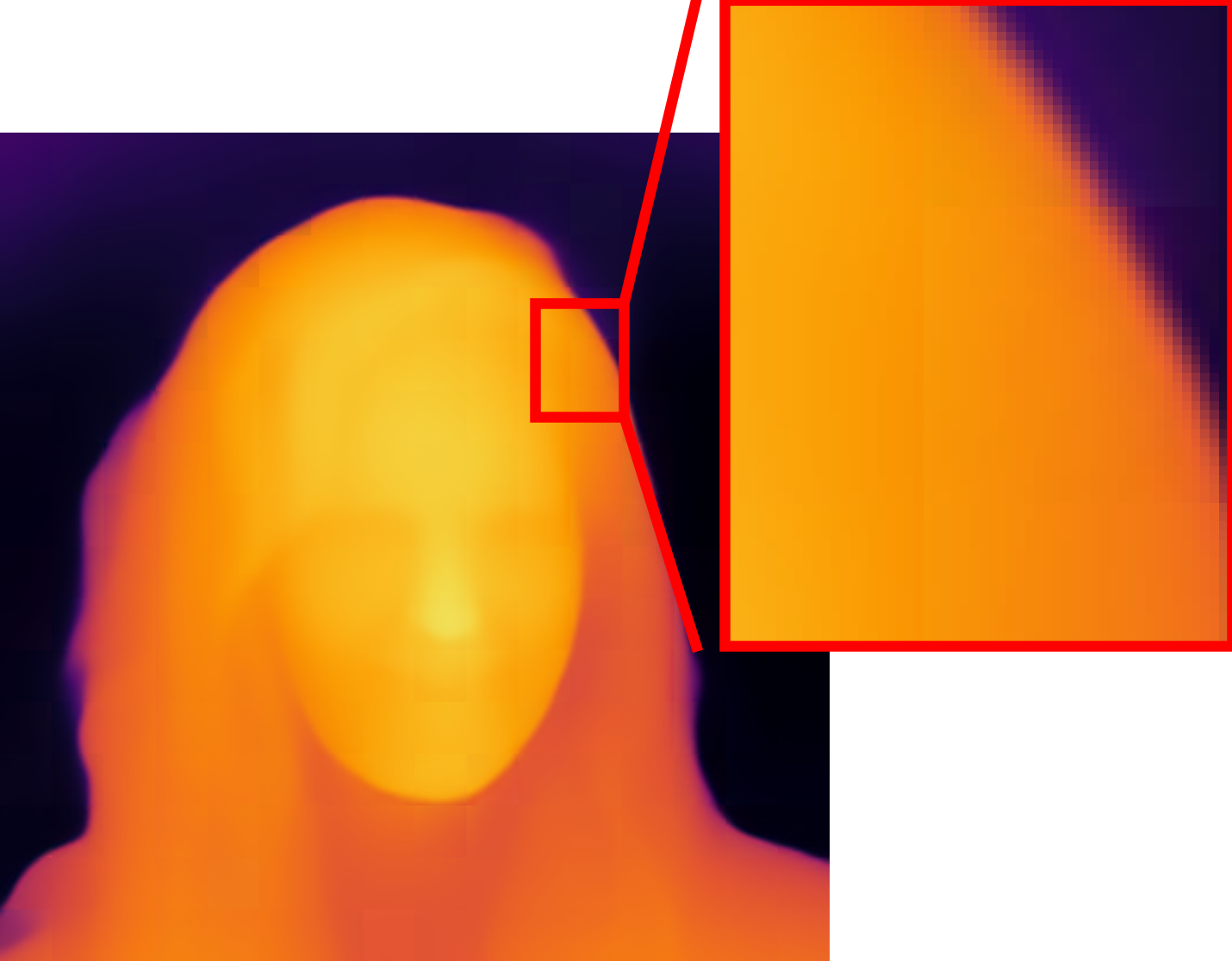}
        \caption{Using $\mathcal{L}_{Depth}$}
    \end{subfigure}\hfill
    \begin{subfigure}[t]{0.4\columnwidth}
        \centering
        \includegraphics[width=\linewidth]{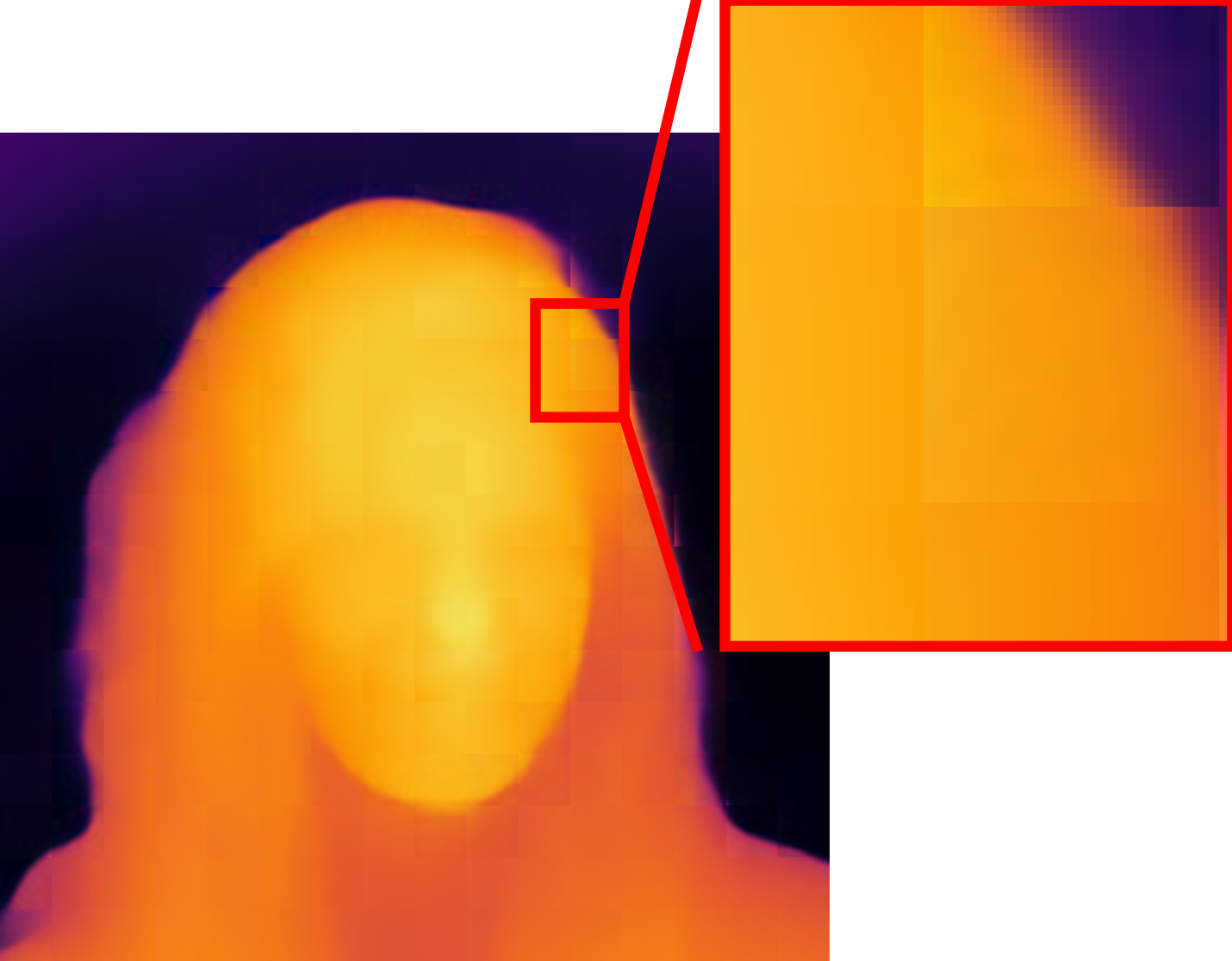}
        \caption{Using $\mathcal{L}_{RGB}$}
    \end{subfigure}\hfill
    \vspace{-1em}
    \caption{Depth reconstruction using (a) $\mathcal{L}_{\text{Depth}}$ suppresses patch artifacts more effectively than (b) $\mathcal{L}_{\text{RGB}}$.}
    \label{fig:loss_comp}

\vspace{-1em}
\end{figure}

\vspace{-1em}
\subsubsection{\textbf{Modality-Specific Optimization}}
Although both RGB and depth exhibit spatio-temporal correlations, they encode fundamentally different information: RGB captures color and texture, whereas depth encodes geometric structure. Our goal is to improve the QoE of the final rendered RGB-D content by effectively reconstructing both color and geometry, so that the projected RGB-D video exhibits minimal artifacts. Consequently, we use separate models for RGB and depth, each trained with different objectives.

\noindent\textbf{RGB loss:}
We optimize RGB reconstruction using a hybrid objective~\cite{liu2022devrf} consisting of pixel-wise mean squared error and total variation (TV) regularization:
{\setlength{\abovedisplayskip}{0pt}
\setlength{\belowdisplayskip}{0pt}
\begin{align}
\mathcal{L}_{\text{RGB}}  = MSE + \alpha * TV
\label{eq:l_rgb}
\end{align}
The MSE term ensures accurate pixel-level reconstruction, preserving fine spatial details and color fidelity. The TV term serves as a spatial smoothness prior by penalizing high-frequency discontinuities between neighboring pixels.

\noindent\textbf{Depth loss:}   
Applying $\mathcal{L}_{RGB}$ directly is suboptimal for depth recovery. Empirically, we observe that training ViViT with $\mathcal{L}_{RGB}$ produces noticeable discontinuities at patch boundaries in the depth maps (Figure~\ref{fig:loss_comp}), which manifest as geometric artifacts in rendered views and can significantly degrade QoE. To address this, we utilize a geometry-aware objective \cite{paul2022edge}:
\begin{align}
    \mathcal{L}_{Depth} = \alpha_r \mathcal{L}_{recon} + \alpha_e \mathcal{L}_{edge} + \alpha_s \mathcal{L}_{SSIM}.
    \label{loss_func}
\end{align}

$\mathcal{L}_{recon}$ is an $\mathcal{L}_{1}$ norm-based reconstruction loss that ensures pixel-wise depth accuracy, and $\mathcal{L}_{SSIM}$ is a structural similarity index loss that preserves local structure. The key term is $\mathcal{L}_{edge}$, a gradient-based loss that evaluates the mean difference in gradient magnitudes between the predicted and ground-truth depth values. By penalizing incorrect gradient predictions, $\mathcal{L}_{edge}$ effectively suppresses the artificial high-gradient transitions at patch borders, yielding a smoother, artifact-free continuous depth function (Figure~\ref{fig:loss_comp}). Consequently, $\mathcal{L}_{Depth}$ considerably improves depth reconstruction and overall QoE compared to using the RGB objective.

\vspace{-1em}
\subsubsection{\textbf{Robustness Across Codecs}}

The mapping from compressed bitstreams to decoded pixel space depends on the codec and its parameters. Packet loss in the bitstream is translated into complex, widespread pixel-level artifacts after decoding, which vary across codecs. To ensure robustness, we adopt a two-stage training pipeline that separates generic spatio-temporal learning from codec-specific adaptation.
In the first \emph{pretraining stage}, models are trained on videos with randomly masked pixels, forcing them to learn general spatio-temporal correlations and motion patterns independent of codec behavior. In this stage, we use $\mathcal{L}_{RGB}$ for both RGB and depth, as the primary goal of pretraining is to establish strong spatio-temporal context priors; learning geometric structure is deferred to the finetuning stage. In the second \emph{finetuning stage}, we simulate packet loss directly in encoded bitstreams. Corrupted bits are replaced with zeros to preserve alignment and ensure decodability, producing realistic artifact patterns and error propagation after decoding. Fine-tuning on these samples allows the model to adapt to codec-specific distortions.


This two-stage training design enables a single neural recovery architecture to operate across both traditional (e.g., H.264, H.265) and neural codecs (e.g., DCVC-RT) without requiring system-level modifications. Pretraining is performed once per modality. During the fine-tuning stage, while \sysname{} currently trains separate models for each codec, the underlying ViViT architecture remains unchanged.

%% file: 4-implementation.tex
\vspace{-1em}
\section{\sysname{} Implementation}
\label{sec:implementation}

We implement \sysname{} in $\sim$2500 lines of Python. \sysname{} is built on WebRTC~\cite{webrtc} (\emph{aiortc}~\cite{aiortc} library), and consists of a sender, a receiver, and a signaling server. At session startup, the sender and receiver exchange SDP \emph{offer} and \emph{answer} messages via the signaling server to establish a peer connection, after which they communicate directly.

\noindent\textbf{Sender.} The sender creates two WebRTC data channels over SCTP for RGB and depth packets, respectively. To meet real-time constraints, both channels allow out-of-order delivery and disable retransmissions. We use \emph{zfec}~\cite{zfec} library for FEC. 

At session start, the sender and the receiver agree on frame dimensions, frame rate, the data packet length, GoP length, and the bitrate via an \emph{INIT} message. We set the GoP length to 30 (one GoP per second). RGB (resp. depth) packet lengths is 1024 (resp. 512) bytes. The bitrate is decided using the codec Quantization Parameter (QP). 
During the session, each frame is preprocessed by cropping to the face region, removing background using VideoMatting~\cite{rvm}, and separating RGB and depth components, after which the frames are encoded using a codec selected by the application.

The sender packetizes an encoded frame, and each packet is annotated with a \emph{DESC} header containing the frame ID, GoP ID, packet count, and encoded frame length. This metadata allows the receiver to detect loss, identify frame boundaries, and track missing portions in real time.

\noindent\textbf{Receiver.} The receiver uses three threads for packet handling, decoding, and display. An asynchronous event loop receives packets and writes them to a shared buffer. A decoder thread reassembles frames using the \emph{DESC} metadata and decodes them using the codec. If a P-frame is partially lost but decodable, the loss recovery module reconstructs the frame. Decoded or reconstructed frames are written to a display buffer. A display thread renders frames at the target frame rate. If a frame is unavailable, the last displayed frame is repeated, causing a temporary freeze. Playback resumes once a valid frame becomes available, skipping intermediate frames.

\noindent\textbf{Evaluated codecs.} \sysname{} is compatible with a wide range of codecs. In this paper, we evaluate \sysname{} with H.264~\cite{h264}, H.265~\cite{h265}, and DCVC-RT~\cite{dcvcrt}. Other codecs can be integrated with \sysname{} with minimal code changes.

\noindent\textbf{Loss Recovery Module Training.} 
We use a 90:10 train/validation split (7200 videos for training and 800 for validation). For both RGB and depth models, the initial codec-agnostic pretraining phase lasts 40 epochs, followed by a 20-epoch finetuning stage using the modality-specific training objectives discussed in \S\ref{sec:neural-loss-recovery}. Each epoch takes 2-3 hours on $16$ Nvidia L40S GPU. The initial learning rate is set to $2e-4$, and we use the Adam optimizer. For the RGB loss ($\mathcal{L}_{RGB}$), we set $\alpha = 0.01$. For the depth loss ($\mathcal{L}_{Depth}$), we set $\alpha_r = 0.1$, $\alpha_e = 1$, and $\alpha_s = 1$ for depth reconstruction, following Paul et al~\cite{paul2022edge}. The finetuning stage is executed separately for each evaluated video codec. We provide additional training details in Appendix~\ref{app:additonal_train_details}.

\noindent\textbf{Bitrate Negotiation and Adaptation.} \textcolor{black}{\sysname{} currently uses a fixed bitrate negotiated at session startup. In practice, videoconferencing systems often adapt bitrate based on network conditions. Although we do not currently implement bitrate adaptation, \sysname{} is compatible with such techniques for two reasons. First, our recovery module operates post-decode and is therefore orthogonal to bitrate adaptation. Second, our models are trained to tolerate loss patterns across multiple compression settings (see Appendix~\ref{app:sec-codec-aware-finetuning}), suggesting that \sysname{} can operate effectively across a range of bitrates. However, fully integrating bitrate adaptation would require additional system design and a careful joint evaluation with loss recovery. We leave this to future work.}

%% file: 5-evaluation.tex
        

\vspace{-1em}
\section{Performance Evaluation}
\label{sec:evaluation}
\vspace{-0.5em}
We now evaluate the resiliency of \sysname{} under packet loss. We first study the impact of our key design choices, then compare \sysname{} against existing state-of-the-art loss recovery baselines, and finally evaluate the generalizability of \sysname{} across different codecs. Broadly, we seek to answer the following five research questions:

\begin{enumerate}[leftmargin=*, itemsep=0pt, topsep=0pt, parsep=0pt]
    \item How do our design choices (cross-layer approach and distinct loss functions for reconstructing RGB and depth frames) affect \sysname{} performance? (\S~\ref{sec:eval-design-choices})
    \item How does our system compare against state-of-the-art loss recovery techniques that operate exclusively at either the network or application layer, in terms of QoE? (\S~\ref{sec:eval-baseline-comparison})
    \item Is \sysname{} robust across different codecs? (\S~\ref{sec:eval-codec-agnostic})
    \item How does \sysname{} perform in real-world settings (\S~\ref{sec:eval-aws-deployment} and \S~\ref{sec:eval-user-study})?
\end{enumerate}

We answer these questions through a series of targeted experiments. We begin by describing the experimental setup, baselines, metrics, and the network loss traces used.


\vspace{-1em}
\subsection{Experimental Setup}
\label{sec:exp-setup}
\noindent \textbf{Testbed.}
Our testbed consists of a sender and a receiver running on separate machines connected via a 1 Gbps LAN. We evaluate \sysname{} on two desktop-grade GPUs: an Nvidia RTX 5070~\cite{rtx-5070} hosted on a machine with an Intel Core Ultra 7 265K CPU (20 cores, 32 GB RAM), and an Nvidia RTX 4070~\cite{rtx-4070} hosted on a second machine with an AMD Ryzen 5 5600X CPU (6 cores, 32 GB RAM). Unless otherwise specified, we report results on RTX 4070 and use H.265 as the codec. 

\noindent \textbf{Baselines.} We compare \sysname{} against six baselines spanning network (L3) and application (L7) layer loss recovery. All baselines except \grace{} use H.265 (GoP = 30, QP = 30), as the underlined video codec.  The baselines follow.

\noindent\textit{(B1) WebRTC-default.} This is standard WebRTC with reactive FEC based on the probed packet loss rate~\cite{webrtc-rfc, holmer2013handling}. If frames are lost, playback freezes by repeating the last displayed frame until a correct I-frame arrives, matching default codec behavior. We set the redundancy to 0\%, 50\%, or 100\%.

\noindent\textit{(B2)} \tambur{}. \tambur{}~\cite{tambur} is a system for 2D videos that employs streaming codes at L3, and that dynamically adjusts redundancy using predicted loss rates. Since \tambur{}’s predictor is not publicly available, we assume \emph{oracle loss knowledge} and set redundancy to 0\%, 50\%, or 100\%.

\noindent\textit{(B3)} \grace{}. \grace{}~\cite{grace} is a loss-resilient system for 2D videos that trains a neural video codec across a spectrum of packet losses. As \grace{} is not well-suited to run on desktop-grade hardware, and it's implementation is not publicly available, we evaluate an \emph{idealized version} that assumes I-frames are never lost and ignores real-time constraints.

\noindent\textit{(B4) \emph{DMVFN}.} \emph{DMVFN}~\cite{hu2023dynamic} is a neural frame extrapolation baseline that reconstructs lost frames using the two most recent frames, without any L3 redundancy. Under consecutive losses, it recursively uses extrapolated frames as inputs.

\noindent\textit{(B5) \sysname{}-L3.} This variant of \sysname{} disables L7 recovery. I-frames use fixed 50\% FEC, while P-frames are sent best effort. When frames are lost, the video freezes until the next correct I-frame is received.

\noindent\textit{(B6) \sysname{}-L7.} This variant disables L3 protection. Lost P-frames are reconstructed by our neural loss recovery module, while I-frame loss causes the entire GoP to be discarded and playback to freeze until the next I-frame arrives.

\noindent \textbf{Metrics.} We evaluate using the three categories of metrics. 
\noindent \textit{(1) Visual quality.} We use structural similarity index (SSIM)~\cite{ssim} and peak signal-to-noise ratio (PSNR)~\cite{psnr}, which compare reconstructed frames to the original, along with PointSSIM (PSSIM)~\cite{alexiou2020towards}, an extension of SSIM for volumetric frames that jointly captures geometric (depth) and color (RGB) distortions. 
\noindent \textit{(2) Playback robustness.} We measure median freeze duration (ms), defined as the median time the receiver’s video remains frozen, and the percentage of non-recoverable frames. 
\noindent \textit{(3) Overhead.} We report bandwidth overhead, defined as the FEC overhead. 
All metrics are computed only over frames corrupted during transmission.


\noindent \textbf{Test Videos.}
We use 30 randomly sampled videos from the TalkingHead~\cite{thead} dataset. Each video is looped to form a 5-minute sequence. For simplicity, all preprocessing to create RGB-D videos are performed offline. 
The resulting videos are then used as input to our system. 

\noindent \textbf{Real world traces.} 
We use real-world network loss traces from Hairpin~\cite{meng2024hairpin} spanning cellular, WiFi, and Ethernet networks. The traces include bandwidth, RTT, and loss rate at 15 ms granularity. We replay them using \emph{tc}~\cite{tc}, with a fixed one-way propagation delay of 40 ms. For each network type, we stream all 30 videos (150 minutes total), corresponding to 270K RGB and 270K depth frames at 30 fps. Each video is paired with a randomly selected trace of the target network. 

\begin{figure}[t]
    \centering
    \begin{subfigure}[b]{0.45\columnwidth}
        \centering
        \includegraphics[width=\linewidth]{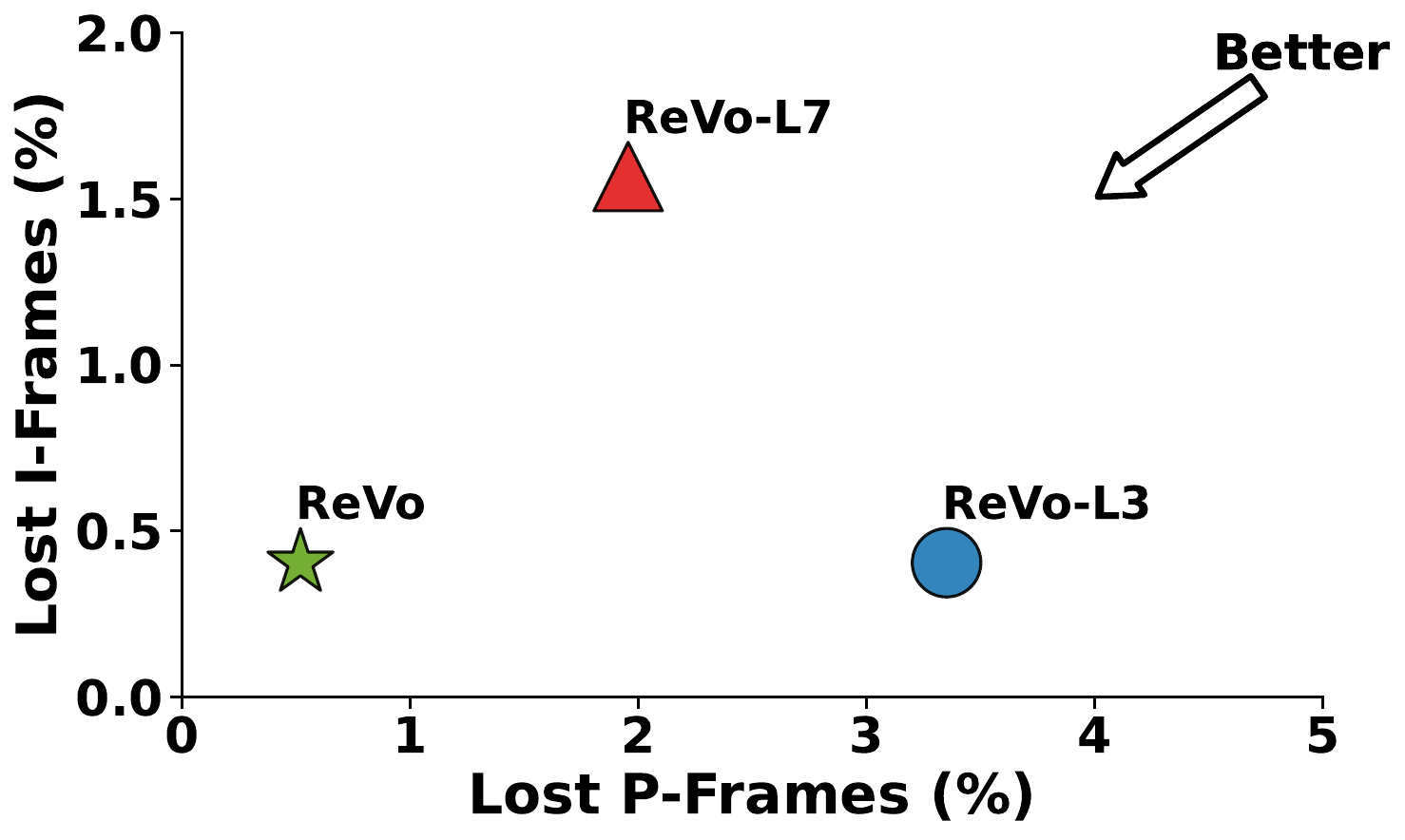}
        \caption{Cross Layer Ablation}
    \end{subfigure}
    \hfill
    \begin{subfigure}[b]{0.45\columnwidth}
        \centering
        \includegraphics[width=\linewidth]{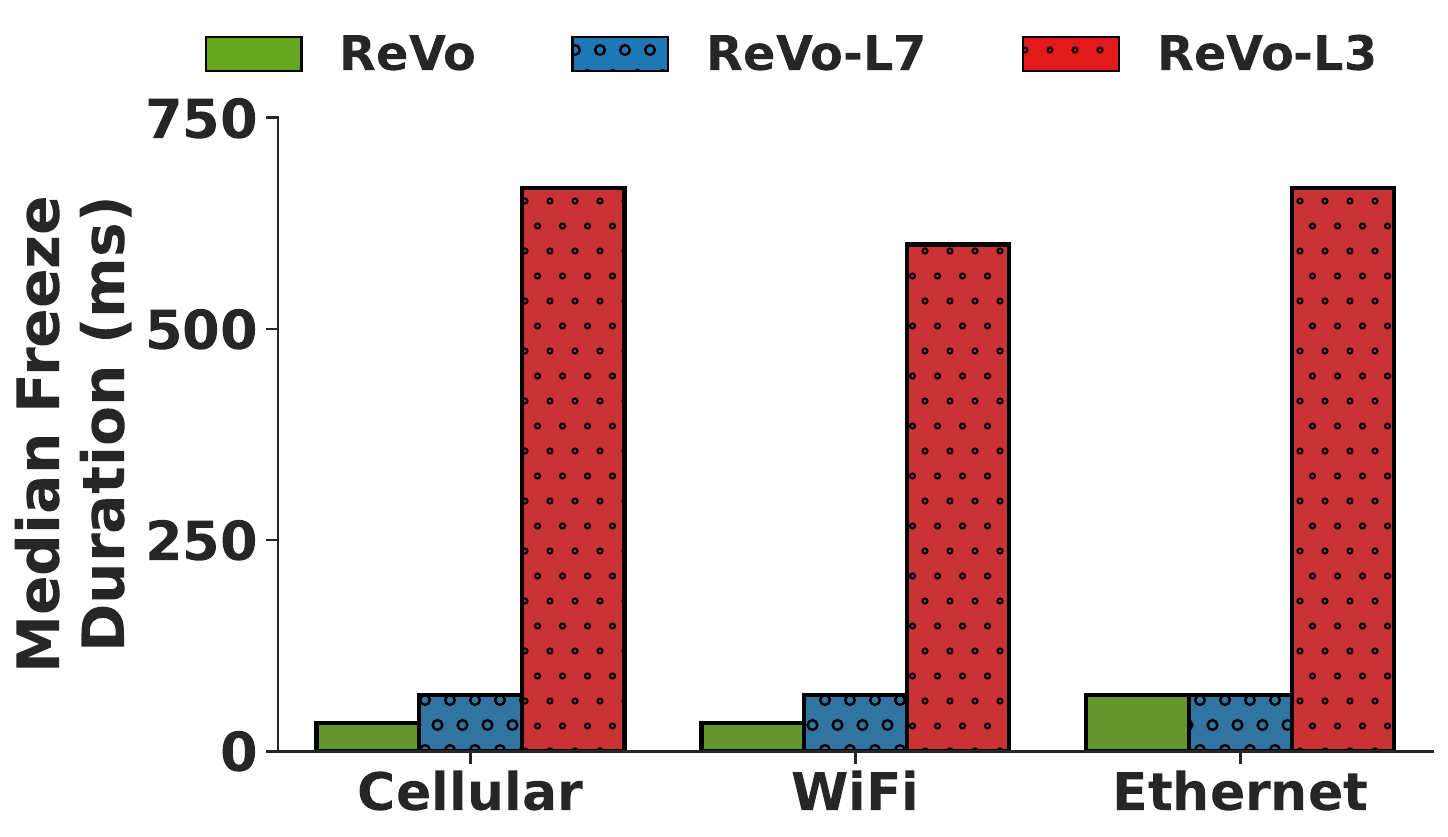}
        \caption{Med. Freeze Duration}
    \end{subfigure}

    \vspace{2pt}
    \centering
    \includegraphics[width=0.6\linewidth]{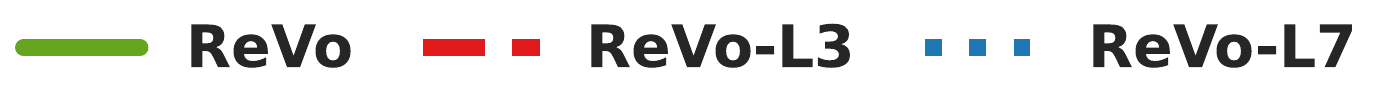}
    \begin{subfigure}[t]{0.45\columnwidth}
        \centering
        \includegraphics[width=\linewidth]{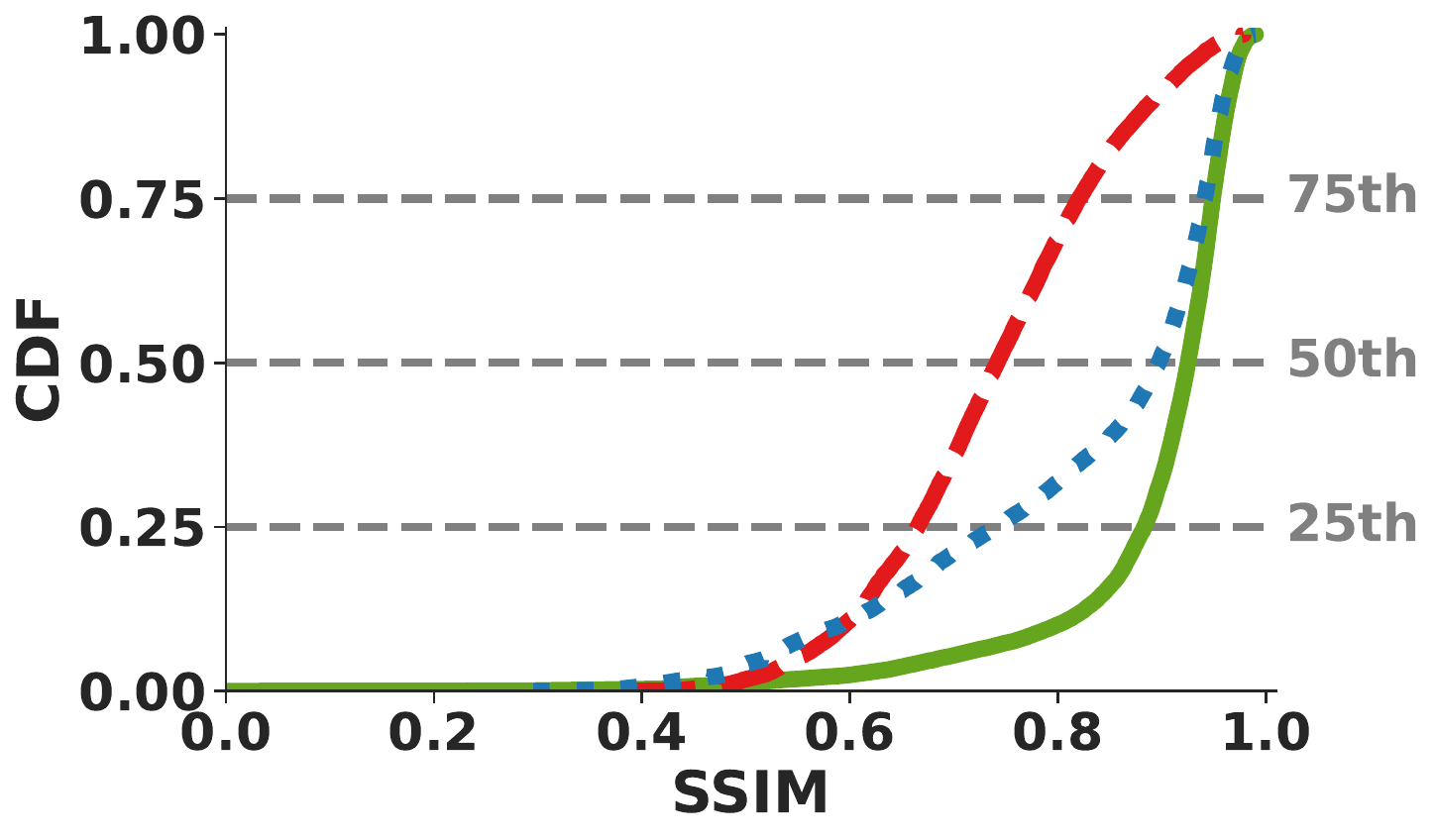}
        \caption{RGB Frames}
    \end{subfigure}
    \hfill
    \begin{subfigure}[t]{0.45\columnwidth}
        \centering
        \includegraphics[width=\linewidth]{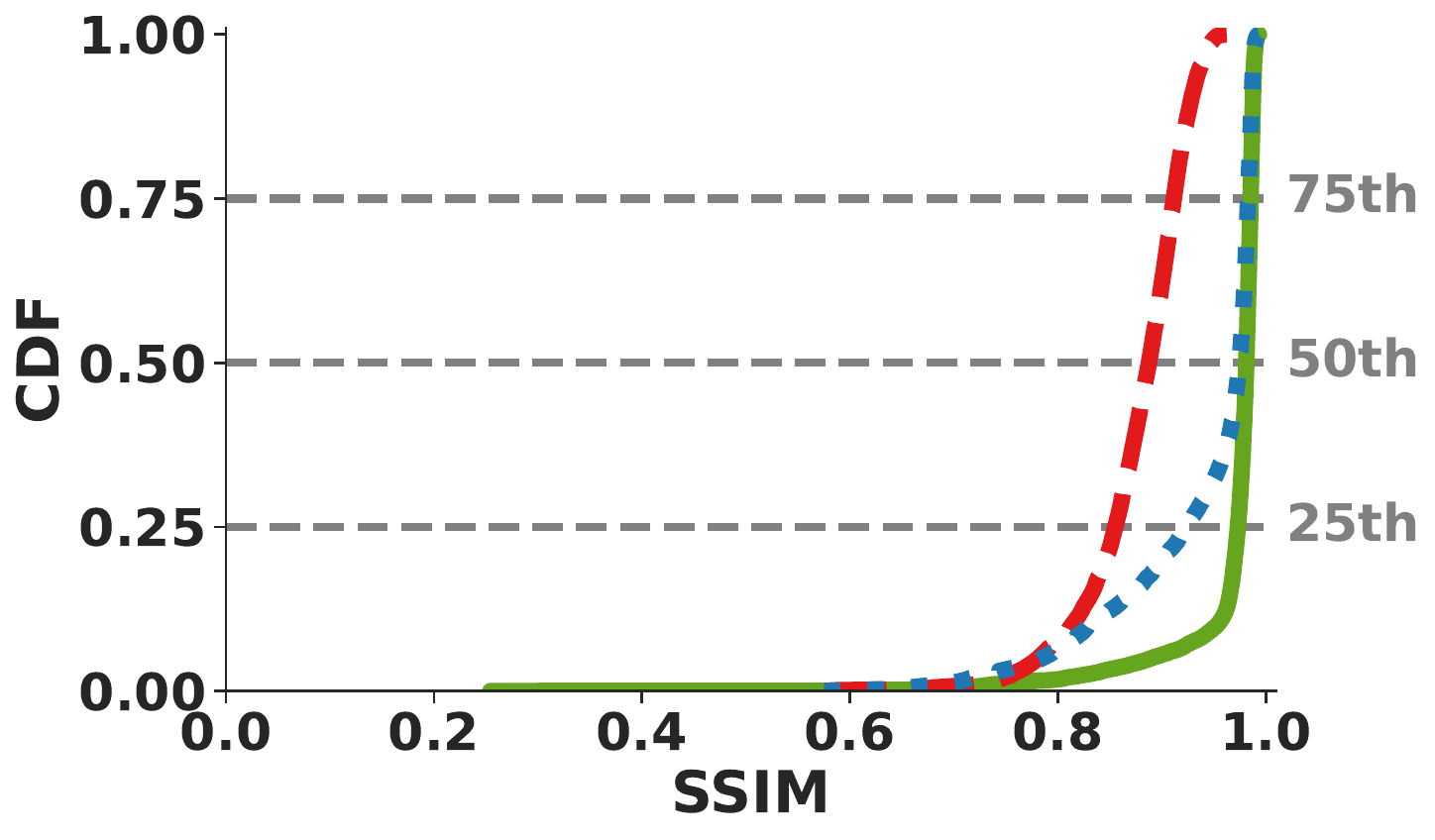}
        \caption{Depth Frames}
    \end{subfigure}
    \caption{\sysname{}’s cross-layer recovery (a) minimizes frame losses, (b) reduces video freezes, and delivers higher-quality reconstructions for (c) RGB and (d) depth.}

    \label{fig:cross_layer_motivation}
    \vspace{-1em}
\end{figure}

\begin{table}[t]
\centering
\begin{tabular}{c|cc}
\hline
\textbf{Loss function} & \textbf{Avg. PSNR (dB)} & \textbf{Avg. SSIM} \\
\hline
$\mathcal{L}_{RGB}$ & 32.56 &  0.963  \\
$\mathcal{L}_{Depth}$  &  33.75  & 0.971 \\
\hline
\end{tabular}
\caption{Comparison of Depth reconstruction quality under different loss functions.}
\label{tab:loss_psnr_ssim}
\vspace{-1em}
\end{table}

\begin{figure}[t]
\centering
\includegraphics[width=0.8\linewidth]{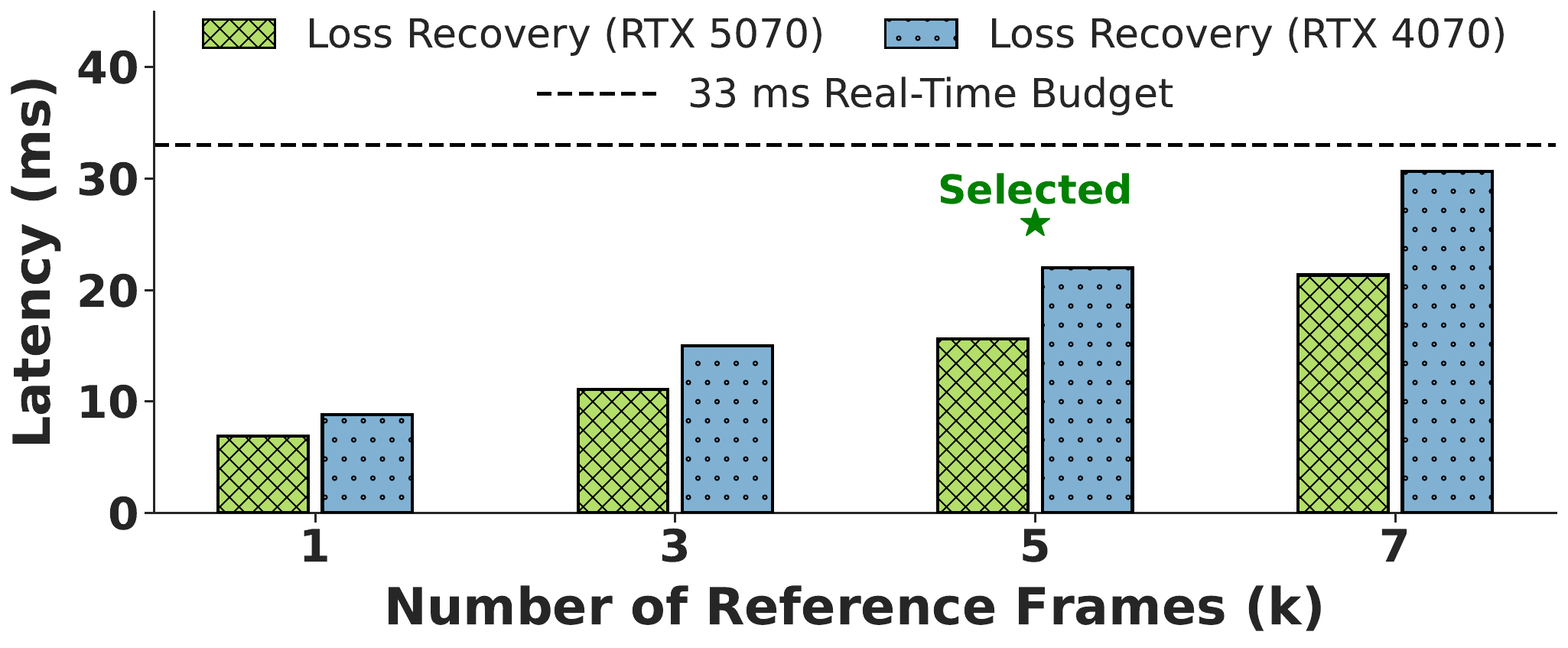}
\vspace{-1em}
\caption{Inference latency versus reference frames}
\label{fig:benchmark_lossrec_latency}
\vspace{-1em}
\end{figure}

\begin{figure}[t]
    \centering

    \begin{subfigure}[t]{0.25\columnwidth}
        \centering
        \includegraphics[width=\linewidth]{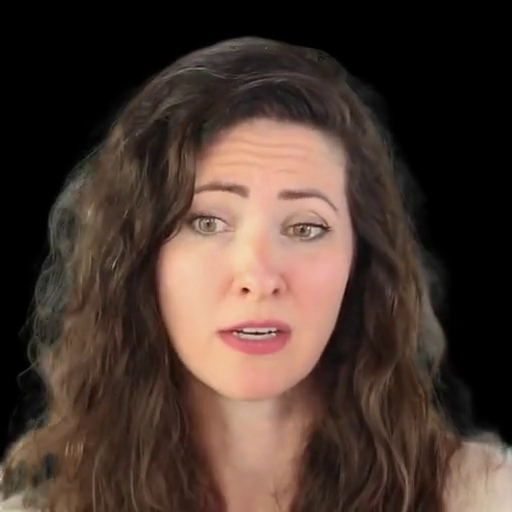}
        \caption{GT}
    \end{subfigure}\hfill
    \begin{subfigure}[t]{0.25\columnwidth}
        \centering
        \includegraphics[width=\linewidth]{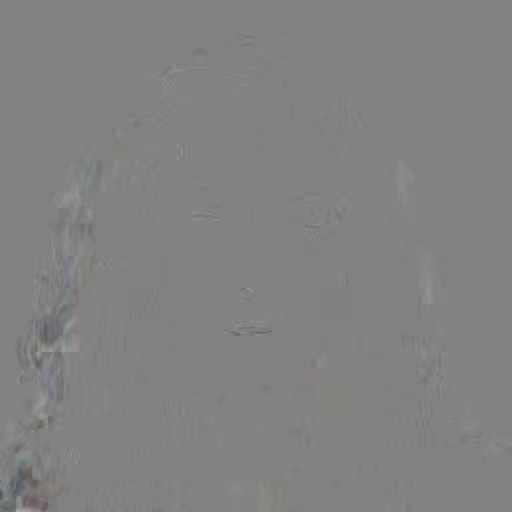}
        \caption{Corrupted}
    \end{subfigure}\hfill
    \begin{subfigure}[t]{0.25\columnwidth}
        \centering
        \includegraphics[width=\linewidth]{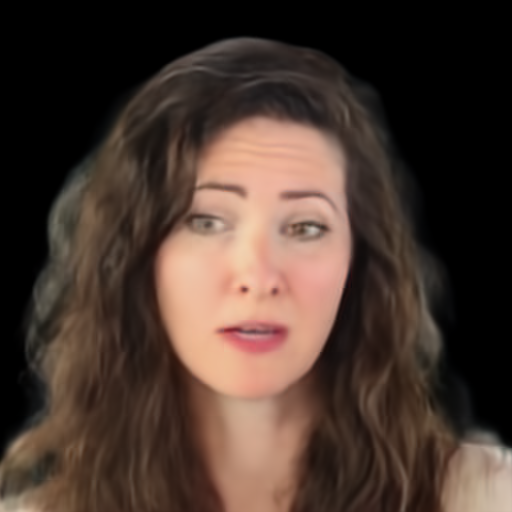}
        \caption{Recovered}
    \end{subfigure}

    \vspace{1mm}

    \begin{subfigure}[t]{0.25\columnwidth}
        \centering
        \includegraphics[width=\linewidth]{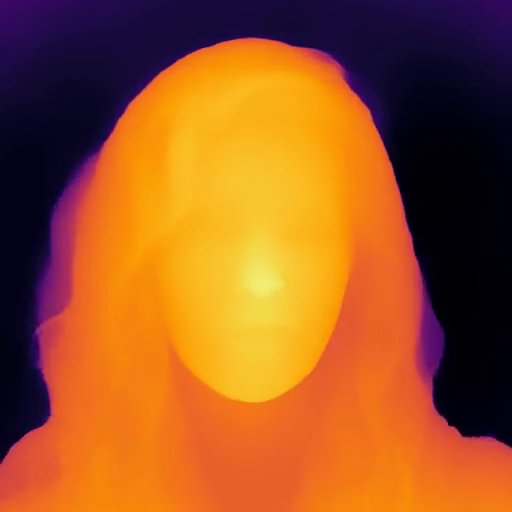}
        \caption{GT}
    \end{subfigure}\hfill
    \begin{subfigure}[t]{0.25\columnwidth}
        \centering
        \includegraphics[width=\linewidth]{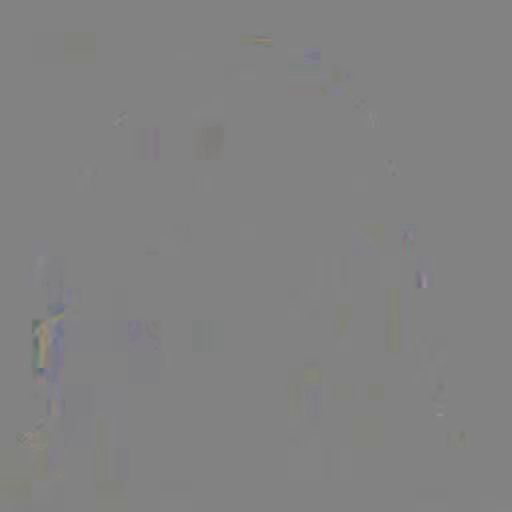}
        \caption{Corrupted}
    \end{subfigure}\hfill
    \begin{subfigure}[t]{0.25\columnwidth}
        \centering
        \includegraphics[width=\linewidth]{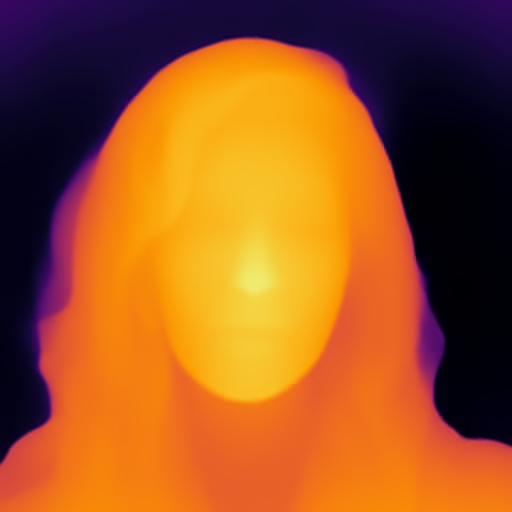}
        \caption{Recovered}
    \end{subfigure}
    \vspace{-1em}
    \caption{Visual comparison of frame recovery under packet loss (top row: RGB, bottom row: depth).}
    \label{fig:visual_rgbd_recovery}
    \vspace{-1em}
\end{figure}

\begin{figure}[t]
    \centering
    \begin{subfigure}[b]{0.32\linewidth}
        \centering
        \includegraphics[width=\linewidth]{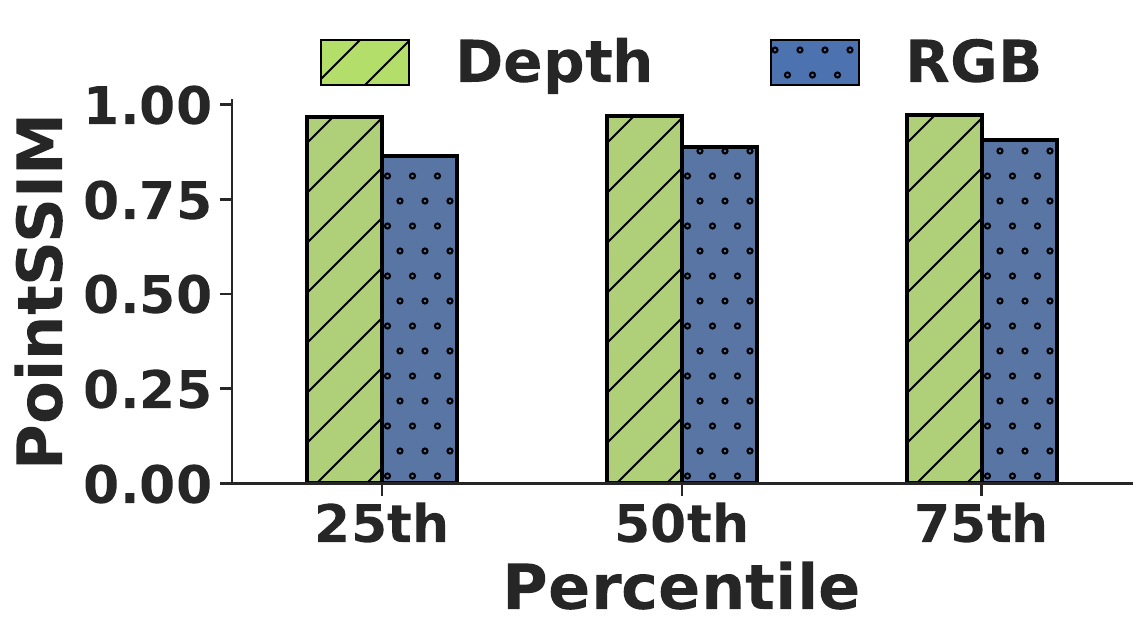}
        \caption{Cellular}
        \label{fig:rgbd_cell}
    \end{subfigure}
    \hfill
    \begin{subfigure}[b]{0.32\linewidth}
        \centering
        \includegraphics[width=\linewidth]{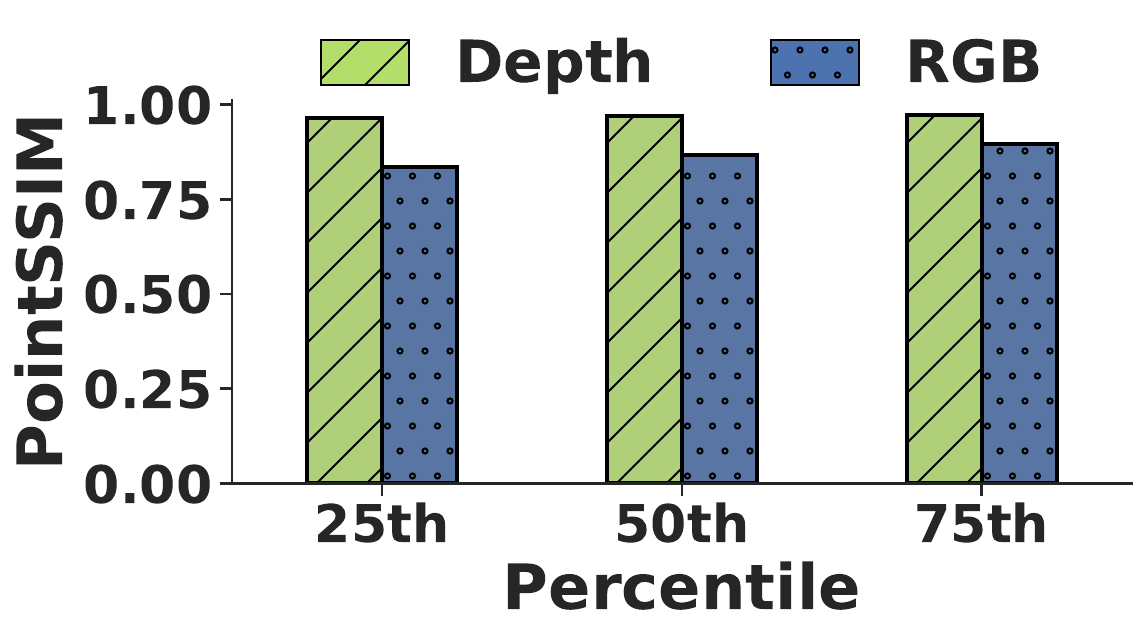}
        \caption{WiFi}
        \label{fig:rgbd_wifi}
    \end{subfigure}
    \hfill
    \begin{subfigure}[b]{0.32\linewidth}
        \centering
        \includegraphics[width=\linewidth]{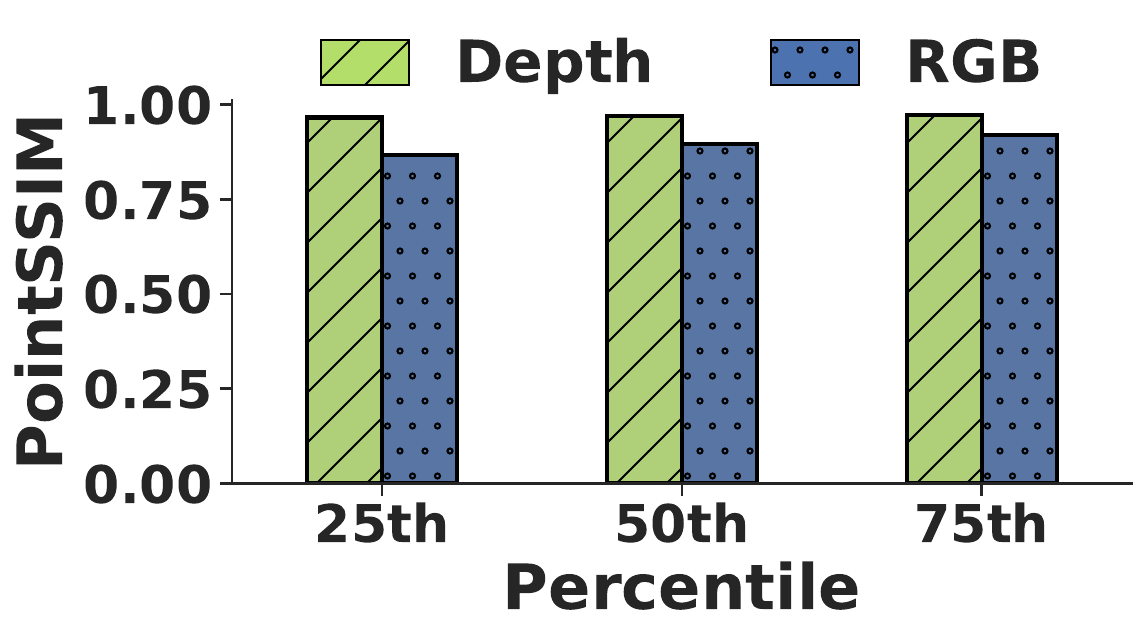}
        \caption{Ethernet}
        \label{fig:rgbd_eth}
    \end{subfigure}
    \vspace{-1em}
    \caption{\textbf{PointSSIM performance of \sysname{}.} We report the 25th, 50th, and 75th percentiles across the traces.}
    \label{fig:overall_eval}
\end{figure}

\subsection{Evaluation of Our Design Choices}
\label{sec:eval-design-choices}
We begin by evaluating the design choices that enable \sysname{} to recover from packet loss while maintaining high QoE. Specifically, we assess the benefits of our cross-layer design, the impact of our loss functions on reconstruction quality, and our choice of temporal context for neural recovery.

\vspace{-1em}
\subsubsection{\textbf{Benefits of Cross-Layer Design}}
\label{sec:cross-layer-benefits}
Figure~\ref{fig:cross_layer_motivation} shows how \sysname{}'s cross-layer recovery reduces both I- and P-frame losses compared to \sysname{}-L3 and \sysname{}-L7. During videoconferencing, both frame types may be lost. Losing an I-frame causes all dependent P-frames in the GoP to become unusable (\emph{GoP collapse}): 29 P-frames per I-frame when GoP is 30, while P-frames may also be lost independently. Since \sysname{}-L3 protects I-frames with FEC, it achieves an I-frame loss rate of just $0.40\%$. However, it cannot recover independently lost P-frames. In contrast, \sysname{}-L7 can reconstruct such P-frame losses but lacks I-frame protection, resulting in a $1.57\%$ I-frame loss rate ($3.9\times$ higher than \sysname{}-L3). We further observe that independent P-frame losses dominate the overall P-frame loss. Consequently, \sysname{}-L3 incurs a $3.35\%$ P-frame loss rate, $1.7\times$ higher than \sysname{}-L7 ($1.96\%$). Although \sysname{}-L7 suffers from more GoP collapses, its ability to recover independent P-frame losses keeps its overall P-frame loss lower.

\sysname{} combines the complementary strengths of both approaches to achieve the lowest loss rates for both frame types. $0.40\%$ of I-frames are lost, matching \sysname{}-L3 due to identical I-frame protection. At the same time, independently lost P-frames are recovered by our neural loss recovery module, reducing the P-frame loss rate to 0.52\%. This is a $6.4\times$ (resp. $3.7\times$) reduction over \sysname{}-L3 (resp. \sysname{}-L7). Thus, our cross-layer approach preserves substantially more usable frames and enables \sysname{} to maintain high QoE even under severe packet loss.


\vspace{-1em}
\subsubsection{\textbf{Efficacy of Modality-Aware Neural Recovery}}
\label{sec:neural-recovery-analysis}
We next evaluate the effect of our loss functions on neural recovery quality. 
Recall that we use $\mathcal{L}_{RGB}$ and $\mathcal{L}_{Depth}$ for recovering RGB and depth content, respectively. Figure~\ref{fig:loss_comp} (\S\ref{sec:neural-loss-recovery}) highlights the improvement qualitatively: $\mathcal{L}_{Depth}$ suppresses patch-boundary discontinuities, whereas these artifacts remain visible with $\mathcal{L}_{RGB}$. In this section, we report the improvement in depth frame reconstruction quantitatively (Table~\ref{tab:loss_psnr_ssim}). Averaged across all test videos, SSIM (resp. PSNR) increases from 0.963 (resp. 32.56 dB) to 0.971 (resp. 33.75 dB). These gains indicate that $\mathcal{L}_{Depth}$ better preserves depth map structures. Thus, while $\mathcal{L}_{RGB}$ is effective for color reconstruction, preserving geometric consistency in depth requires $\mathcal{L}_{Depth}$. Together, they maximize the final RGB-D QoE.


\begin{figure}[t]
\centering
\includegraphics[width=\linewidth]{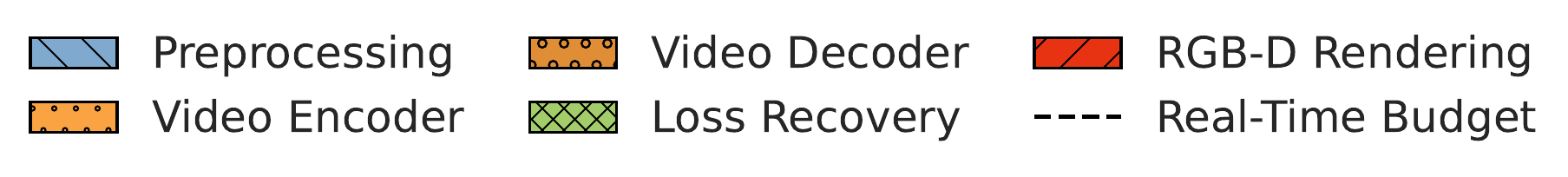}
\includegraphics[width=\linewidth]{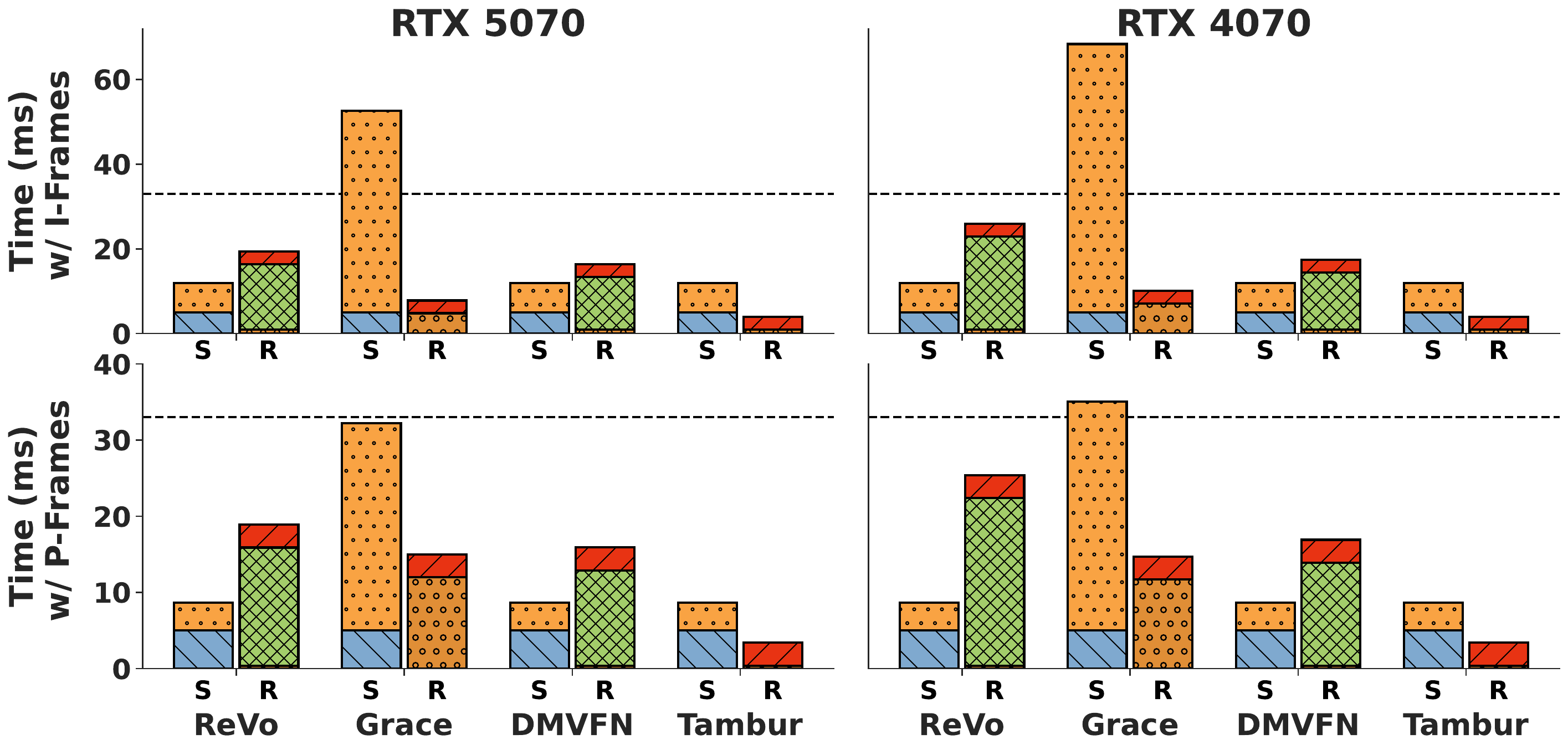}
\vspace{-1em}
\caption{Microbenchmarking sender/receiver latencies of \sysname{} and baselines (S: Sender, R: Receiver).}
\label{fig:microbenchmarking}
\vspace{-2em}
\end{figure}

\vspace{-1em}
\subsubsection{\textbf{Optimal Temporal Context for Inference}}


We now validate our choice of temporal context ($k$), that is, the number of clean reference frames provided to the recovery model. Inference latency is sensitive to $k$: increasing $k$ can improve accuracy but also raise latency. 
Figure~\ref{fig:benchmark_lossrec_latency} shows the inference latencies at different values of $k$. Based on these results, we set $k=5$, balancing accuracy while preserving a sufficient computational budget for other pipeline components, such as video decoding and RGB-D rendering. Higher values of $k$ cause \sysname{} to exceed the 33 ms deadline for 30 fps streaming, while lower values of $k$ reduce reconstruction SSIM. For example, at $k=7$, although the inference latency with RTX 4070 is $30.6$ ms, the total receiver latency exceeds 33ms in RTX 4070, violating the real-time budget. Additional ablation results are in Appendix~\ref{app:paramter_ablation}.

\vspace{-1em}
\subsubsection{\textbf{System Performance with Our Design Choices}}

Figure~\ref{fig:visual_rgbd_recovery} qualitatively shows how \sysname{} accurately reconstructs corrupted frames to provide high-quality videos. It compares the Ground Truth (GT), corrupted RGB and depth frames, and the corresponding reconstructions. Despite substantial degradation in the corrupted inputs, our model removes artifacts and recovers both RGB and depth effectively, yielding high-quality RGB-D video.

Figure~\ref{fig:overall_eval} reports the $25$th percentile, median, and $75$th percentile PointSSIM values across the three network types. \sysname{} consistently achieves high PointSSIM for both RGB and depth across the percentiles and network types. In particular, median PointSSIM reaches up to $0.90$ (resp. $0.97$) for RGB (resp. depth) across network types, with depth consistently achieving higher reconstruction quality. These results show that our cross-layer, modality-aware design delivers high QoE under diverse packet-loss conditions.


\begin{figure*}[t]
    \centering
    \includegraphics[width=0.7\textwidth]{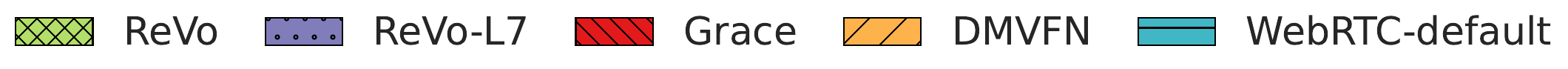}
    \begin{subfigure}[b]{0.30\textwidth}
        \centering
        \includegraphics[width=\textwidth]{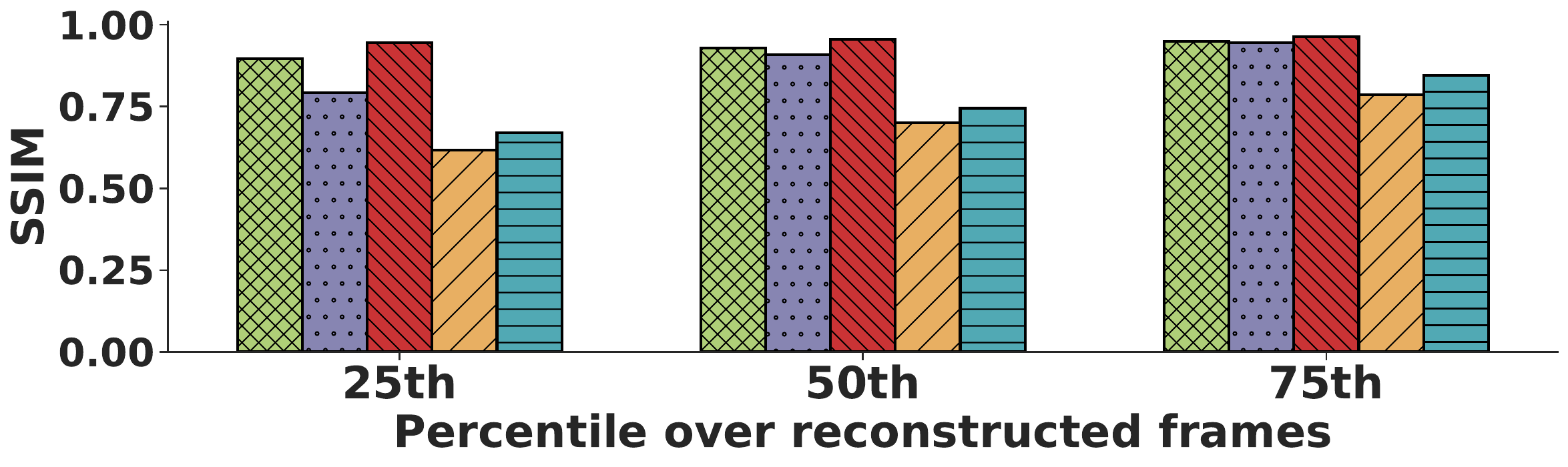}
        \caption{Cellular (RGB frames)}
        \label{fig:cell_trace_quality}
    \end{subfigure}
    \hfill
    \begin{subfigure}[b]{0.30\textwidth}
        \centering
        \includegraphics[width=\textwidth]{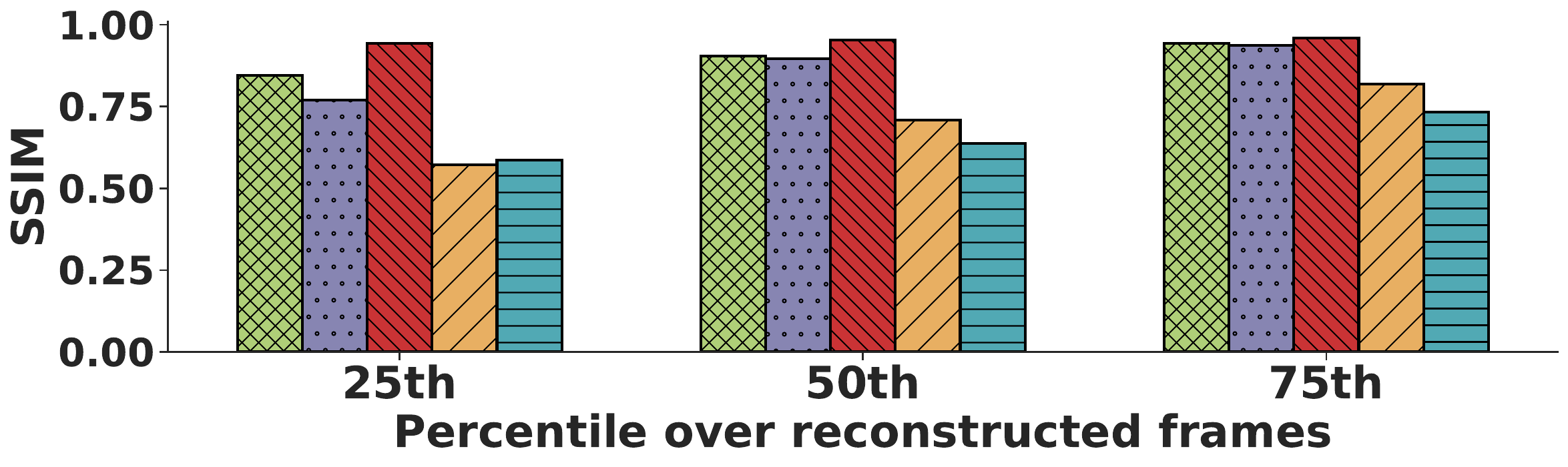}
        \caption{WiFi (RGB frames)}
        \label{fig:wifi_trace_quality}
    \end{subfigure}
    \hfill
    \begin{subfigure}[b]{0.30\textwidth}
        \centering
        \includegraphics[width=\textwidth]{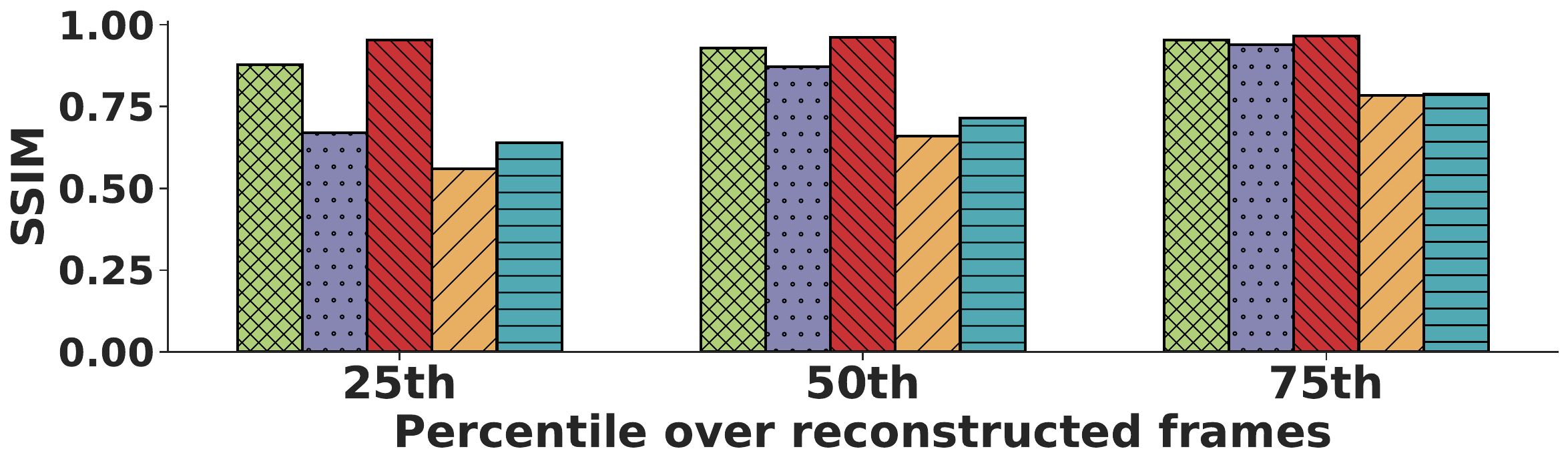}
        \caption{Ethernet (RGB frames)}
        \label{fig:eth_trace_quality}
    \end{subfigure}

    \vspace{0.2em}

    \begin{subfigure}[b]{0.30\textwidth}
        \centering
        \includegraphics[width=\textwidth]{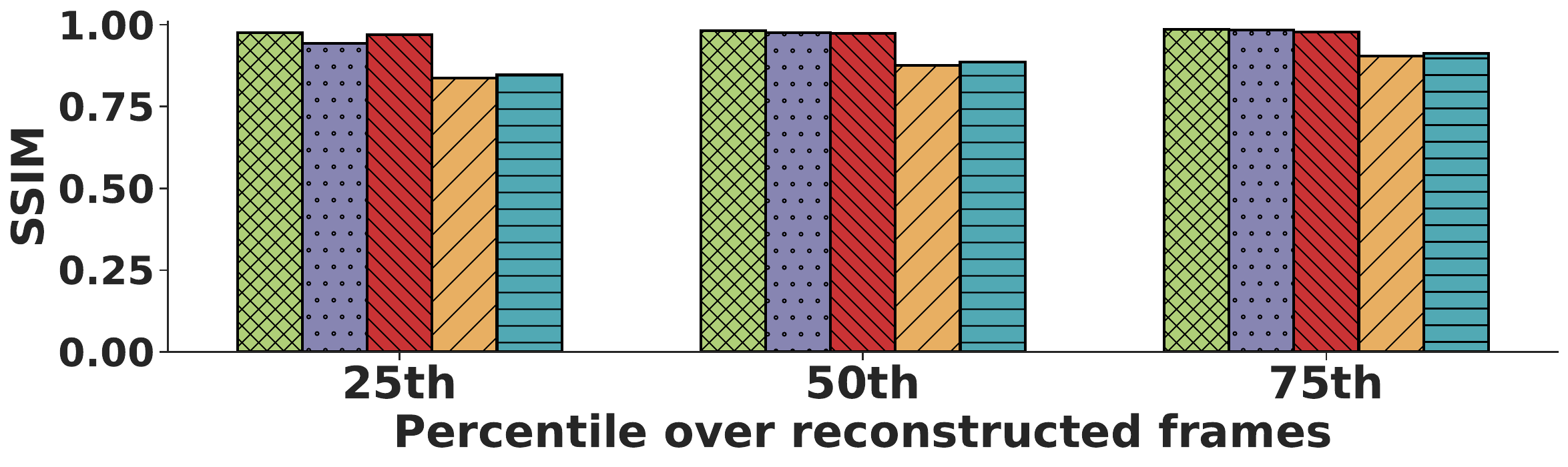}
        \caption{Cellular (Depth frames)}
        \label{fig:cell_trace_quality_depth}
    \end{subfigure}
    \hfill
    \begin{subfigure}[b]{0.30\textwidth}
        \centering
        \includegraphics[width=\textwidth]{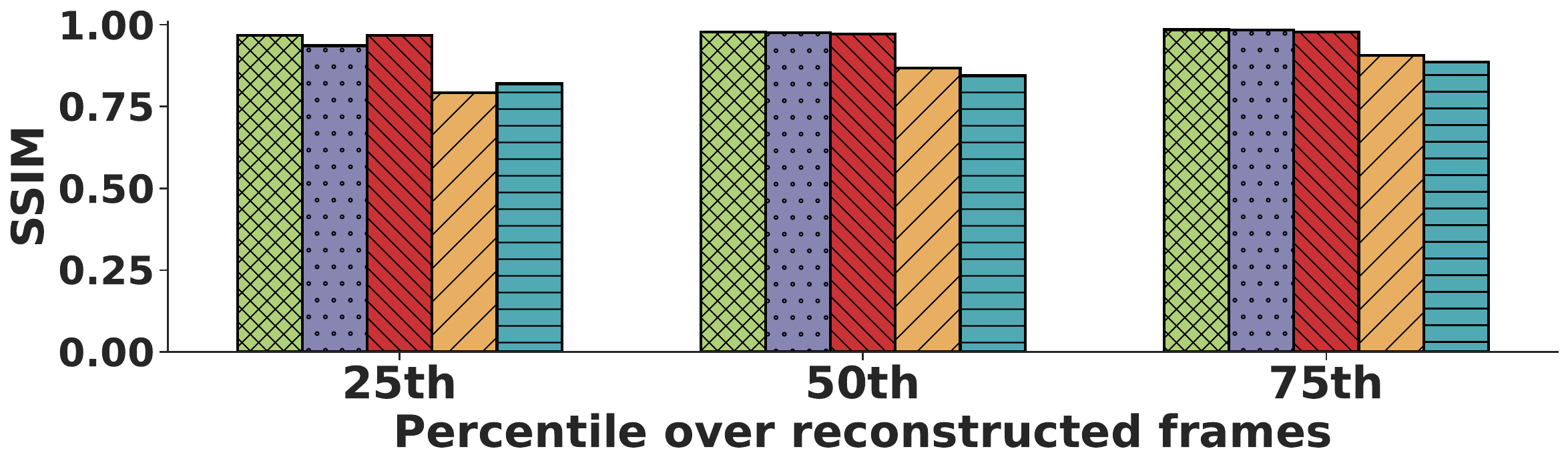}
        \caption{WiFi (Depth frames)}
        \label{fig:wifi_trace_quality_depth}
    \end{subfigure}
    \hfill
    \begin{subfigure}[b]{0.30\textwidth}
        \centering
        \includegraphics[width=\textwidth]{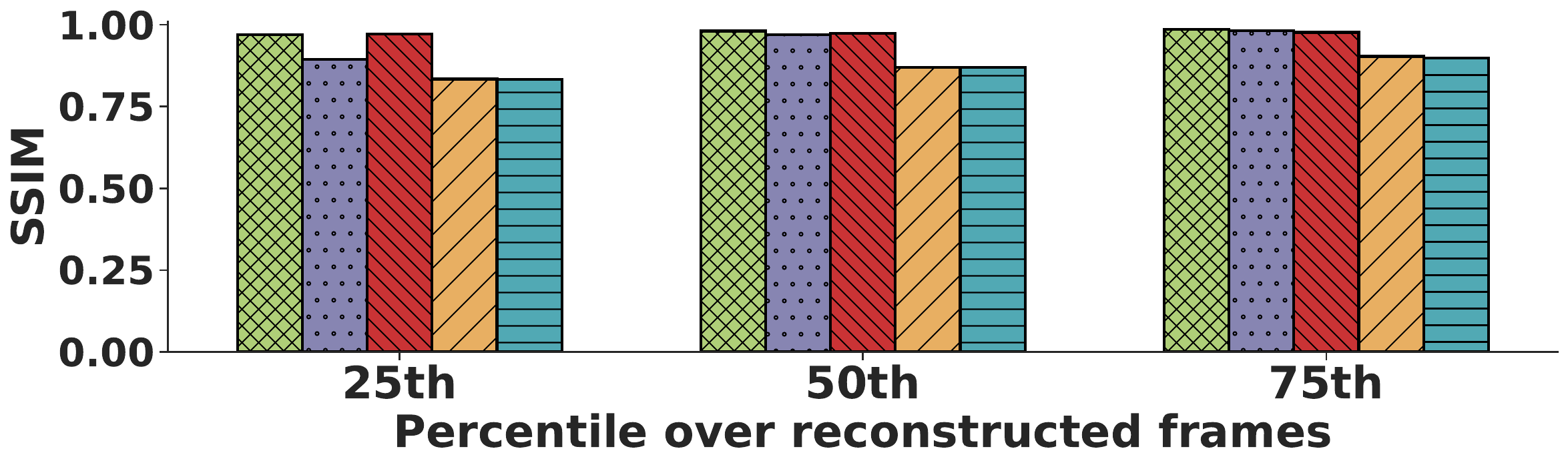}
        \caption{Ethernet (Depth frames)}
        \label{fig:eth_trace_quality_depth}
    \end{subfigure}
    \vspace{-1em}
    \caption{SSIM at the 25th, 50th, and 75th percentiles for RGB ((a)--(c)) and depth frames ((d)--(f)) across cellular, WiFi, and ethernet traces. \sysname{} consistently achieves high SSIM across different network conditions. }
    \label{fig:application_layer_rgb_and_D}
    \vspace{-1em}
\end{figure*}

\vspace{-1em}

\subsection{Comparison with Baselines}
\label{sec:eval-baseline-comparison}

We now compare \sysname{} against baselines that operate solely at L3 or L7. While some methods may deliver high reconstruction quality, they may be infeasible to run in real time on desktop-grade hardware used during videoconferencing. We therefore first assess real-time feasibility in our setup, and then compare performance using the metrics defined in \S~\ref{sec:exp-setup}. Since the baselines are designed for 2D, we apply their RGB recovery mechanisms to depth streams as well. Although such mechanisms are not optimized for depth, this isolates their recovery behavior under identical packet loss.

\vspace{-1em}
\subsubsection{\textbf{Microbenchmarking for Real-Time Feasibility}}

Figure~\ref{fig:microbenchmarking} breaks down the sender/receiver latencies on both GPUs. 
For \sysname{}, at the sender, preprocessing takes $5$~ms, and encoding up to $7$~ms, for a total of $12$~ms per frame. At the receiver, decoding takes up to $0.9$~ms, neural loss recovery upto $22$~ms, and RGB-D rendering $3$~ms, for a total of $25.9$~ms per frame. Both sender and receiver pipelines operate well below the $33.33$~ms per-frame deadline, showing that \sysname{} supports interactive streaming at 30 fps. Results for the other codecs are provided in Appendix~\ref{app:microbenchmark-across-codecs}.

Figure~\ref{fig:microbenchmarking} also reports the latencies of the baselines. \tambur{} and \emph{DMVFN}, both using \emph{H.265}, meet the real-time deadline.\emph{ However,} \grace{} \emph{violates the 33.3~ms budget, with I-frame encoding taking $47.7$~ms on average on RTX 5070 and $63.4$~ms on RTX 4070.} Thus, \sysname{} and all the baselines \emph{except} \grace{} are feasible at interactive frame rates (30+ fps) on desktop-grade hardware, making them practically deployable for volumetric videoconferencing. We omit WebRTC-default since its latency is identical to \tambur{}.


\vspace{-1em}
\subsubsection{\textbf{Comparison with L7 Loss Recovery Baselines}}

We compare \sysname{} against four L7 baselines: \emph{\sysname{}-L7}, \emph{DMVFN}, \emph{WebRTC-default}, and \grace{}. Although \grace{} is not real-time in our setting, we still include it because it is a state-of-the-art method with strong reconstruction accuracy. Figure~\ref{fig:application_layer_rgb_and_D} reports SSIM distributions for both RGB and depth. Across all network types, \emph{DMVFN} and \emph{WebRTC-default} perform moderately, achieving median SSIM between 0.64 and 0.75 (resp. 0.84 and 0.89) for RGB (resp. depth). \emph{\sysname{}-L7} performs better, with median SSIM between 0.87 and 0.91 (resp 0.87 and 0.98) for RGB (resp. depth). Among the baselines, \grace{} achieves the highest reconstruction accuracy, with median SSIM reaching 0.96 (resp. 0.97) for RGB (resp. depth) across different traces.

\sysname{} consistently outperforms all baselines except \grace{}, and attains a median SSIM of $0.93$ (cellular/ethernet) and $0.90$ (WiFi) for RGB. For depth frames, \sysname{} achieves a median SSIM of $0.98$ across all traces. The weaknesses of the L7 baselines are most evident under consecutive frame losses. \emph{DMVFN} degrades rapidly due to error accumulation during extrapolation. \emph{\sysname{}-L7} and \emph{WebRTC-default} suffer severe SSIM degradation after an I-frame loss, since playback freezes for the remainder of the GoP. In contrast, \sysname{} combines neural recovery for independent P-frame losses with selective FEC for I-frames, reducing both error propagation and GoP freezes.

Although \sysname{} does not outperform \grace{}, it achieves comparable SSIM (median SSIM within 5.3\%). \sysname{} also offers two practical advantages. First, unlike GRACE, it is deployable in real time on desktop-grade GPUs. Second, GRACE assumes reliable I-frame delivery and does not handle I-frame loss. Under I-frame loss, GRACE degrades sharply, resulting in 25.37\% lower median SSIM than \sysname{} (Figure~\ref{fig:grace_comp_L7} in Appendix~\ref{app:grace-comparison}).

Overall, \sysname{} is deployable on desktop-grade GPUs, and achieves up to 32\% (resp. 13\%) higher SSIM for RGB (resp. depth) than the feasible L7 baselines. We observe similar results in terms of PSNR (Figure~\ref{fig:psnr_l7_comp_median} in Appendix~\ref{app:additional_comp_l7}).

\begin{figure*}
    \centering
    \includegraphics[width=0.55\textwidth]{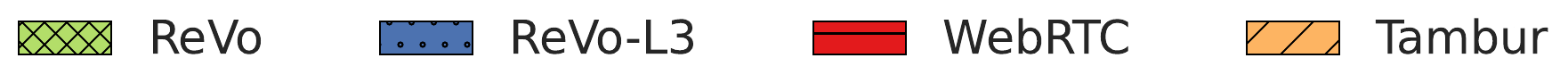}

    \begin{subfigure}[b]{0.3\textwidth}
        \centering
        \includegraphics[width=\linewidth]{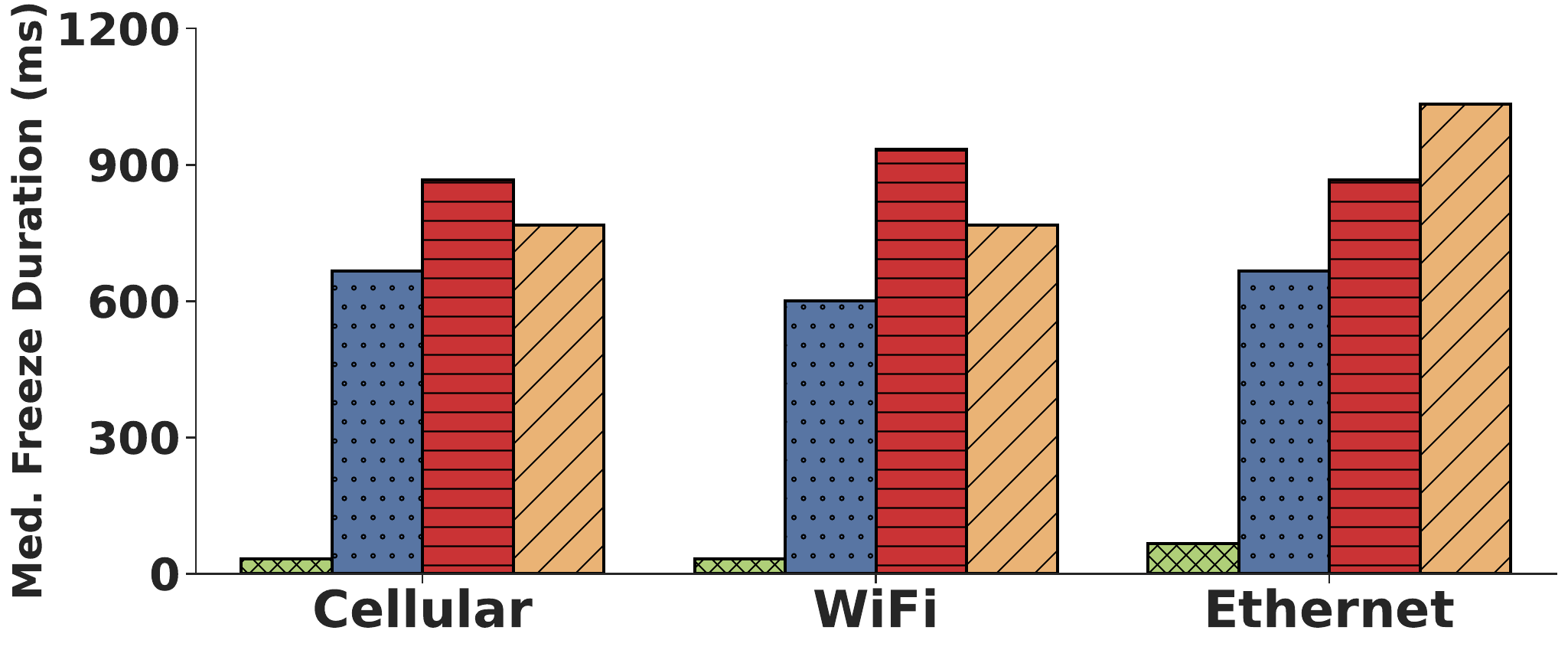}
        \caption{Median Duration of Freezes (ms)}
    \end{subfigure}
    \hfill
    \begin{subfigure}[b]{0.3\textwidth}
        \centering
        \includegraphics[width=\linewidth]{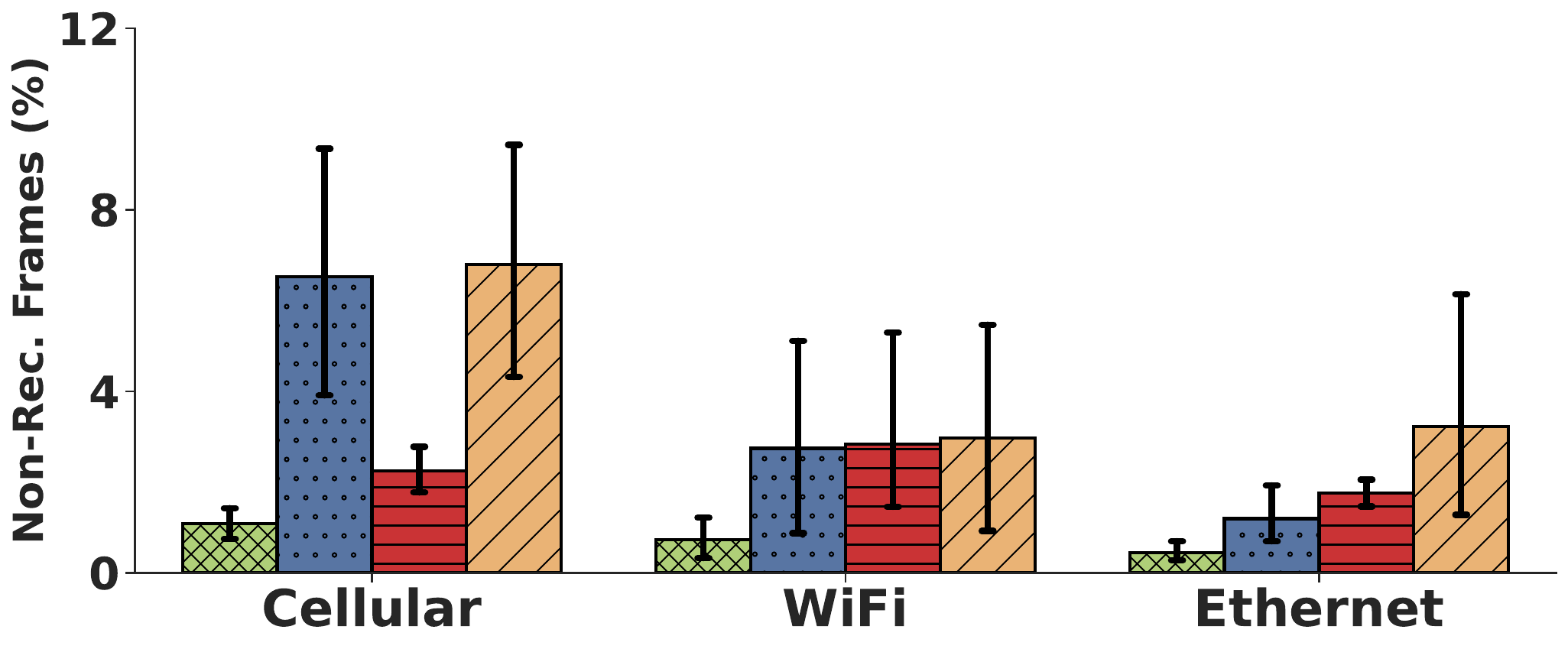}
        \caption{Non-Recovered Frames ($\%$)}
    \end{subfigure}
    \hfill
    \begin{subfigure}[b]{0.3\textwidth}
        \centering
        \includegraphics[width=\linewidth]{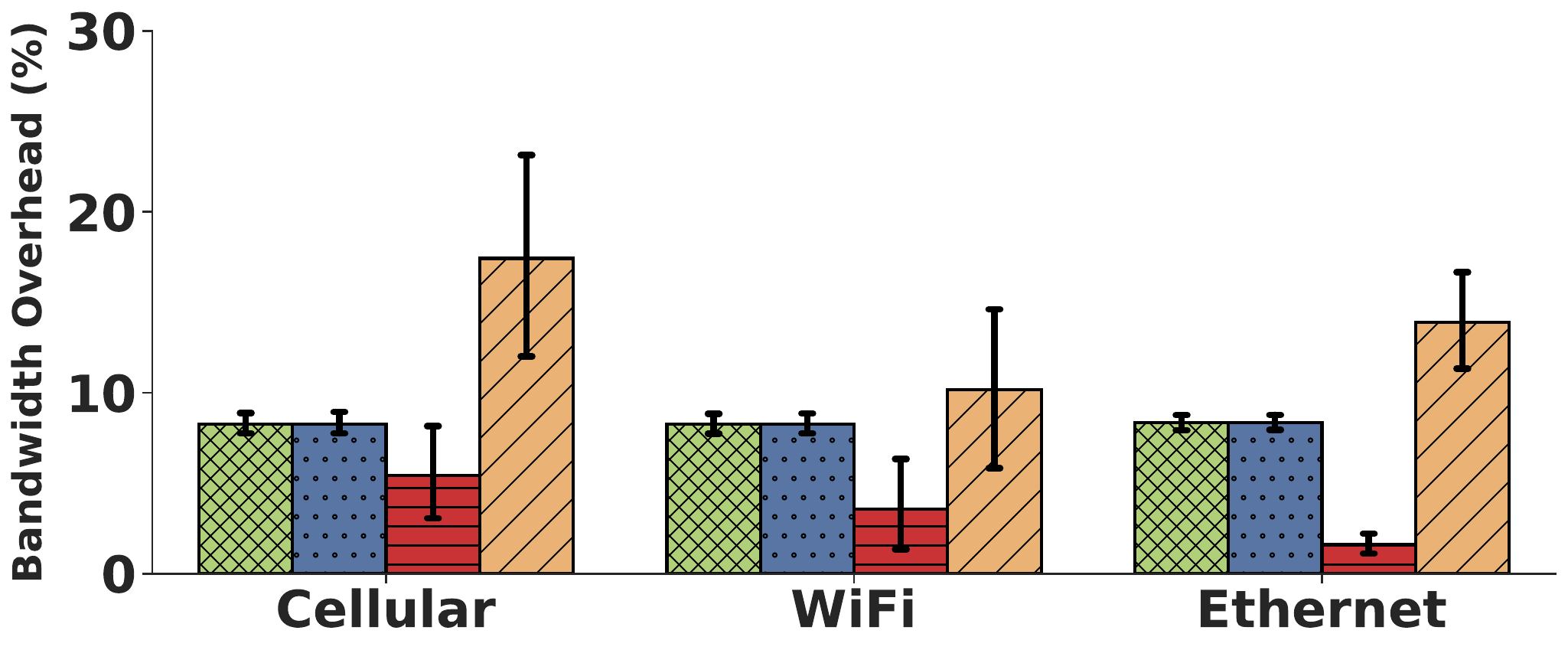}
        \caption{Bandwidth Overhead ($\%$)}
    \end{subfigure}
    \vspace{-1em}
    \caption{(a) \sysname{} achieves lowest median duration of freezing compared to baselines. (b) It also significantly reduces the number of non-recovered frames, and (c) has low bandwidth overhead, showing higher network efficiency.}
    \label{fig:network_layer_comp}

    \vspace{-1.5em}
\end{figure*}
\vspace{-1em}
\subsubsection{\textbf{Comparison with L3 Loss Recovery Baselines}} 
\label{sec:l3_comp}
Next, we compare \sysname{} against L3 baselines: \tambur{}, and \emph{\sysname{}-L3}. We again include \emph{WebRTC-default} as a reference. These methods apply different levels of FEC redundancy, and any I- or P-frame not received due to insufficient redundancy or network delay causes playback to freeze for the remainder of the GoP, consistent with default codec behavior. 

Figure~\ref{fig:network_layer_comp} summarizes the median freeze duration, percentage of non-recovered frames, and bandwidth overhead. The baselines experience considerable freezing, with median freeze duration reaching up to  $1.03 s$ (\tambur{} on Ethernet). In contrast, median freeze duration in \sysname{} is 33 ms. This is because when a P-frame is lost independently, \sysname{} still enables the codec to decode subsequent corrupted frames for reconstruction, effectively limiting freezes to a single frame. Overall, \sysname{} reduces the median freeze duration by up to $95.7\%$ 
The percentage of non-recovered frames can reach up to $6.79\%$ (\tambur{} in Cellular) among the baselines, substantially degrading QoE. In contrast, \sysname{} recovers all partially lost P-frames, reducing the percentage of non-recovered-frame in cellular to 1.08\%. Across all network types, we observe \sysname{} reduces the percentage of non-recovered by up to $86\%$.

\sysname{} and \sysname{}-L3 also incur up to $52.5\% $ lower bandwidth overhead than \tambur{}. While \tambur{} dynamically adapts redundancy, it uses a long lookahead window to set the redundancy rate for all frames, which tends to overpredict loss and yields consistently high overhead. \tambur{} can still suffer losses under congestion or when the loss rate exceeds what its maximum 100\% redundancy can tolerate. In contrast, \sysname{} and \sysname{}-L3 apply FEC only to I-frames, resulting in only $8.3\%$ bandwidth overhead. Interestingly, WebRTC-default incurs lower bandwidth overhead than \sysname{} because it reacts to loss and often applies no redundancy. However, because loss events are typically short, this reactive strategy often adds FEC after losses subside, leading to longer freezes and more non-recovered frames.

Overall, \sysname{} reduces median freeze duration by up to $95.7\%$ and non-recovered frames by $86\%$ compared to L3-only baselines, achieving higher QoE and lower bandwidth overhead through cross-layer recovery.


\vspace{-1em}
\subsubsection{\textbf{Error Propagation and Reconstruction Analysis}}
\begin{figure}
    \centering
    \includegraphics[width=0.8\linewidth]{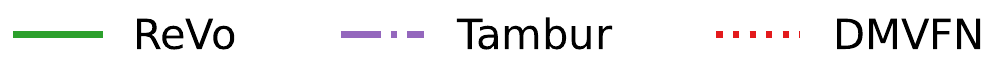}
    \includegraphics[width=0.85
    \linewidth]{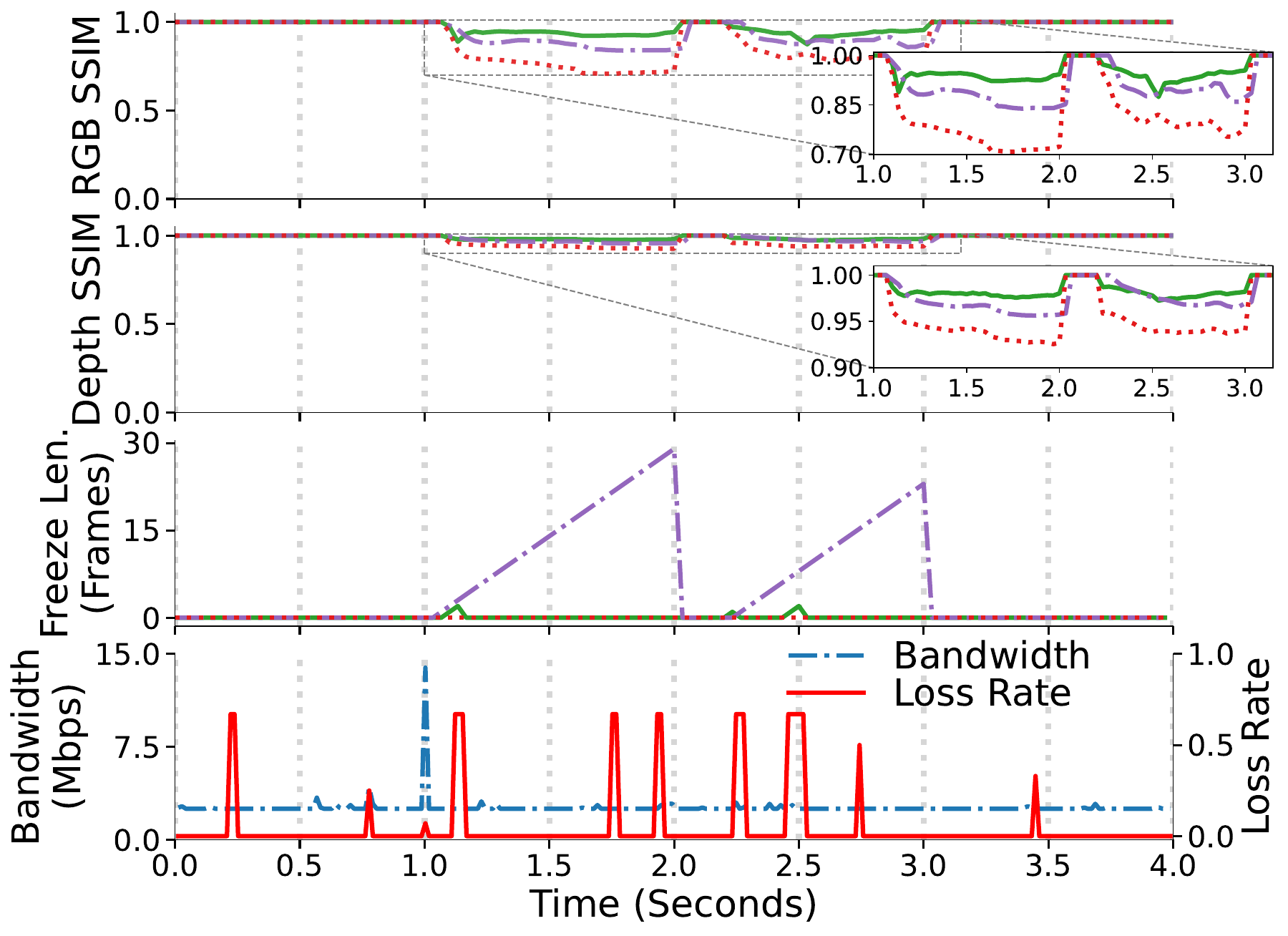}
    \vspace{-1em}
    \caption{Under bursty losses, \sysname{}’s cross-layer recovery reconstructs consecutive packet losses while maintaining low freeze duration.}
    \label{fig:time-series-comp}
\end{figure}

\vspace{-0.5em}
To understand how different recovery strategies behave under bursty loss, we analyze a four-second trace snippet in Figure~\ref{fig:time-series-comp}. During this interval, available bandwidth varies from $2.5~\mathrm{Mbps}$ to $13.9~\mathrm{Mbps}$, while packet loss ranges from $0\%$ to $67\%$. We evaluate video quality using freeze duration and SSIM over this time window. For brevity, we include one representative L3 baseline (\tambur{}) and one representative L7 baseline (\emph{DMVFN}) which are feasible in real time.

During a bursty loss event at 1.03 s, the first P-frame in a GoP is lost. \tambur{} freezes playback for the remainder of the GoP, while \sysname{} freezes only for the lost P-frame and reconstructs subsequent frames. \emph{DMVFN} extrapolates the frame, avoiding any freezes.
Interestingly, \tambur{} attains higher SSIM than \emph{DMVFN} despite freezing. A likely reason is that the TalkingHead dataset has limited motion, making a repeated frame closer to the ground truth, while \emph{DMVFN}'s extrapolation errors accumulate over the GoP.

In contrast, \sysname{} maintains a stable SSIM by reconstructing each corrupted frame, preventing error propagation. Overall, \sysname{} achieves the highest SSIM for both RGB and depth, while maintaining substantially shorter freeze durations than \tambur{}. 
Thus, by limiting video freezes and recovering corrupted frames throughout an error-propagation event, \sysname{} achieves higher SSIM and better playback continuity than the baselines.


\begin{figure}[!t]
    \centering
        \centering
        \includegraphics[width=0.8\linewidth]{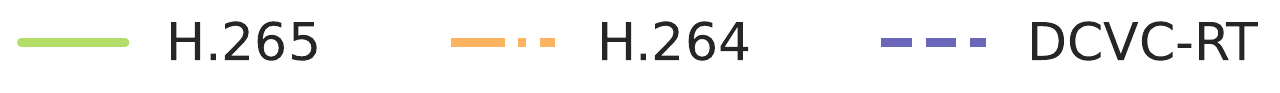}
        \begin{subfigure}{0.45\linewidth}
            \centering
            \includegraphics[width=\linewidth]{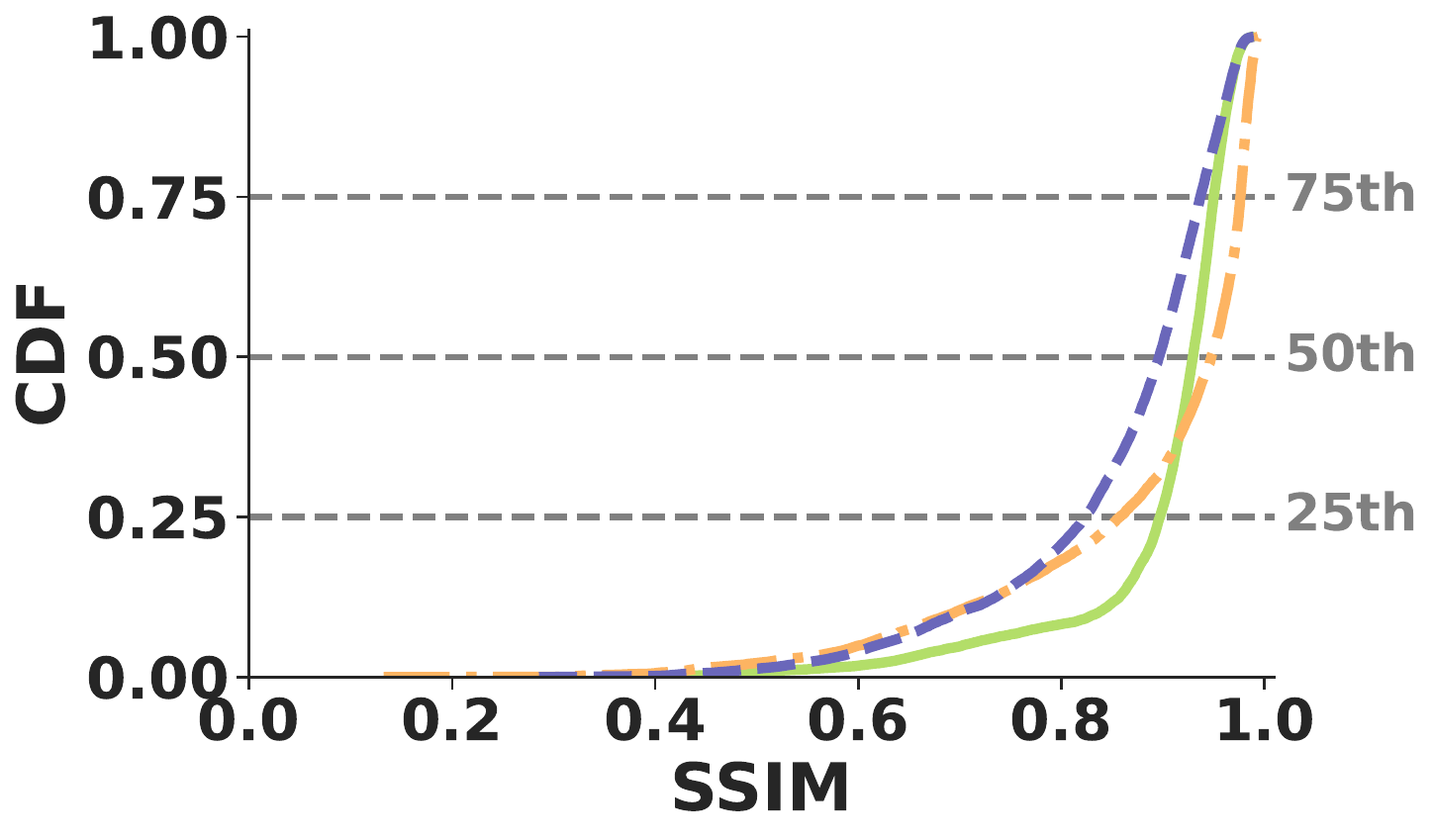}
            \caption{RGB}
            \label{fig:cdf-rgb}
        \end{subfigure}\hfill
        \begin{subfigure}{0.45\linewidth}
            \centering
            \includegraphics[width=\linewidth]{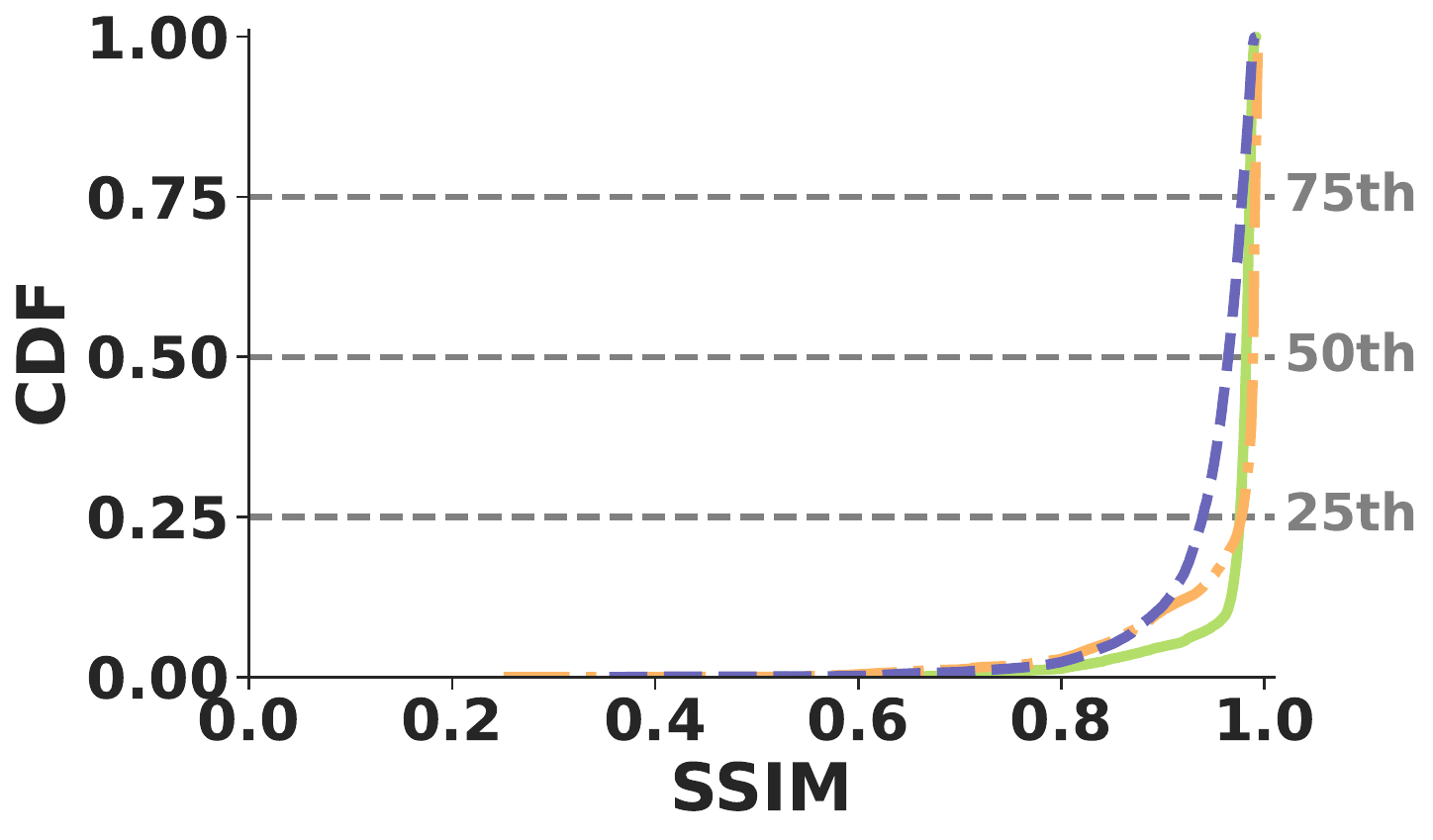}
            \caption{Depth}
            \label{fig:cdf-depth}
        \end{subfigure}
        \caption{\sysname{} SSIM CDF across codecs.}
        \label{fig:codec-agnostic}
    \vspace{-1em}
\end{figure}


\begin{figure}[!t]
    \centering
    \includegraphics[width=0.8\linewidth]{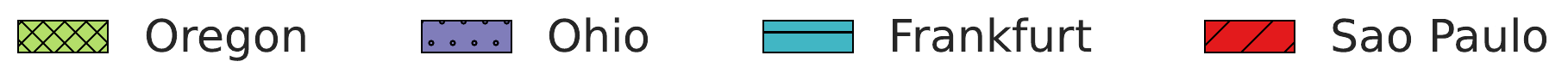}
    \begin{subfigure}[t]{0.45\columnwidth}
        \centering
        \includegraphics[width=\linewidth]{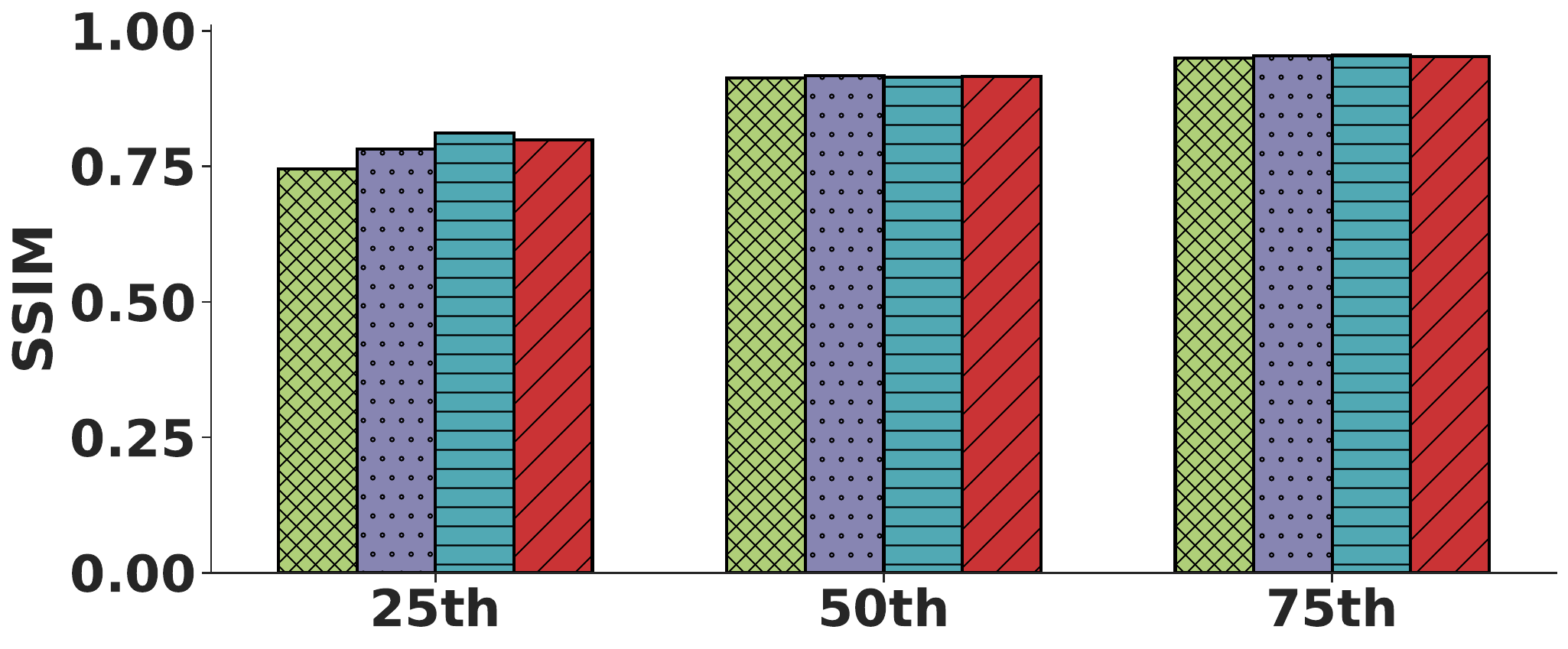}
        \caption{RGB (percentile)}
    \end{subfigure}\hfill
    \begin{subfigure}[t]{0.45\columnwidth}
        \centering
        \includegraphics[width=\linewidth]{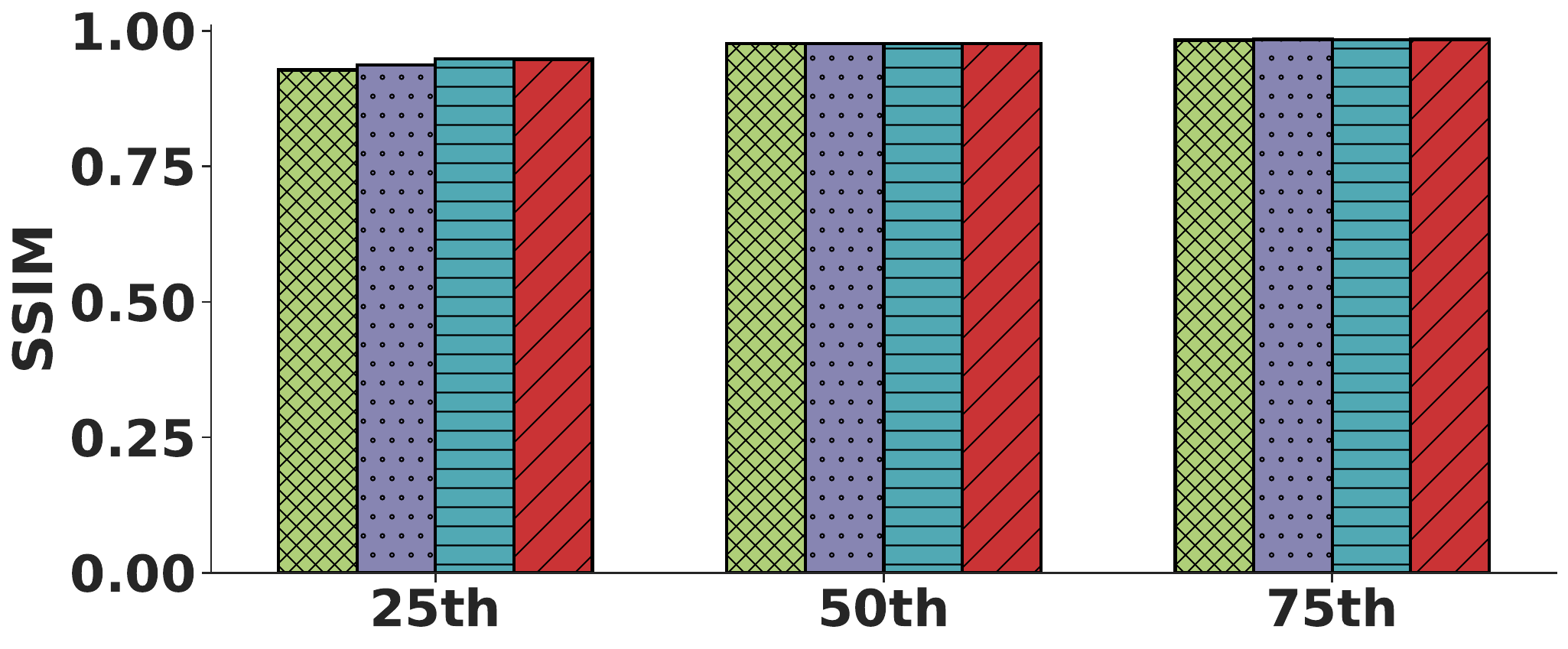}
        \caption{Depth (percentile)}
    \end{subfigure}\hfill
    \vspace{-1em}
    \caption{AWS deployment results spanning four regions.}
    \label{fig:aws_deployment}

\end{figure}

\subsection{Evaluation With Different Codecs} 
\label{sec:eval-codec-agnostic}
We now report \sysname{} performance across three different video codecs: \emph{H.264}~\cite{h264} and \emph{H.265}~\cite{h265}, and \emph{DCVC-RT}~\cite{dcvcrt}. Since DCVC-RT requires the sender and receiver to use the same GPU model, we run the DCVC-RT experiments only on Nvidia RTX 5070. We evaluate all codecs under identical network conditions using representative cellular traces, with packet loss rates ranging from $0\%$ to $75\%$ and an average available bandwidth of $4.16~\mathrm{Mbps}$.

Figure~\ref{fig:codec-agnostic} compares reconstruction quality across codecs using SSIM. With \emph{H.264}, \sysname{} achieves a median SSIM of $0.95$ (resp. $0.99$) for RGB (resp. depth). With \emph{H.265}, median SSIM is $0.93$ (resp. $0.98$) for RGB (resp. depth). With \emph{DCVC-RT}, \sysname{} achieves a median SSIM of $0.90$ (resp. $0.96$) for RGB (resp. depth). One reason for slightly lower SSIM with DCVC-RT may be that it compresses the frames more aggressively than other codecs. Consequently, more portions of the frames are corrupted at similar losses, leaving fewer recoverable cues at high loss rates. Similar to Figure~\ref{fig:application_layer_rgb_and_D}, SSIM degrades at the tail when I-frames are lost, and the entire GoP collapses, leading to video freezes.
Overall, these results show that \sysname{} can work effectively with multiple codecs, consistently achieving median SSIMs of $> 0.9$.
\subsection{Performance in Wide Area Networks}
\label{sec:eval-aws-deployment}

We now evaluate \sysname{} over real-world Wide Area Networks (WANs) by deploying it on AWS. Our receiver runs locally in the US-East region on a machine equipped with Nvidia RTX 4070, while the sender runs on an AWS \textit{r6i.xlarge} EC2 instance. To assess generality across WANs, we deploy the sender in four AWS regions: Ohio (us-east-2), Oregon (us-west-2), Frankfurt (eu-central-1), and Sao Paulo (sa-east-1).

For each deployment, we measure the SSIM of reconstructed frames (Figure~\ref{fig:aws_deployment}). Across all regions, \sysname{} maintains a median SSIM above $0.91$ for RGB and $0.97$ for depth. At the tail ($25$th percentile), RGB SSIM is lower due to unrecoverable I-frame losses, consistent with the trace-driven results in Figure~\ref{fig:application_layer_rgb_and_D}. Thus, \sysname{} remains effective under real-world WAN conditions. Its consistently high median SSIM across different deployments shows both robustness and practical viability for volumetric videoconferencing.


\subsection{\sysname{} User Study}
\label{sec:eval-user-study}
\begin{figure}[t]
    \centering
        \includegraphics[width=0.8\linewidth]{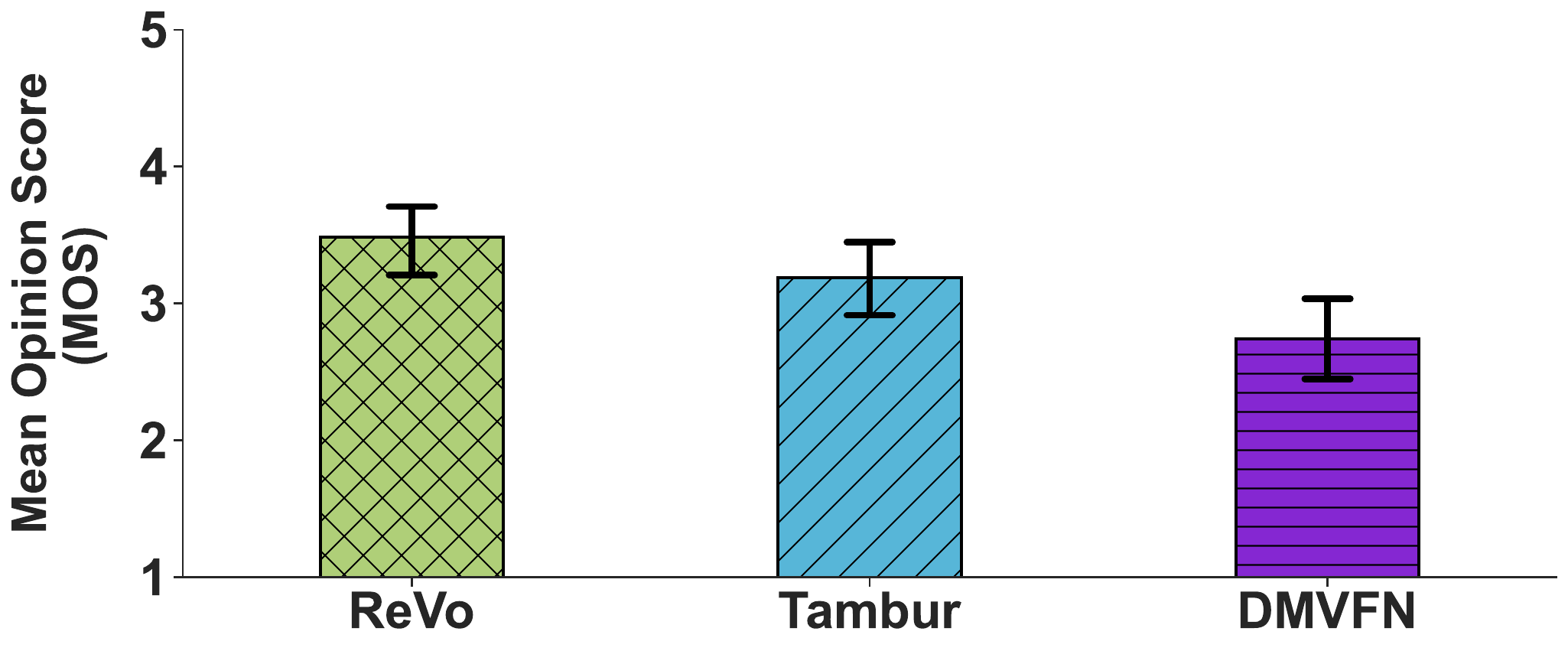}
    \vspace{-1em}
    \caption{Mean Opinion Score across systems.}
    \label{fig:user_study_mos}
\vspace{-1em}

\end{figure}

Finally, we conduct an IRB-approved user study to evaluate the perceptual quality of \sysname{}. We recruit $28$ participants via the crowd-sourced \emph{Prolific} platform, and administer the study using Qualtrics~\cite{qualtrics}.
For evaluation, we randomly sample a 30-second video clip from the TalkingHead dataset~\cite{thead}, and stream it using \sysname{}, and two real-time baselines \tambur{}~\cite{tambur} and \emph{DMVFN}~\cite{hu2023dynamic}, over a $30$-second segment of a cellular trace. During this interval, packet loss varies from $0\%$ to $67\%$, while the bandwidth ranges from $2.5$ to $21.41$ Mbps. Participants watch the videos from each system and rate both visual quality and temporal smoothness on a 5-point scale (1: poor, 5: excellent). We then compute the Mean Opinion Score (MOS) by averaging these two ratings. Figure~\ref{fig:user_study_mos} presents the resulting MOS values. \sysname{} achieves the highest MOS ($3.48$) among all the methods (up to $1.3\times$ higher than baselines), showing that \sysname{} delivers the best perceived quality among all evaluated systems and indicating a user preference for our approach.

\vspace{-1em}





%% file: 6-related-work.tex
\section{Related Work}
\label{sec:related-work}
 
\noindent \textbf{Reliable Real-Time 2D Video Streaming.} Several systems have addressed packet loss in real-time video streaming. Tambur~\cite{tambur} uses streaming codes to add redundancy and dynamically adapts the redundancy rate based on predicted network conditions to reduce bandwidth overhead. Hairpin~\cite{meng2024hairpin} jointly optimizes redundancy and retransmissions to achieve reliability with low overhead, including under tail loss. Grace~\cite{grace} and Reparo~\cite{reparo} develop loss-resilient neural codecs trained across a wide range of packet loss rates, enabling robust decoding under severe loss. However, these works target 2D videos and operate at a single layer (either the L3 or L7). In contrast, \sysname{} is a cross-layer loss recovery system for volumetric videos. Voxel~\cite{voxel} is also a cross-layer system, similar to our work, and aims to reduce the probability of losing critical content during transmission. However, it does not explicitly reconstruct lost content, as \sysname{} does.

\noindent \textbf{Volumetric Video Streaming.} Prior work has explored volumetric video streaming using explicit representations such as point clouds \cite{han2020vivo, guan2023metastream} or RGB-D~\cite{ghosh2025livo, lee2020groot}, as well as implicit neural representations \cite{wu2025nevo, li2025gifstream4dgaussianbasedimmersive, yuan20251000fps4dgaussian}. These systems primarily focus on reducing transmitted data to lower bandwidth requirements and enabling real-time volumetric applications. For example, LiVo~\cite{ghosh2025livo} dynamically adapts bandwidth allocation between RGB and depth content to enable videoconferencing in congested networks. 4DGS-1K~\cite{yuan20251000fps4dgaussian} and GIFStream~\cite{li2025gifstream4dgaussianbasedimmersive} use Gaussian splatting to stream high-quality volumetric videos in real time. GROOT~\cite{lee2020groot} employs parallelism to reduce volumetric video encoding and decoding latency, enabling streaming in mobile devices. Holoportation~\cite{orts2016holoportation} uses LZ4-based compression~\cite{lz4} to reduce transmission overhead. MetaStream~\cite{guan2023metastream} reduces the number of points in a captured point cloud. However, these works focus on bandwidth efficiency and latency under reliable delivery, and do not explicitly handle packet loss as \sysname{}. Recently, NeVo~\cite{wu2025nevo} proposed a lightweight learning-based recovery module for packet loss. However, NeVo is designed for video-on-demand streaming, while our work targets volumetric videoconferencing, which imposes significantly stricter end-to-end latency constraints.

%% file: 8-conclusions.tex
\vspace{-1em}
\section{Conclusion}
\label{sec:conclusions}
\vspace{-1em}
We present \sysname{}, a cross-layer system for loss-resilient volumetric videoconferencing that combines selective L3 protection with neural L7 recovery. By decoupling RGB and depth streams and using modality-specific reconstruction, \sysname{} tolerates packet loss under strict real-time constraints. Across real-world network traces, \sysname{} improves median SSIM by up to $32\%$ (resp. 13\%) for RGB (resp. depth) content and reduces video freezes by up to $95.7\%$ compared to existing techniques.

\noindent\textbf{Ethical concerns ---} This work does not raise any ethical issues.


%% file: A2-appendix.tex
\newpage
\section{\sysname{} Performance Across Codecs}
\label{app:perform_Across_codecs}
To evaluate the generalizability of our neural loss recovery module, we analyze its performance across three distinct video codecs: \emph{H.264}, \emph{H.265}, and \emph{DCVC-RT}. We assess both the qualitative visual characteristics of packet loss and the quantitative improvements in structural similarity (SSIM) provided by our reconstruction.
\vspace{-1em}

\subsection{Analysis of Codec-Specific Loss Artifacts}
\label{app:bitstream-map}
\begin{figure}[!ht]
    \centering

    \begin{subfigure}[t]{0.48\columnwidth}
        \centering
        \includegraphics[width=\linewidth]{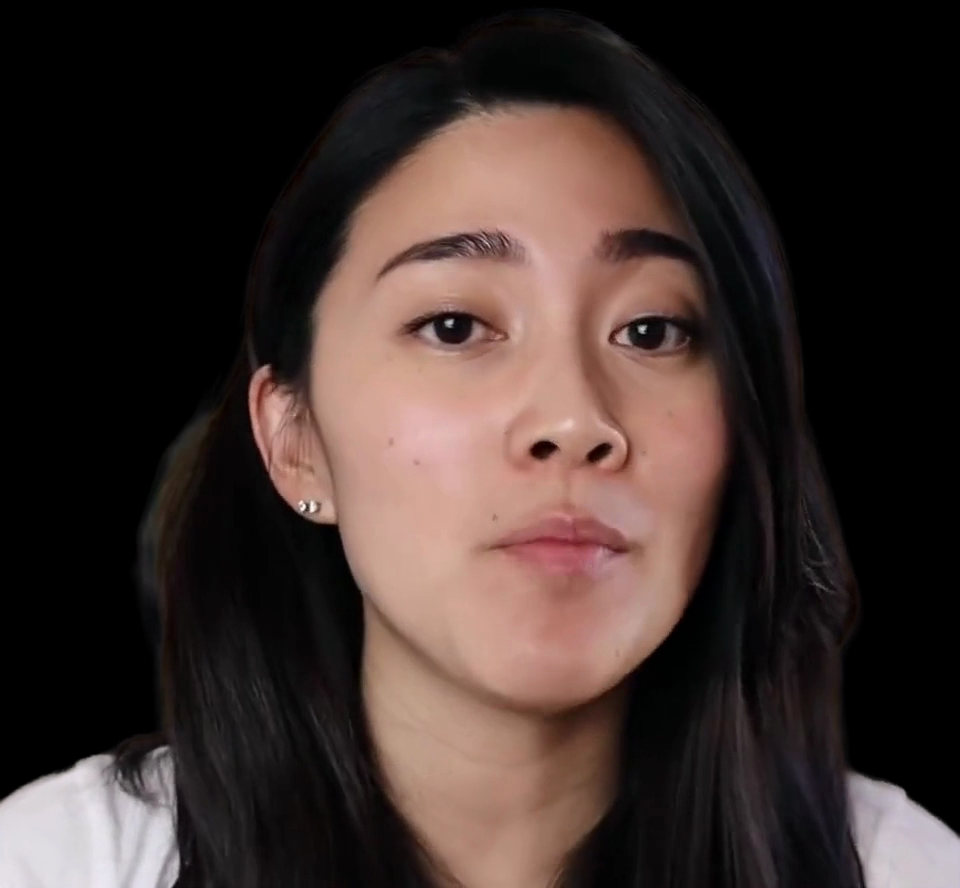}
        \caption{RGB Ground Truth}
        \label{fig:gt_rgb}
    \end{subfigure}
    \hfill
    \begin{subfigure}[t]{0.48\columnwidth}
        \centering
        \includegraphics[width=\linewidth]{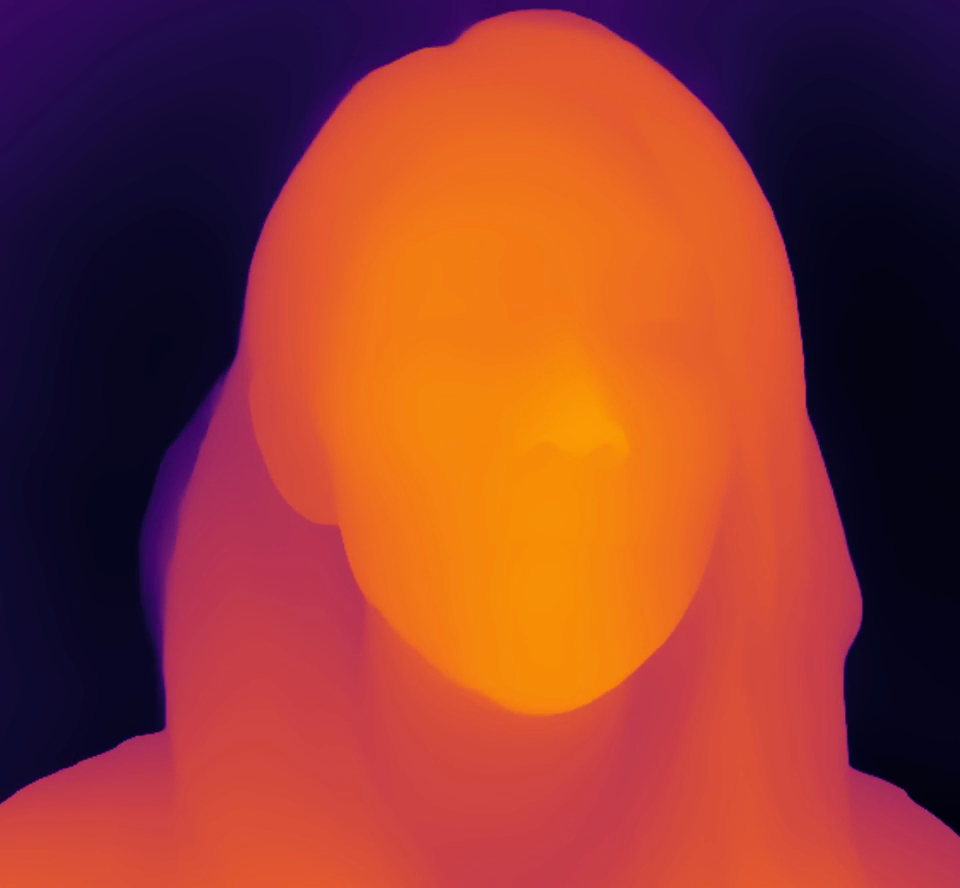}
        \caption{Depth Ground Truth}
        \label{fig:gt_depth}
    \end{subfigure}

    \vspace{0.5em}

    \begin{subfigure}[t]{0.48\columnwidth}
        \centering
        \includegraphics[width=\linewidth]{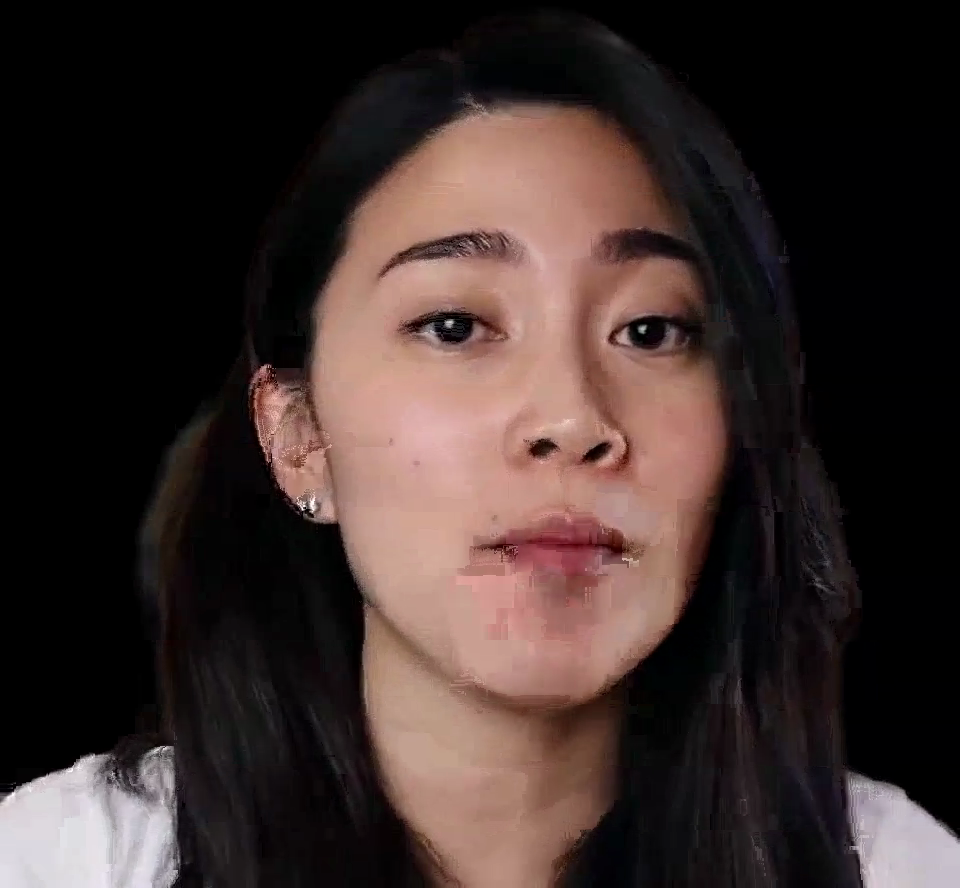}
        \caption{H.264 RGB}
        \label{fig:h264_rgb}
    \end{subfigure}
    \hfill
    \begin{subfigure}[t]{0.48\columnwidth}
        \centering
        \includegraphics[width=\linewidth]{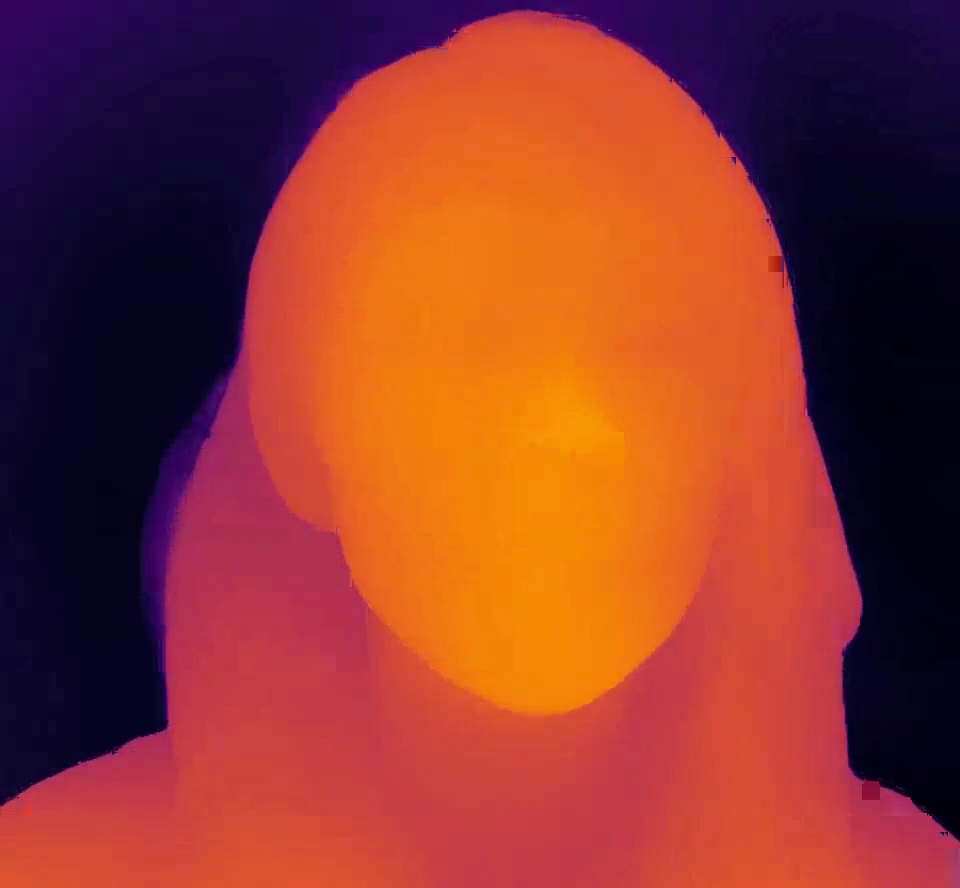}
        \caption{H.264 Depth}
        \label{fig:h264_depth}
    \end{subfigure}

    \vspace{0.5em}

    \begin{subfigure}[t]{0.48\columnwidth}
        \centering
        \includegraphics[width=\linewidth]{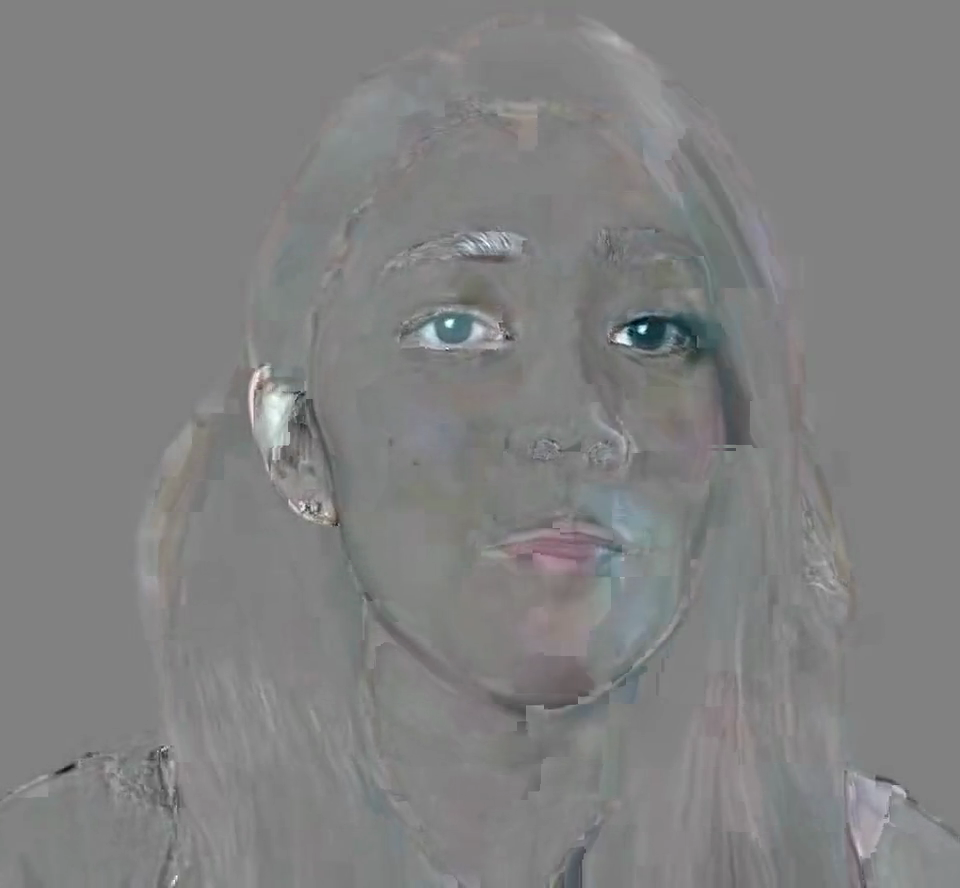}
        \caption{H.265 RGB}
        \label{fig:h265_rgb}
    \end{subfigure}
    \hfill
    \begin{subfigure}[t]{0.48\columnwidth}
        \centering
        \includegraphics[width=\linewidth]{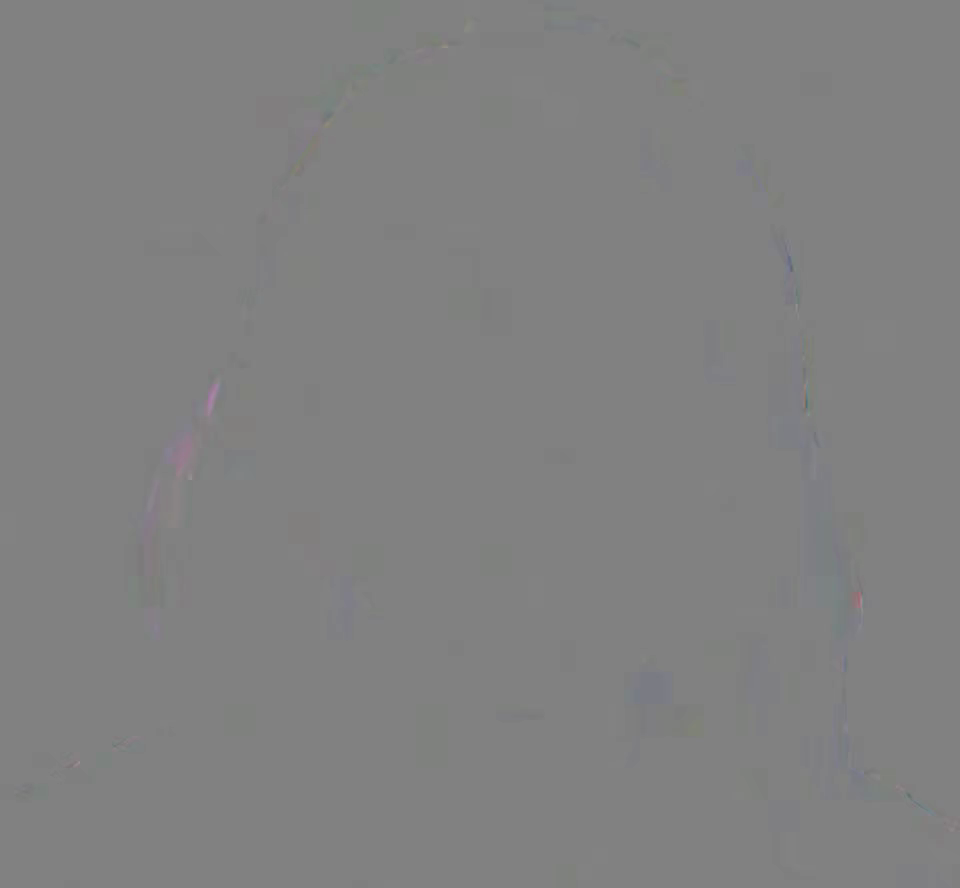}
        \caption{H.265 Depth}
        \label{fig:h265_depth}
    \end{subfigure}

    \vspace{0.5em}

    \begin{subfigure}[t]{0.48\columnwidth}
        \centering
        \includegraphics[width=\linewidth]{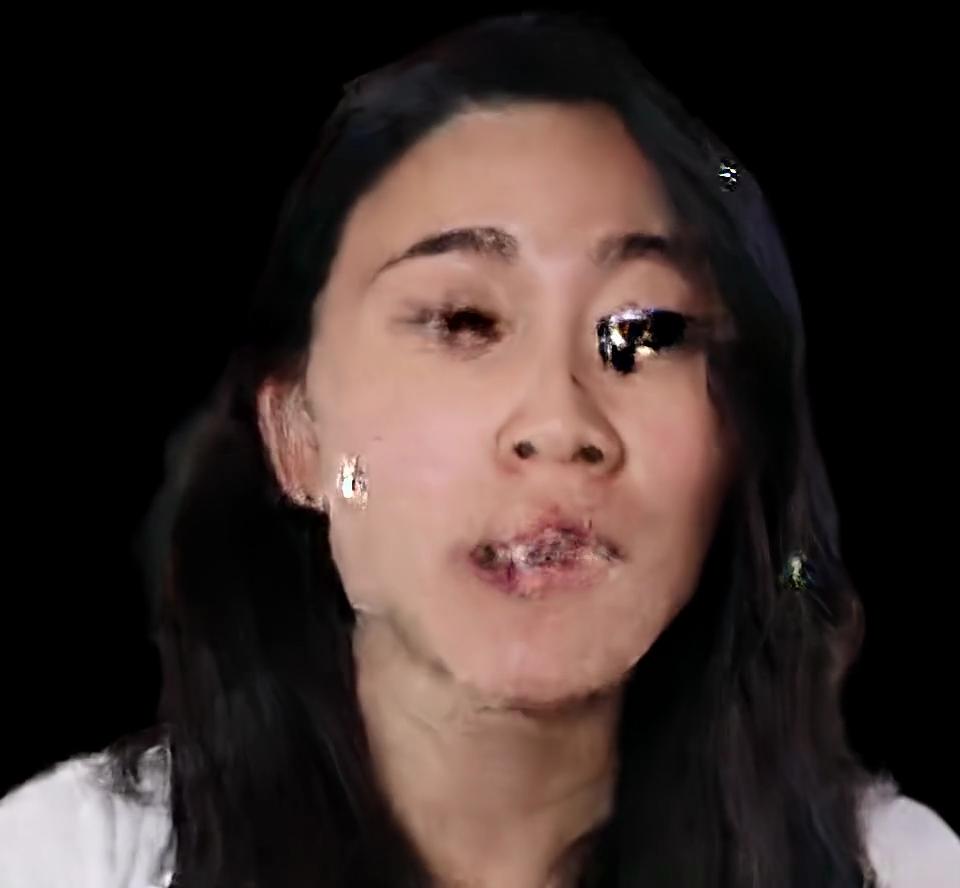}
        \caption{DCVC-RT RGB}
        \label{fig:dcvcrt_rgb}
    \end{subfigure}
    \hfill
    \begin{subfigure}[t]{0.48\columnwidth}
        \centering
        \includegraphics[width=\linewidth]{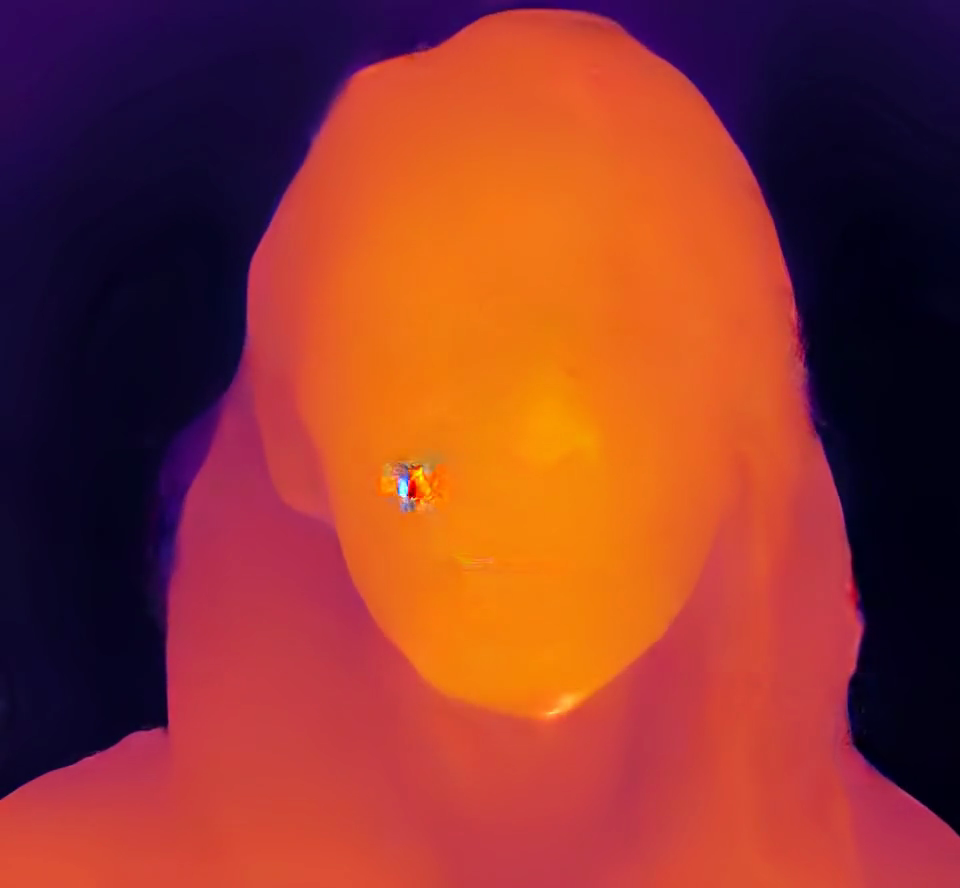}
        \caption{DCVC-RT Depth}
        \label{fig:dcvcrt_depth}
    \end{subfigure}

    \caption{Impact of packet loss on RGB and depth frames across H.264, H.265, and DCVC-RT streams, highlighting the unique visual artifacts produced by each compression scheme compared to the ground truth.}
    \label{fig:loss_across_codecs}
    \vspace{-2em}
\end{figure}

To understand the challenge of loss recovery across different compression schemes, Figure~\ref{fig:loss_across_codecs} provides a visual comparison of the reconstruction artifacts produced across codecs under packet loss. Traditional block-based codecs, such as \emph{H.264} and \emph{H.265}, tend to produce sharp, rectangular ``blocking'' artifacts. In contrast, \emph{DCVC-RT} exhibits more spatially contained distortions across the frame. \sysname{} adapts to codec-specific behaviors by finetuning it separately over such artifacts. 



\subsection{CDF of Reconstruction Accuracy}
\begin{figure*}[!hbtp]
    \centering
    \resizebox{0.8\textwidth}{!}{%
    \begin{minipage}{\textwidth}
    \centering
    
    \includegraphics[width=0.4\linewidth]{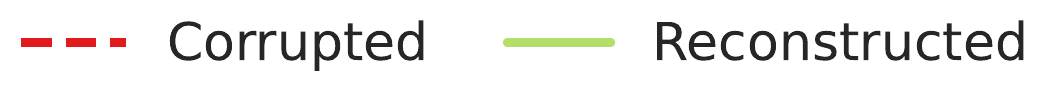}
    \vspace{2mm}

    \begin{subfigure}[b]{0.25\linewidth}
        \centering
        \includegraphics[width=0.9\textwidth]{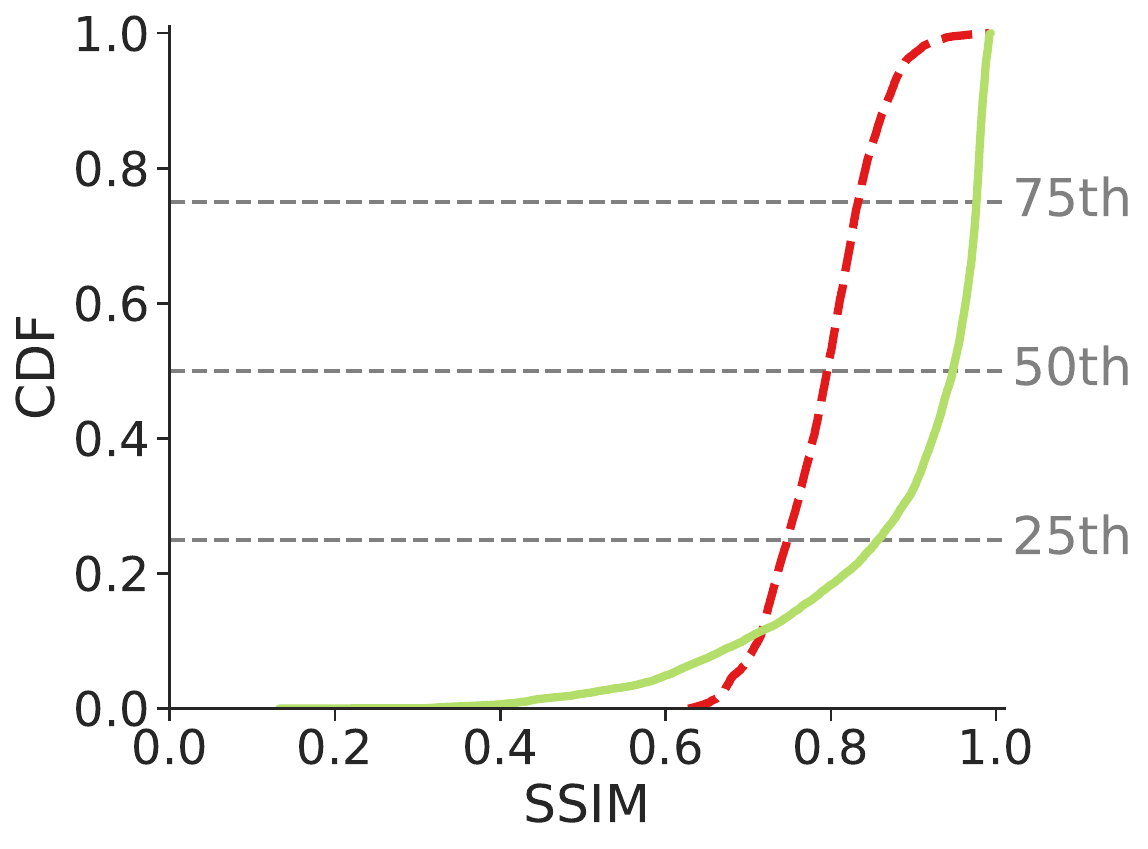}
        \caption{H.264}
        \label{fig:h264_rgb_corr_vs_recon}
    \end{subfigure}
    \hfill
    \begin{subfigure}[b]{0.25\linewidth}
        \centering
        \includegraphics[width=0.9\textwidth]{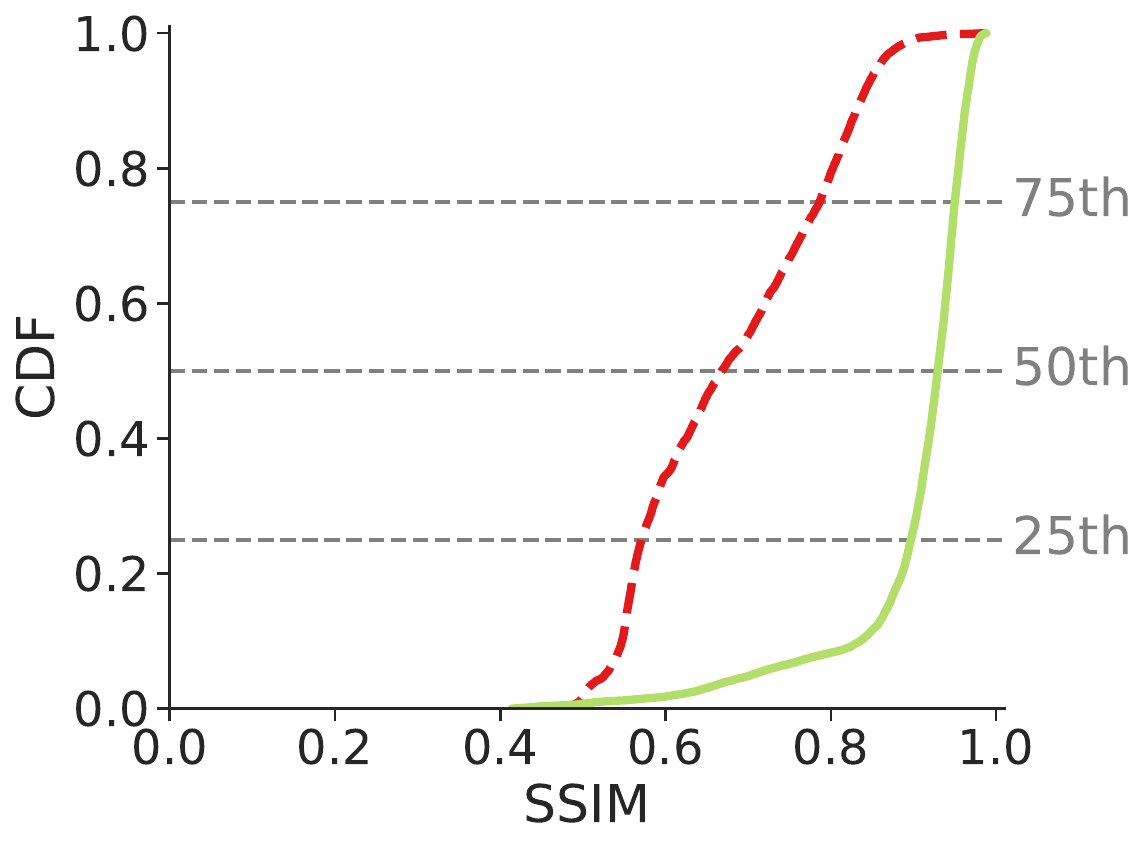}
        \caption{H.265}
        \label{fig:h265_rgb_corr_vs_recon}
    \end{subfigure}
    \hfill
    \begin{subfigure}[b]{0.25\linewidth}
        \centering
        \includegraphics[width=0.9\textwidth]{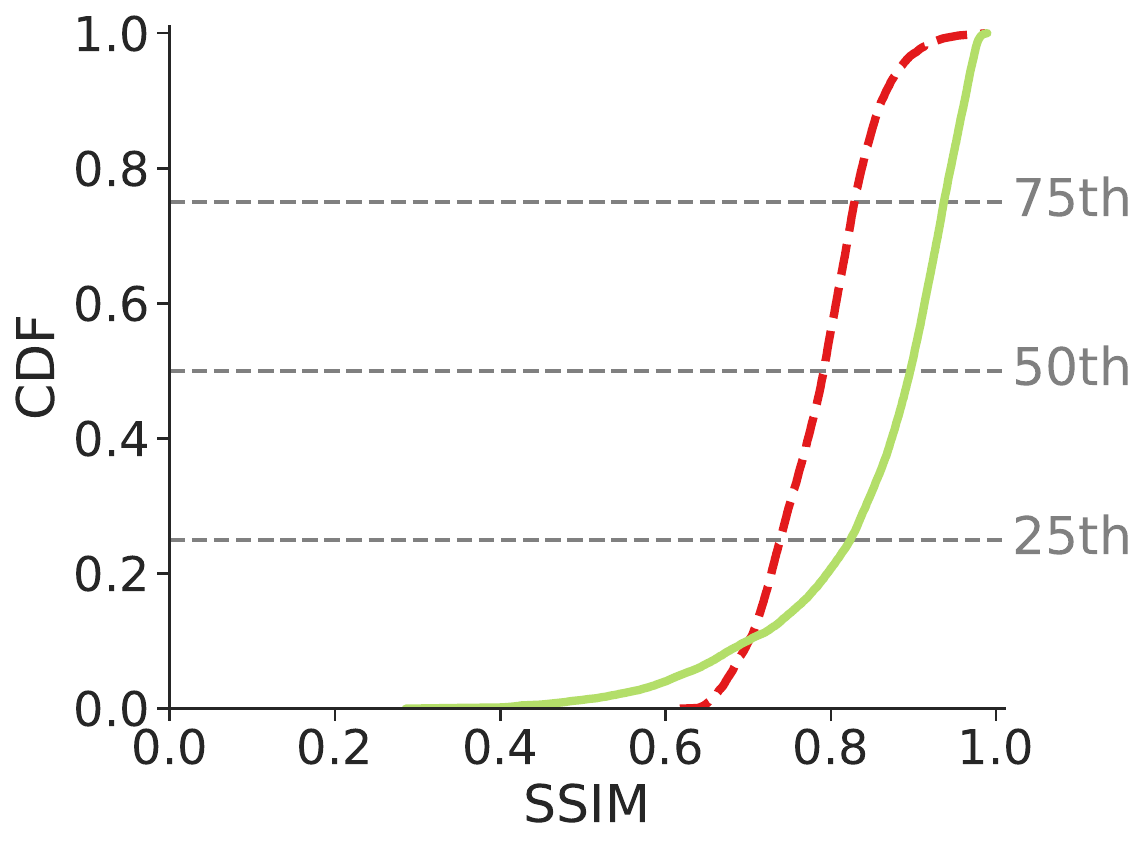}
        \caption{DCVC-RT}
        \label{fig:dcvcrt_rgb_corr_vs_recon}
    \end{subfigure}

    \begin{subfigure}[b]{0.25\linewidth}
        \centering
        \includegraphics[width=0.9\textwidth]{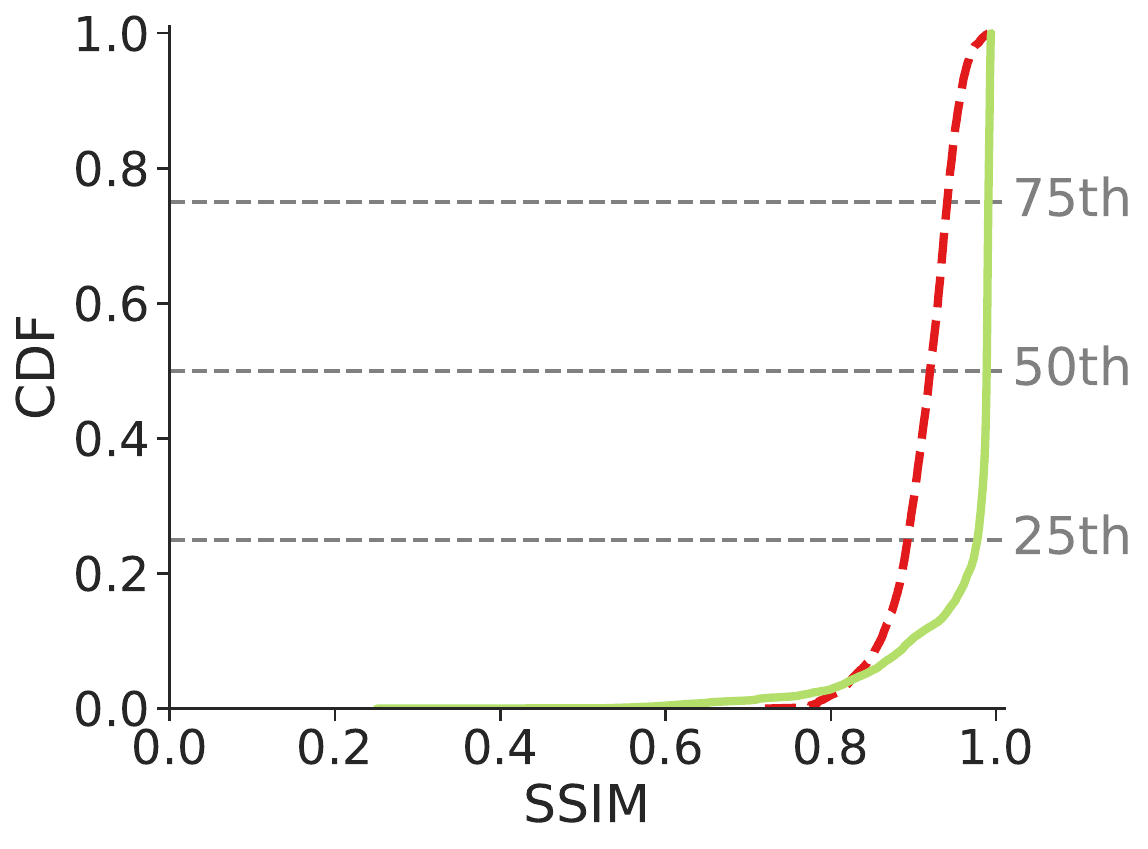}
        \caption{H.264}
        \label{fig:h264_depth_corr_vs_recon}
    \end{subfigure}
    \hfill
    \begin{subfigure}[b]{0.25\linewidth}
        \centering
        \includegraphics[width=0.9\textwidth]{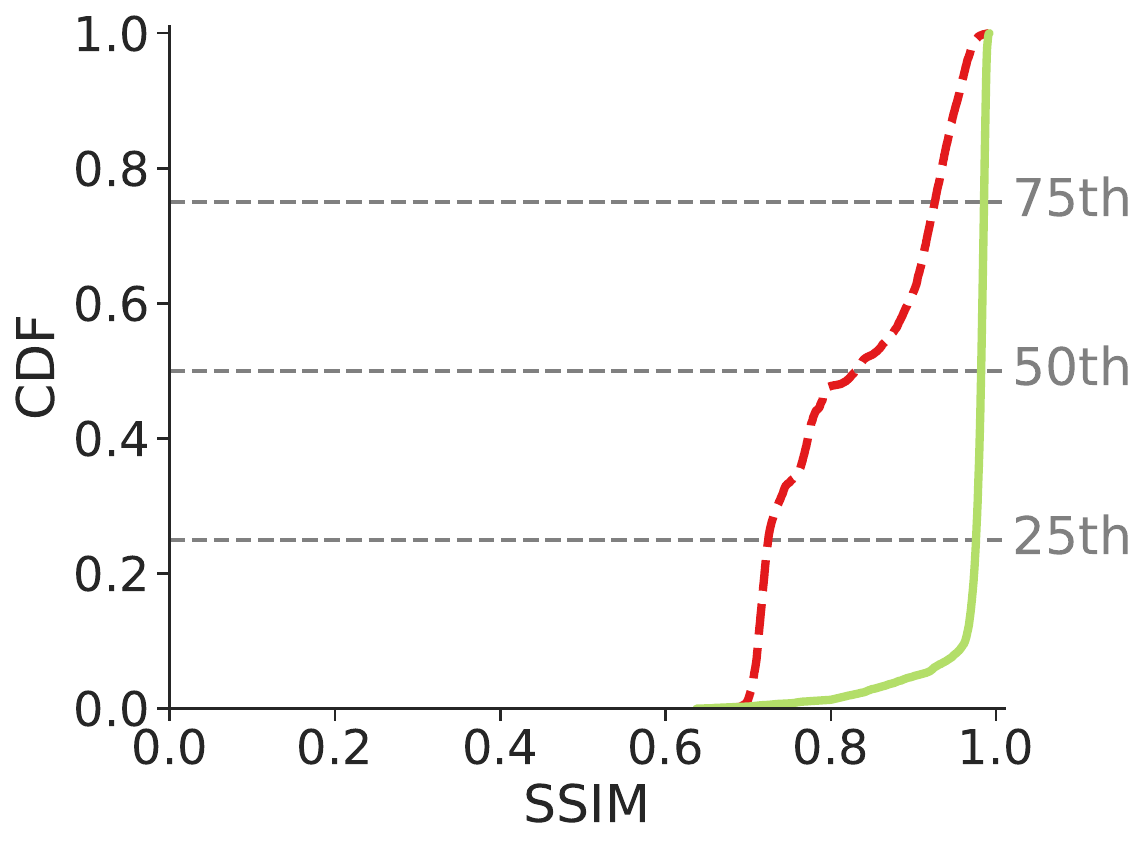}
        \caption{H.265}
        \label{fig:h265_depth_corr_vs_recon}
    \end{subfigure}
    \hfill
    \begin{subfigure}[b]{0.25\linewidth}
        \centering
        \includegraphics[width=0.9\textwidth]{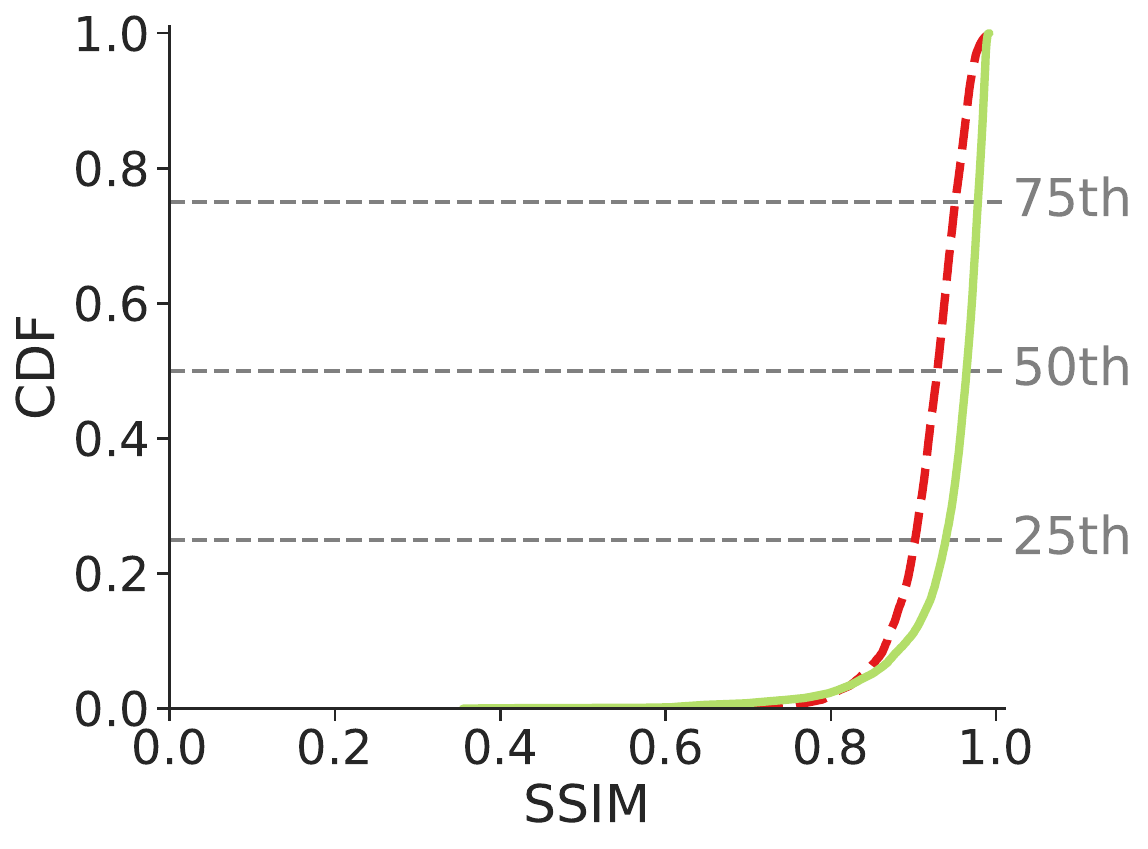}
        \caption{DCVC-RT}
        \label{fig:dcvcrt_depth_corr_vs_recon}
    \end{subfigure}
    \end{minipage}%
    }

    \caption{CDF of SSIM comparing corrupted and reconstructed streams across codecs and modalities. The top row shows RGB results, while the bottom row shows depth results.}
    \label{fig:codec_comparison}
\end{figure*}

Despite the varied and complex artifact characteristics produced by each codec, \sysname{} demonstrates robust, high-quality restoration. Figure~\ref{fig:codec_comparison} compares the SSIM CDF of the corrupted and reconstructed streams. Our results demonstrate that the recovery module consistently shifts the SSIM distribution toward higher quality, effectively suppressing both blocky and diffused distortions.

For RGB frames, we observe a significant improvement in median SSIM: \emph{H.264} improves from $0.80$ to $0.95$ (+$0.15$), \emph{H.265} improves from $0.67$ to $0.93$ (+$0.26$), and \emph{DCVC-RT} improves from $0.79$ to $0.90$ (+$0.11$). The gains are equally consistent for depth information. 
\emph{H.264} median SSIM increases from $0.92$ to $0.99$ (+$0.07$), \emph{H.265} from $0.83$ to $0.98$ (+$0.15$), and \emph{DCVC-RT} from $0.93$ to $0.96$ (+$0.03$). This highlights that \sysname{} effectiveness across different codecs.

%% file: A3-appendix.tex
\section{Additional Training Details of \sysname{}}
\label{app:additonal_train_details}
\vspace{-0.5em}
We further elaborate on our training objective and the rationale behind our design choices.
\vspace{-1em}
\subsection{Training Setup}
As described in \S\ref{sec:neural-loss-recovery}, our neural loss recovery model is built upon Video Vision Transformers (ViViT)~\cite{videomae}, a state-of-the-art family of architectures known for effectively modeling long-range spatio-temporal dependencies in videos. Our design differs from the original VideoMAE~\cite{videomae} architecture in five key ways:
\textbf{1. Architecture:} We use the same encoder-decoder architecture across both pretraining and finetuning. In contrast, \emph{VideoMAE} uses encoder-decoder during pretraining and encoder-only architecture during finetuning. 
\textbf{2. Pixel-Level Masking} We adopt \emph{pixel-level masking} instead of the patch-based masking strategy used in VideoMAE~\cite{videomae}.
\textbf{3. Pixel Exposure:} We feed both the visible and corrupted pixels to the model during training, rather than only the visible regions.
\textbf{4. Optimized Flash Attention:} We replace standard multi-head self-attention with an optimized scaled dot-product attention mechanism~\cite{dao2022flashattention} within the Transformer blocks. This significantly reduces the quadratic memory footprint and computational overhead, enabling efficient processing of long spatio-temporal video sequences.
\textbf{5. Elimination of Token Shuffling:} Because our encoder processes both visible and corrupted patches concurrently, we completely eliminate the complex token shuffling and un-shuffling operations required by standard masked autoencoders before the decoding phase. This streamlines the forward pass and preserves the inherent structural layout of the frames throughout the entire network.

These design choices are fundamentally tailored to our neural loss recovery task. Unlike standard masked autoencoders that simulate missing data by cleanly dropping entire image patches, our setting deals with real-world bitstream-level packet loss. As shown in Figure~\ref{fig:loss_across_codecs}, this type of degradation typically corrupts only a subset of pixels within a given patch. If we were to mask out and drop any patch that contained an error, we would needlessly discard the perfectly intact pixels residing within that same patch. By instead feeding the entire frame containing both the intact and the corrupted pixels into the network, we enable the model to leverage fine-grained spatial context from the surrounding local area, alongside temporal redundancy from prior frames, to accurately reconstruct the specific damaged pixels.

Furthermore, during pretraining, providing the model with access to all pixels encourages it to learn rich spatio-temporal relationships, enabling it to better understand motion dynamics and how visual content evolves spatio-temporal dimensions. This stronger prior is critical for accurately reconstructing corrupted video regions during downstream loss recovery.

\vspace{-1em}
 \subsection{Codec-Aware Finetuning Strategy}
 \label{app:sec-codec-aware-finetuning}
To effectively bridge the gap between our neural loss recovery module and practical video streaming systems, we tailor our finetuning process to the specific characteristics of standard video codecs. This involves two key strategies:

\noindent \textbf{Multi-QP Training:} Packet loss manifests differently depending on the underlying compression level. To ensure our model generalizes across varying network conditions and video qualities, we finetune the model across multiple Quantization Parameters (QPs). This exposes the model to a diverse spectrum of compression artifacts and loss patterns, making it robust to the variable bitrates typically encountered in real-world streaming. We believe our system could be easily extended to include adaptive bit rate algorithms. 

\noindent \textbf{Offline Loss Simulation:} Incorporating standard video codecs into the training loop to simulate bitstream-level packet loss on-the-fly introduces a severe computational bottleneck. To alleviate this, we decouple the loss simulation from the network training. We preprocess our entire training dataset by passing the videos through the codecs, injecting simulated network losses, and storing the resulting corrupted videos offline. The network is then finetuned directly on this pre-computed dataset, significantly accelerating overall training throughput without sacrificing the realism of the codec-induced degradation.

\vspace{-1em}
\subsection{\sysname{} Parameter Ablation}
\label{app:paramter_ablation}

In this section, we further discuss the parameter selections for the neural loss recovery module. Recall from \S\ref{sec:neural-loss-recovery} that our architecture adapts the ViViT backbone~\cite{videomae}. In this architecture, a "tubelet" acts as the spatiotemporal equivalent of a 2D image patch. Rather than processing individual frames independently, the model groups a specific number of consecutive frames together, denoted as the tubelet size ($T$)--into a single 3D block to form one input token.

The rendering latency of our ViViT backbone is fundamentally tied to the transformer's architecture. Specifically, the computational complexity of the self-attention mechanism scales directly with the number of spatiotemporal tokens generated from the input. This token count is heavily influenced by two key parameters: the total number of reference frames ($k$) and the temporal tubelet size ($T$). Decreasing $k$ limits the overall input sequence size, while increasing $T$ compresses the temporal dimension by packing more frames into a single token. Both adjustments significantly reduce computational overhead.

\begin{figure}
    \centering
    \includegraphics[width=0.7\linewidth]{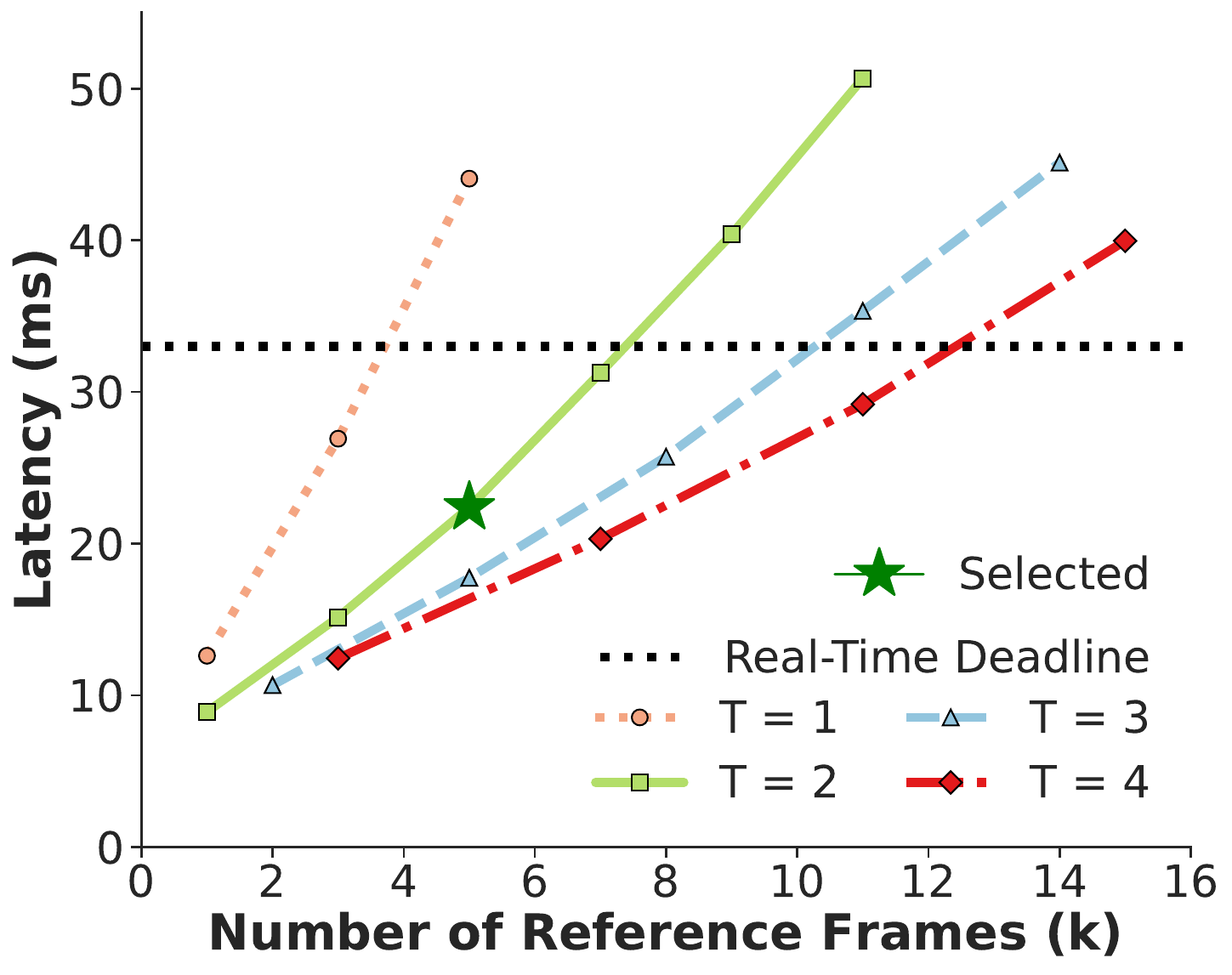}
    \caption{\sysname{} neural loss recovery ablation}
    \label{fig:ablate_lossrec}
    \vspace{-1em}
\end{figure}

To quantify this architectural impact, we evaluate the per-frame inference latency across different combinations of $k$ and $T$, as shown in Figure~\ref{fig:ablate_lossrec}. While the data shows that latency can be aggressively minimized by reducing $k$ or increasing $T$, doing so inherently degrades reconstruction accuracy due to a loss of fine-grained temporal context. Through empirical evaluation, we find that the choice of $k=5$ and $T=2$ enables this optimal balance. It achieves an inference latency of approximately $22$ ms per frame, comfortably satisfying the real-time deadline, while (1) retaining sufficient temporal context to maximize the reconstruction quality of corrupted frames, and (2) leaving enough room for video decoding and RGB-D rendering.

\subsection{Microbenchmarking Across Codecs}
\label{app:microbenchmark-across-codecs}
\begin{figure}
    \centering
    \includegraphics[width=0.8\linewidth]{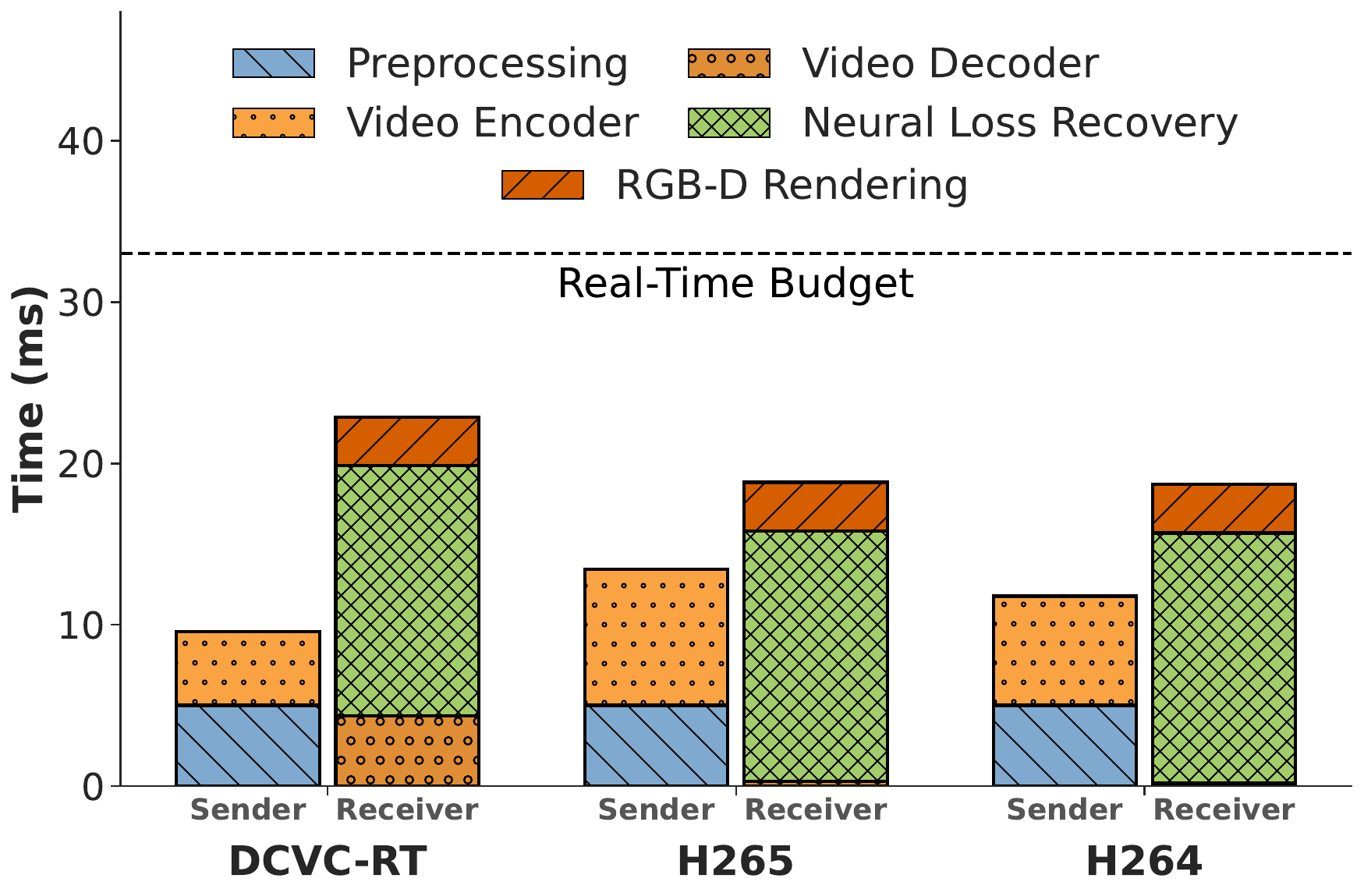}
    \caption{\sysname{} Micro-benchmarking across different underlying codecs on Nvidia RTX 5070. }
    \label{fig:microbenchmarking_volu}
    \vspace{-1em}
\end{figure}

\vspace{-0.5em}
Figure~\ref{fig:microbenchmarking_volu} presents a detailed breakdown of the end-to-end latency of \sysname{}. We report the results across DCVC-RT~\cite{dcvcrt}, H.265~\cite{h265}, and H.264~\cite{h264}  as the underlying video codec.
At the sender, preprocessing (including face cropping and background removal) takes $5.0$,ms, while encoding requires $4.57$–$8.43$,ms across codecs, resulting in a total of $9.57$–$13.43$ms. At the receiver, decoding takes $0.19$–$4.35$ms, the neural loss recovery module requires $15.5$ms, and RGB-D rendering takes $3.02$ms, for a total of $18.71$–$22.87$ms. Overall, the end-to-end pipeline supports streaming at over 30 FPS, demonstrating the practicality of \sysname{} for real-time volumetric videoconferencing.

\begin{figure*}[!hbtp]
    \centering
    \begin{subfigure}[b]{0.27\textwidth}
        \centering
        \includegraphics[width=1\textwidth]{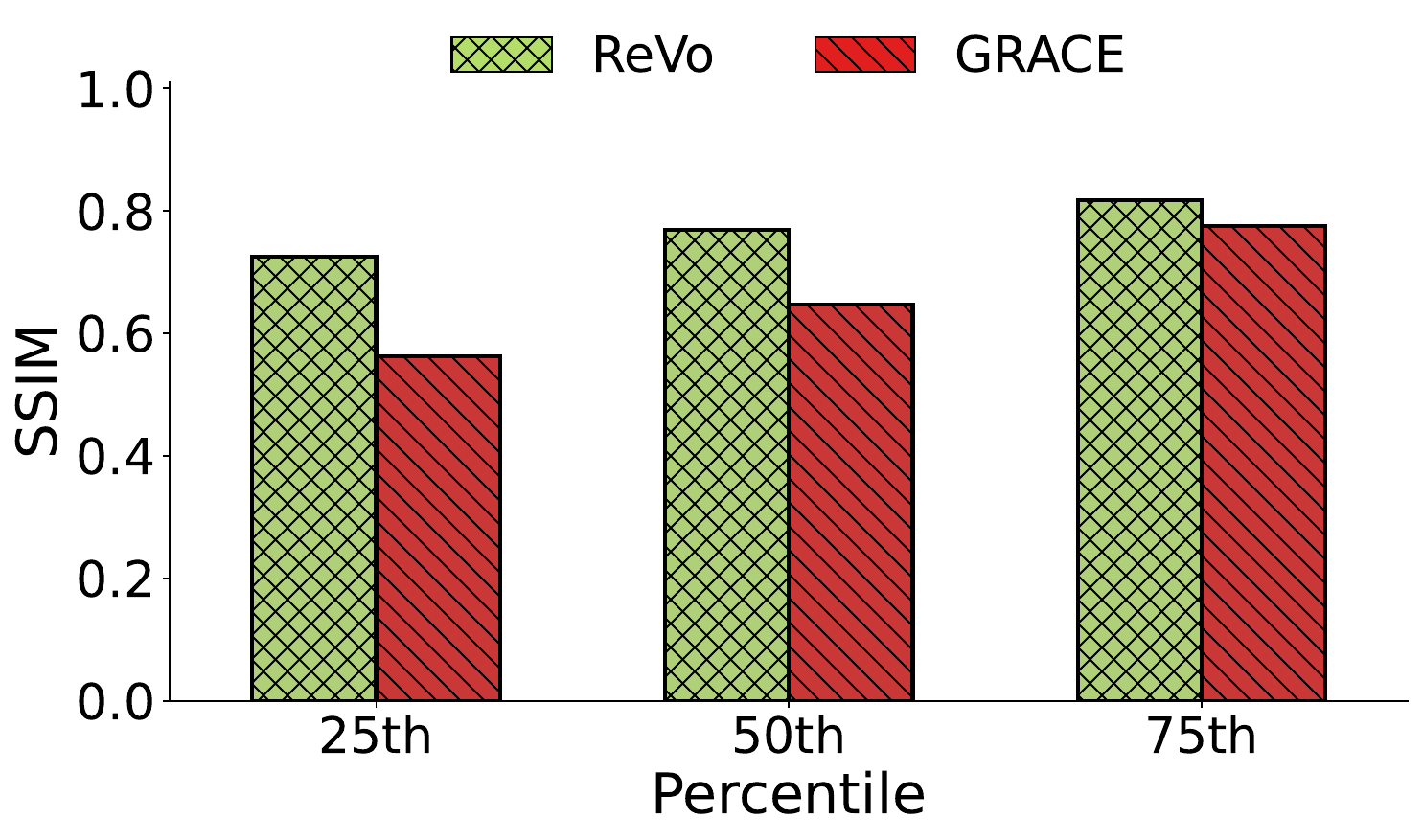}
        \caption{Cellular (RGB frames)}
        \label{fig:cell_comp_quality}
    \end{subfigure}
    \hfill
    \begin{subfigure}[b]{0.27\textwidth}
        \centering
        \includegraphics[width=\textwidth]{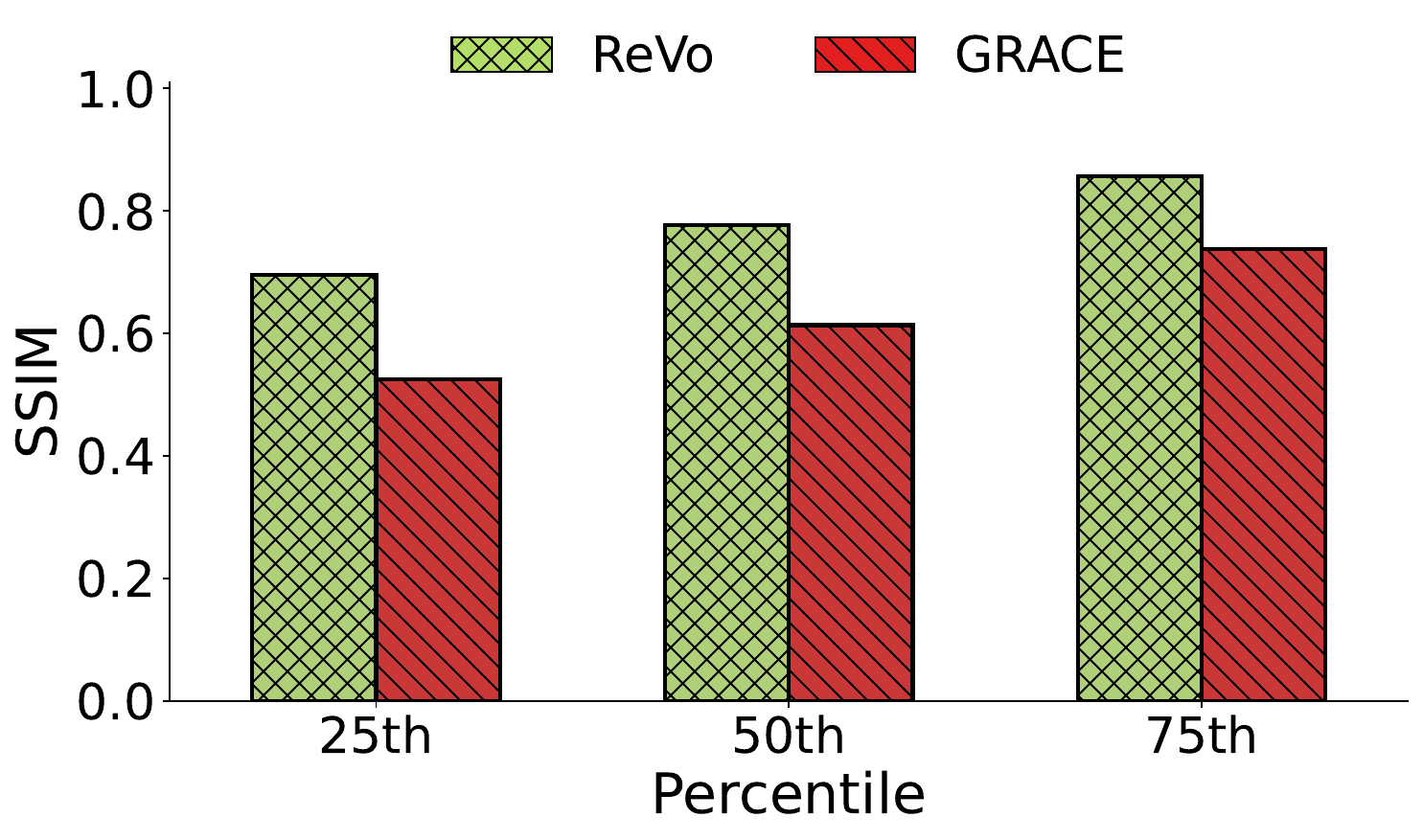}
        \caption{WiFi (RGB frames)}
        \label{fig:wifi_comp_quality}
    \end{subfigure}
    \hfill
    \begin{subfigure}[b]{0.27\textwidth}
        \centering
        \includegraphics[width=\textwidth]{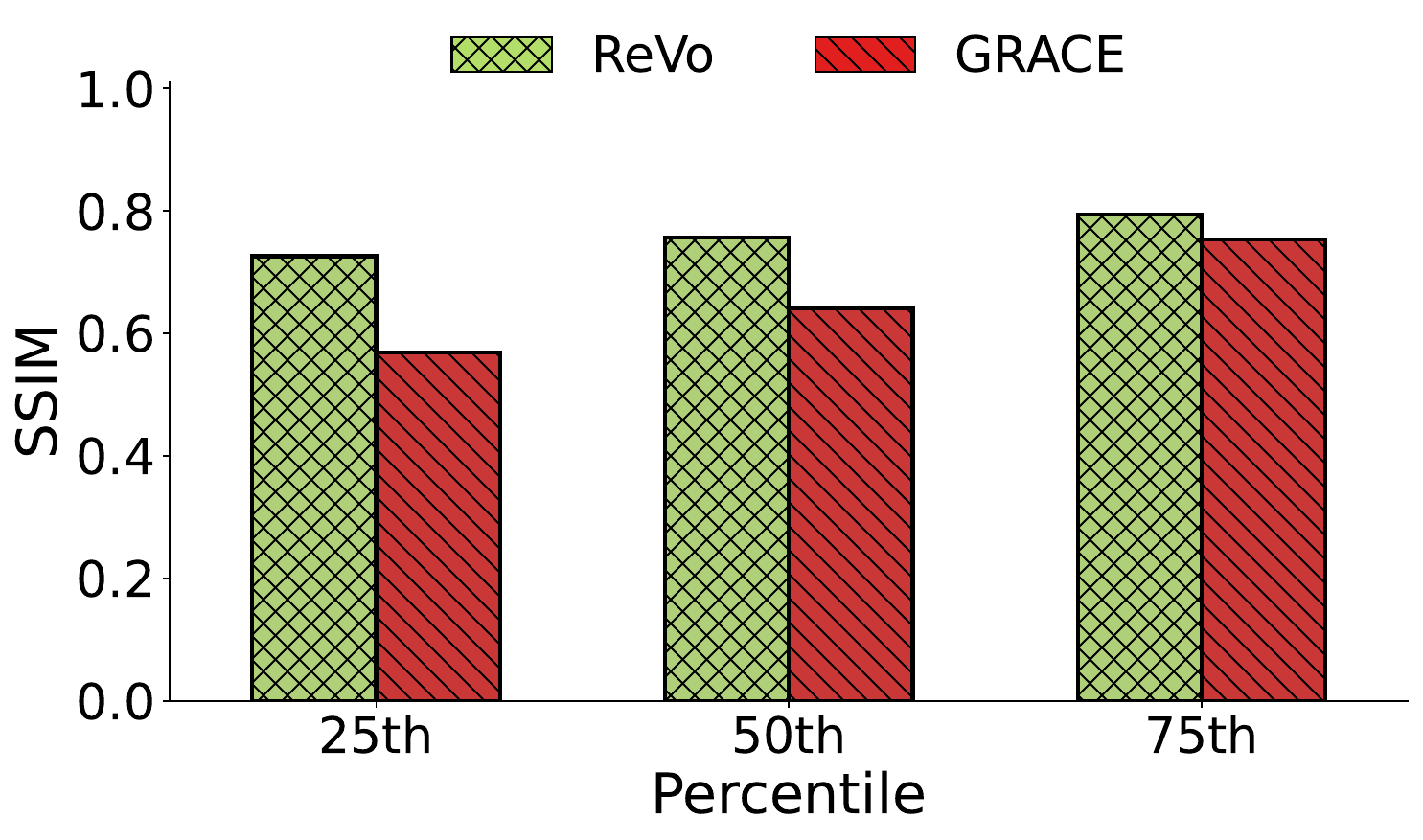}
        \caption{Ethernet (RGB frames)}
        \label{fig:eth_comp_quality}
    \end{subfigure}

    \vspace{0.2em}

    \begin{subfigure}[b]{0.27\textwidth}
        \centering
        \includegraphics[width=\textwidth]{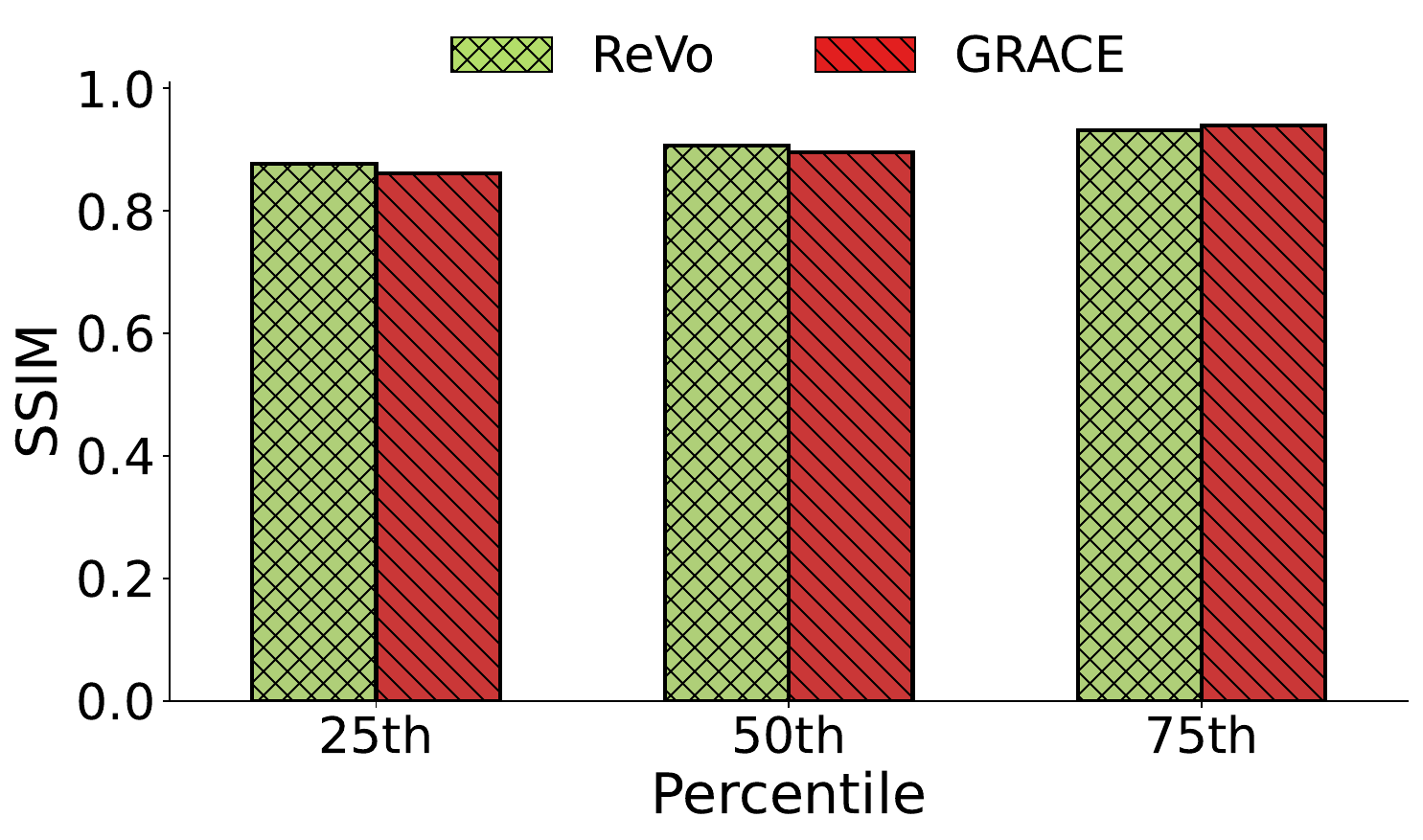}
        \caption{Cellular (Depth frames)}
        \label{fig:cell_comp_quality_depth}
    \end{subfigure}
    \hfill
    \begin{subfigure}[b]{0.27\textwidth}
        \centering
        \includegraphics[width=\textwidth]{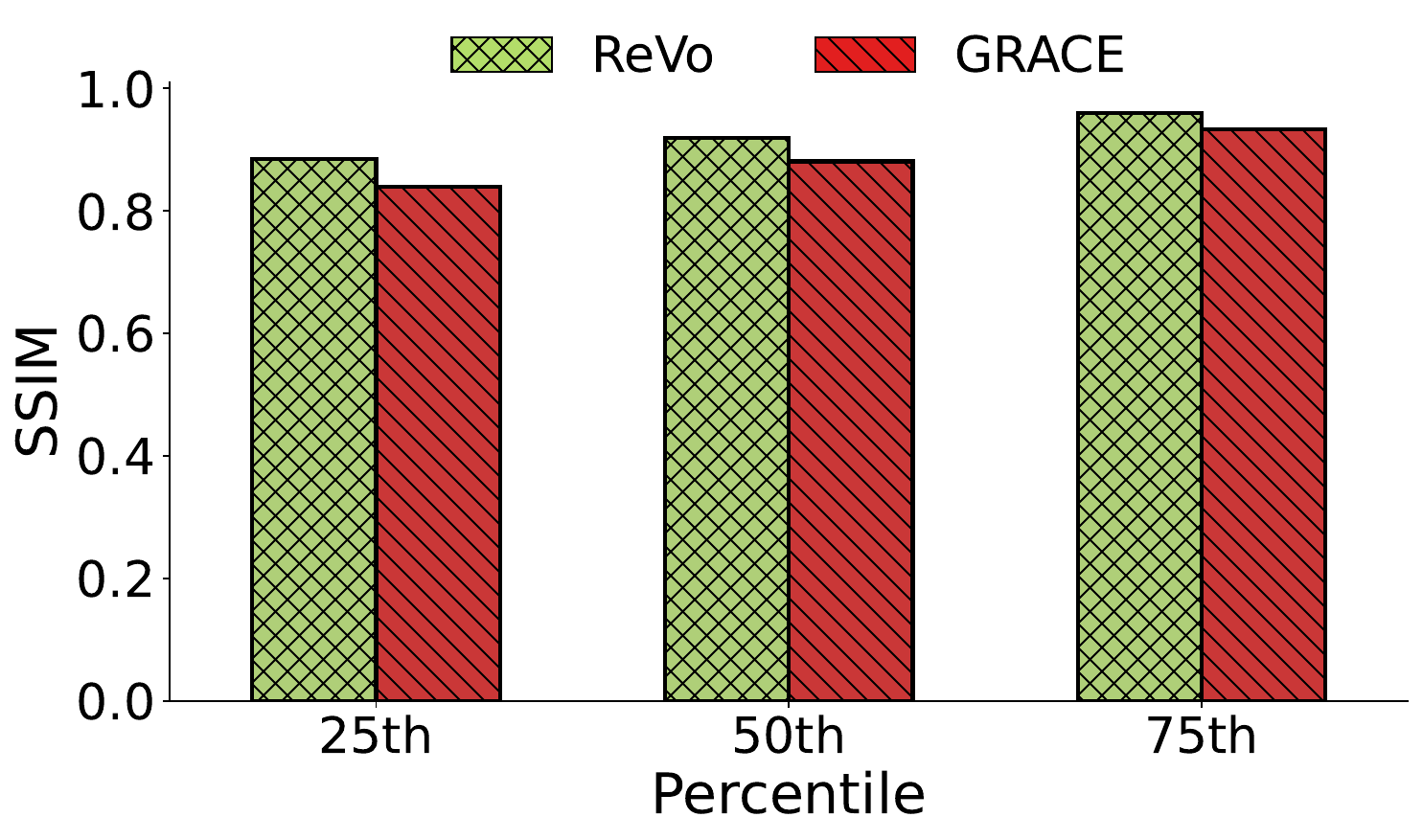}
        \caption{WiFi (Depth frames)}
        \label{fig:wifi_comp_quality_depth}
    \end{subfigure}
    \hfill
    \begin{subfigure}[b]{0.27\textwidth}
        \centering
        \includegraphics[width=\textwidth]{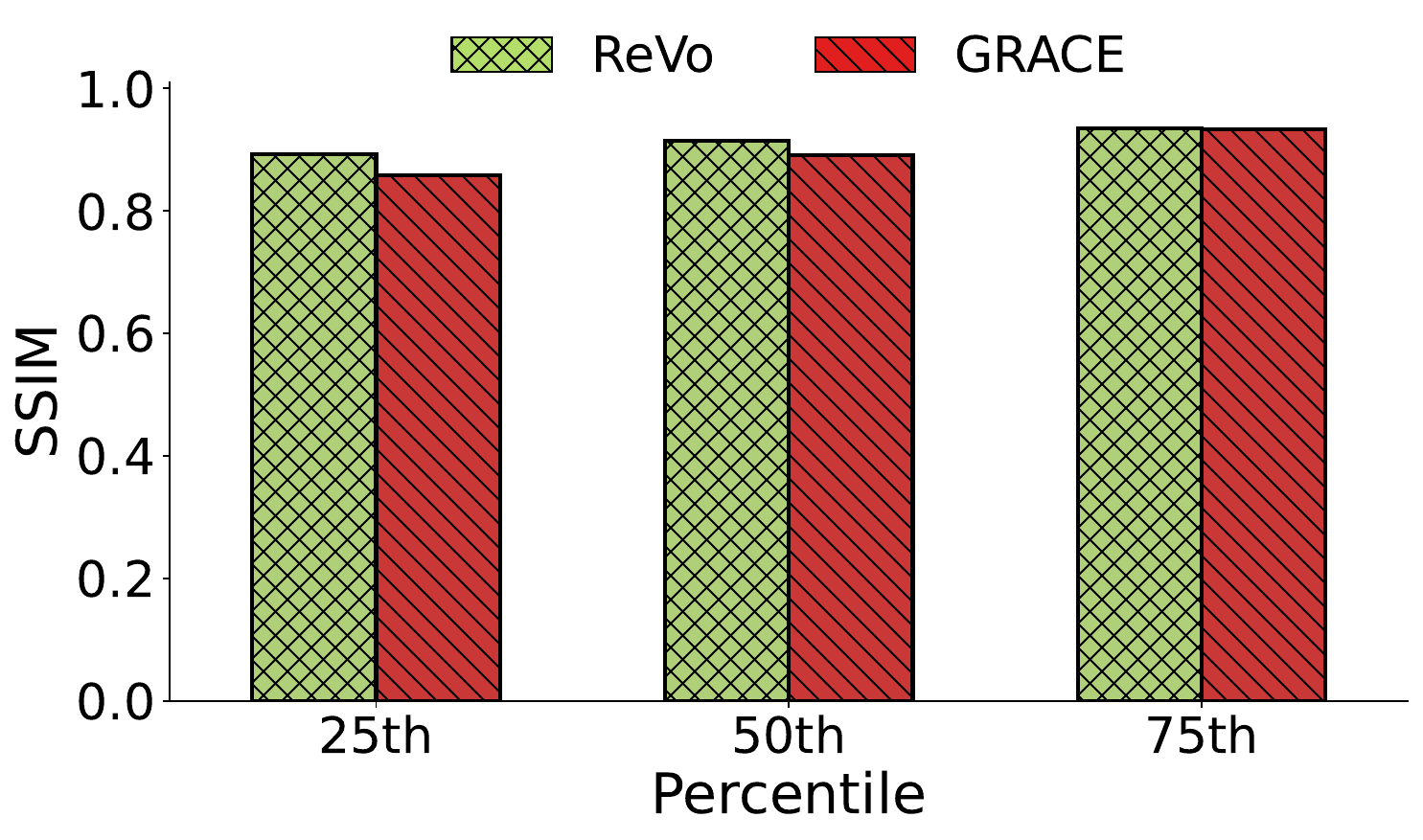}
        \caption{Ethernet (Depth frames)}
        \label{fig:eth_comp_quality_depth}
    \end{subfigure}
    \caption{SSIM comparison between \sysname{} and Grace at different positions of I-frame loss, measured at the 25th, 50th, and 75th percentiles. Results are reported for RGB frames ((a)--(c)) and depth frames ((d)--(f)) across cellular, WiFi, and Ethernet traces. \sysname{} consistently achieves higher reconstruction quality than Grace, demonstrating robustness to I-frame loss across diverse network conditions.}
    \label{fig:grace_comp_L7}
    \vspace{-1em}
\end{figure*}

%% file: A4-appendix.tex
\vspace{-1em}
\section{Additional Comparison with GRACE}
\label{app:grace-comparison}
\vspace{-0.5em}
As discussed in \S~\ref{sec:eval-baseline-comparison}, we implement \emph{GRACE} under the assumption that encoding and decoding can be performed in real time on an Nvidia RTX 5070 GPU. Additionally, to ensure stable operation we assume that I-frames are never lost, as their loss would cause the \emph{Grace} codec to fail. To approximate the effect of I-frame loss, we simulate this scenario by dropping all encoded P-frames within a Group of Pictures (GoP), i.e., applying $100\%$ loss to the P-frames following an I-frame (total $29$ P-Frames for GoP$=30$). We then compare the reconstruction quality of \emph{GRACE} and \emph{\sysname{}} at the time indices corresponding to such loss events.

Figure~\ref{fig:grace_comp_L7} shows the SSIM at the $25$th, $50$th, and $75$th percentiles for both RGB and depth frames across cellular, WiFi, and Ethernet traces. Across these network conditions, \sysname{} achieves median SSIM of $0.77$, $0.82$, and $0.76$ for RGB, and $0.91$, $0.93$, and $0.91$ for depth on cellular, WiFi, and Ethernet traces, respectively. In comparison, \emph{GRACE} achieves lower median SSIM of $0.65$, $0.61$, and $0.64$ for RGB, and $0.90$, $0.88$, and $0.89$ for depth across the same settings. Notably, the gap is more pronounced for RGB, where \sysname{} improves median SSIM by up to $0.21$ (on WiFi), while consistently matching or exceeding \emph{GRACE} for depth. These results demonstrate that \sysname{}’s cross-layer loss recovery significantly outperforms the application-layer ($L7$-only) recovery mechanism employed by \emph{GRACE}.


%% file: A5-appendix.tex
\begin{figure*}[!ht] 
    \centering
    \includegraphics[width=0.55\textwidth]{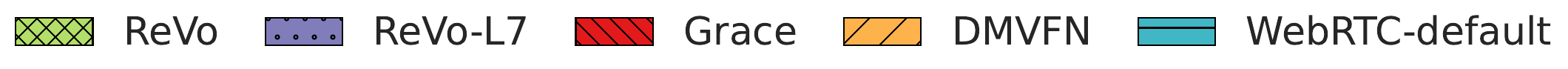}

    \begin{subfigure}[b]{0.3\textwidth}
        \centering
        \includegraphics[width=\linewidth]{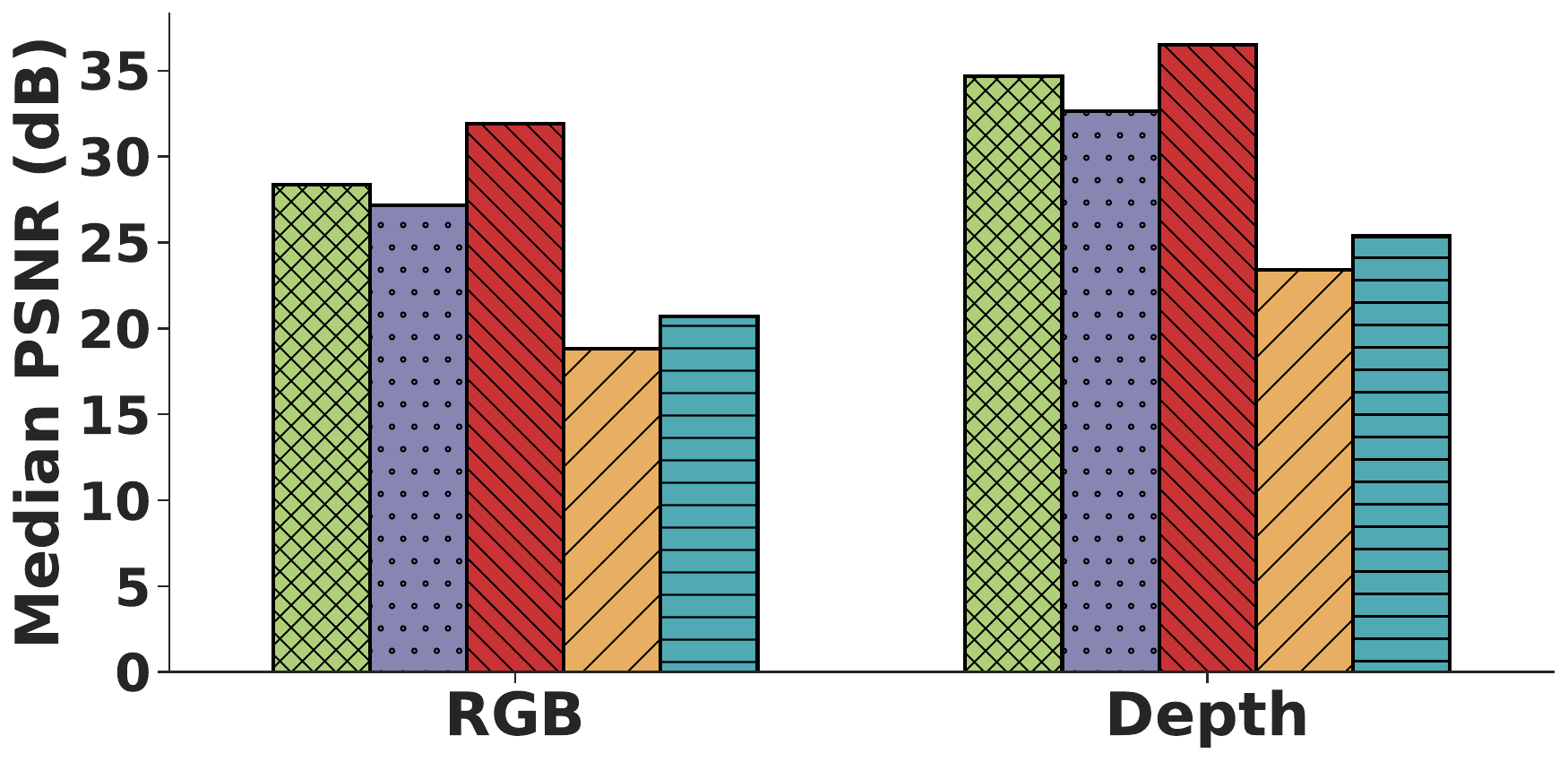}
        \caption{Cellular}
    \end{subfigure}
    \hfill
    \begin{subfigure}[b]{0.3\textwidth}
        \centering
        \includegraphics[width=\linewidth]{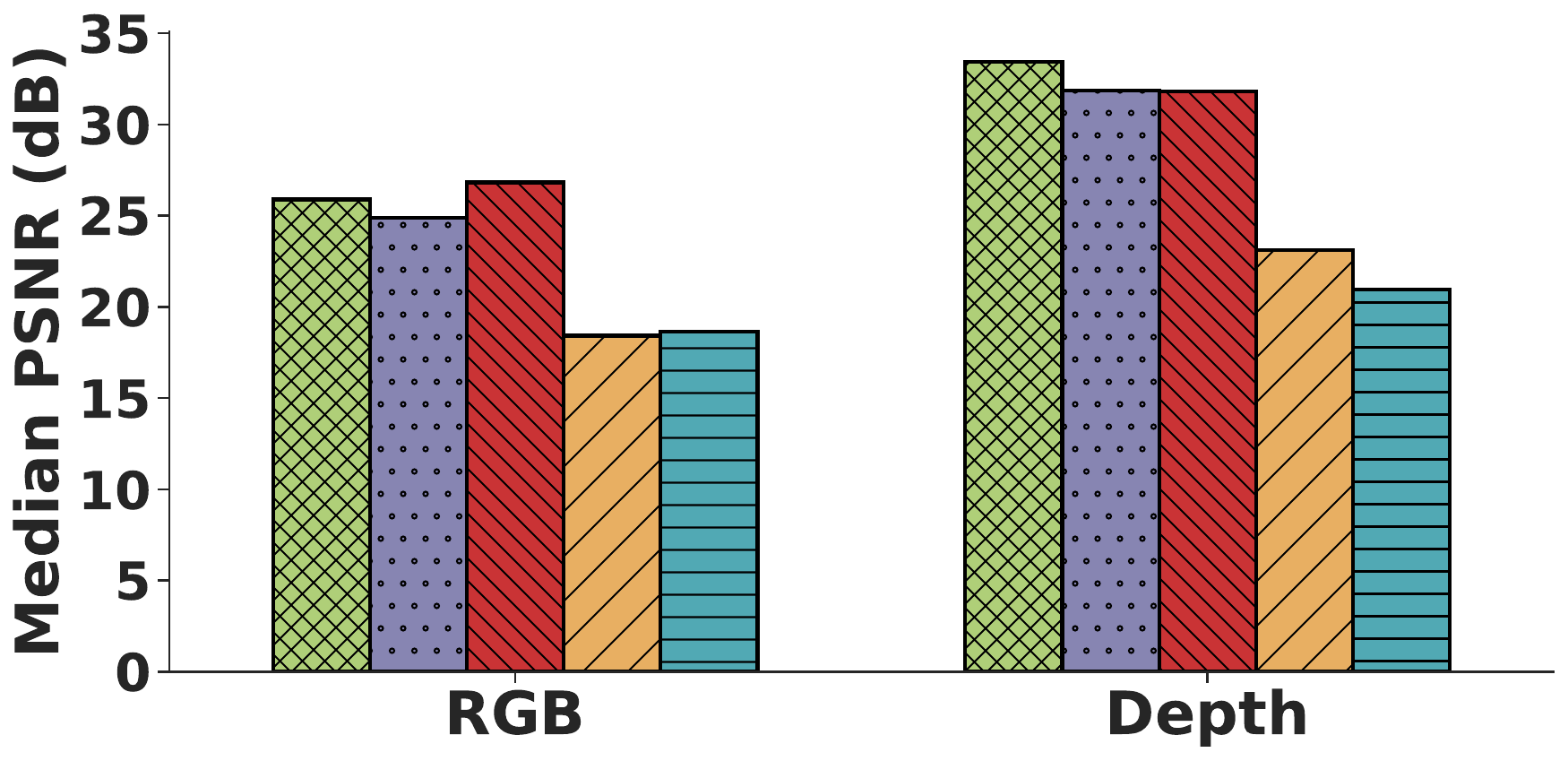}
        \caption{WiFi}
    \end{subfigure}
    \hfill
    \begin{subfigure}[b]{0.3\textwidth}
        \centering
        \includegraphics[width=\linewidth]{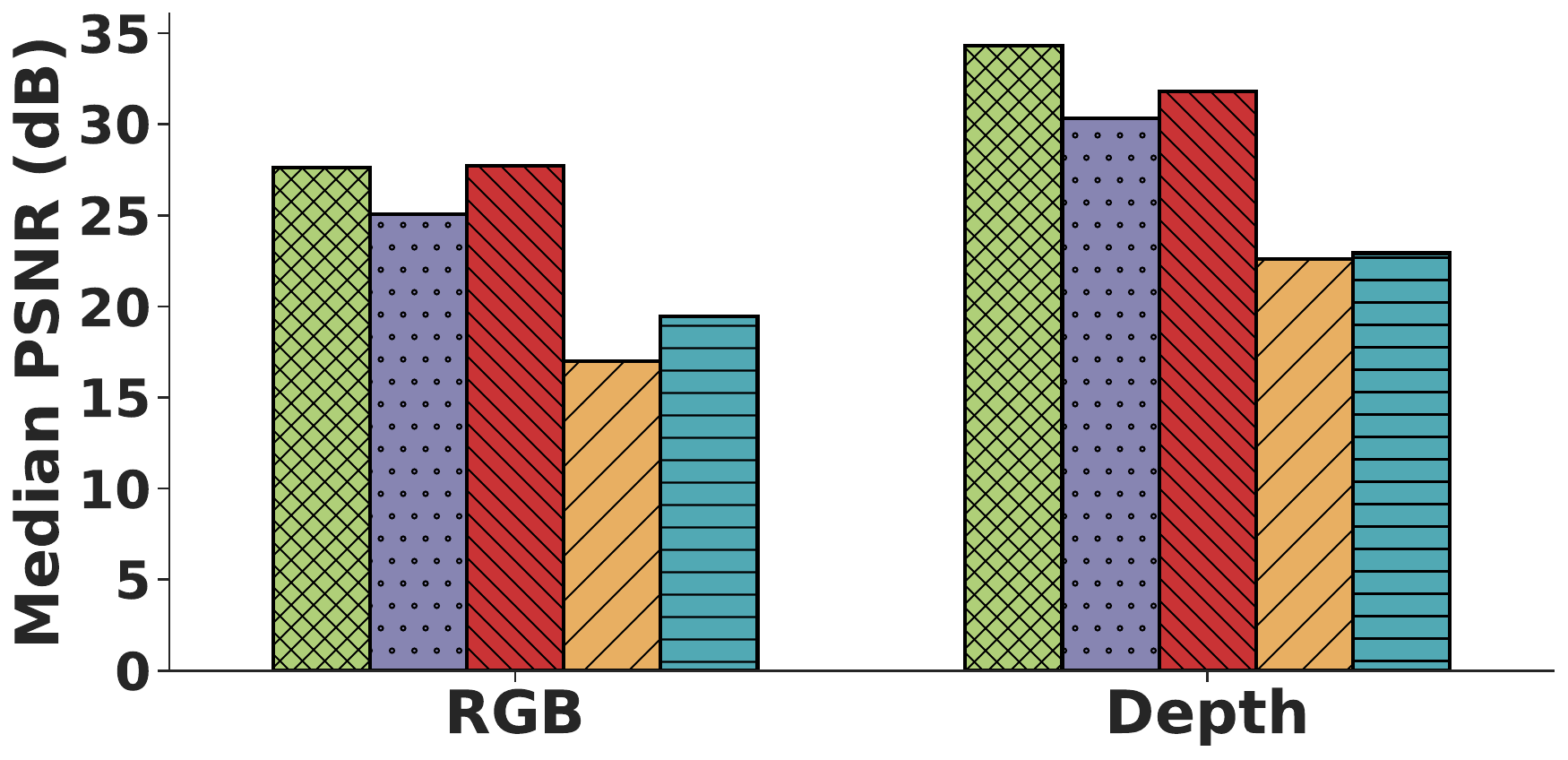}
        \caption{Ethernet}
    \end{subfigure}
    \vspace{-1em}
    \caption{Median PSNR (dB) RGB amd Depth Frames across (a) Cellular (b) WiFi and (c) Ethernet Traces. \sysname{} consistently achieves high PSNR(dB) across different network conditions.}
    \label{fig:psnr_l7_comp_median}

\end{figure*}

\section{Additional Comparison with L7 Baselines}
\label{app:additional_comp_l7}

We additionally compare \sysname{} against four L7 baselines in terms of the Peak Signal-to-Noise Ratio (PSNR)~\cite{psnr} metric: \emph{\sysname{}-L7}, \emph{DMVFN}, \emph{WebRTC-default}, and \grace{}. 

Similar to our SSIM results, \emph{DMVFN} and \emph{WebRTC-default} yield the lowest reconstruction quality across all network types, with median PSNRs between$ 16.98$~dB and $20.72$~dB for RGB, and $20.97$~dB to $25.37$~dB for depth. \emph{\sysname{}-L7} offers a moderate improvement, achieving median PSNRs between $24.89$~dB and $27.18$~dB for RGB, and $30.33$~dB to $32.63$~dB for depth. 

\sysname{} consistently outperforms these practical, real-time baselines across all network conditions. For RGB frames, \sysname{} attains median PSNRs of $28.37$~dB on Cellular, $25.9$~dB on WiFi, and $27.6$~dB on Ethernet. For depth frames, it achieves robust median PSNRs of $34.7$~dB, $33.4$~dB, and $34.3$~dB across the respective traces.

While \grace{} achieves the highest overall PSNR in the Cellular trace ($31.91$~dB for RGB, $36.50$~dB for depth), \sysname{} surpasses \grace{} in depth reconstruction quality on both WiFi ($33.44$~dB vs. $31.83$~dB) and Ethernet ($34.3$~dB vs. $31.79$~dB) traces. This confirms that \sysname{} delivers overall high visual quality while being practically deployable on desktop-grade GPUs.